\documentclass[epj]{svjour}
\usepackage{graphics,cite}

\usepackage{graphics} 
\usepackage{amsmath,amssymb} 
\usepackage{dsfont}

\newcommand{\ket}[1]{\ensuremath{|#1\rangle}} 
\newcommand{\bra}[1]{\ensuremath{\langle#1|}}

\newcommand{\braket}[2]{\langle#1|#2\rangle}

\newcommand{\abs}[1]{\left| #1 \right|} 
\newcommand{\avg}[1]{\left\langle #1 \right\rangle} 

\newcommand{\Sref}[1]{Section~\ref{#1}}
\newcommand{\sref}[1]{Sect.~\ref{#1}}

\newcommand{\sandsref}[2]{Sects.~\ref{#1} and \ref{#2}}

\newcommand{\Fref}[1]{Figure~\ref{#1}}
\newcommand{\fref}[1]{Fig.~\ref{#1}}

\newcommand{\frefs}[1]{Figs.~\ref{#1}}

\newcommand{\Eref}[1]{Equation~(\ref{#1})}
\newcommand{\eref}[1]{Eq.~(\ref{#1})}

\newcommand{\eanderef}[2]{Eqs.~(\ref{#1}) and (\ref{#2})}

\begin{document}

\title{Applied Bohmian Mechanics}

\author{Albert Benseny\inst{1} \and Guillem Albareda\inst{2} \and \'Angel S. Sanz\inst{3} \and Jordi Mompart\inst{4} \and Xavier Oriols\inst{5}\thanks{\email{xavier.oriols@uab.cat}}
}
\institute{
 Quantum Systems Unit, Okinawa Institute of Science and Technology Graduate University, 904-0495 Okinawa, Japan
 \and
 Departament de Qu\'imica F\'isica and Institut de Qu\'imica Te\`orica i Computacional, Universitat de Barcelona, 08028 Barcelona, Spain
 \and
 Instituto de F\'{\i}sica Fundamental (IFF-CSIC), Serrano 123, 28006 Madrid, Spain
 \and
 Departament de F\'{\i}sica, Universitat Aut\`{o}noma de Barcelona, 08193 Bellaterra, Spain
 \and
 Departament d'Enginyeria Electr\`{o}nica, Universitat Aut\`{o}noma de Barcelona, 08193 Bellaterra, Spain
 }
\date{Received: \today / Revised version: date}

\abstract{
Bohmian mechanics provides an explanation of quantum phenomena in terms of point-like particles guided by wave functions.
This review focuses on the use of nonrelativistic Bohmian mechanics to address practical problems, rather than on its interpretation.
Although the Bohmian and standard quantum theories have different formalisms, both give exactly the same predictions for all phenomena.
Fifteen years ago, the quantum chemistry community began to study the practical usefulness of Bohmian mechanics.
Since then, the scientific community has mainly applied it to study the (unitary) evolution of single-particle wave functions, either by developing efficient quantum trajectory algorithms or by providing a trajectory-based explanation of complicated quantum phenomena.
Here we present a large list of examples showing how the Bohmian formalism provides a useful solution in different forefront research fields for this kind of problems (where the Bohmian and the quantum hydrodynamic formalisms coincide).
In addition, this work also emphasizes that the Bohmian formalism can be a useful tool in other types of (nonunitary and nonlinear) quantum problems where the influence of the environment or the nonsimulated degrees of freedom are relevant.
This review contains also examples on the use of the Bohmian formalism for the many-body problem, decoherence and measurement processes.
The ability of the Bohmian formalism to analyze this last type of problems for (open) quantum systems remains mainly unexplored by the scientific community.
The authors of this review are convinced that the final status of the Bohmian theory among the scientific community will be greatly influenced by its potential success in those types of problems that present nonunitary and/or nonlinear quantum evolutions.
A brief introduction of the Bohmian formalism and some of its extensions are presented in the last part of this review.
}

\maketitle

\tableofcontents

\section{Introduction: additional routes are helpful}
\label{sec:intro}

Solving a particular physical problem has many similarities with making a trip. First, we have to decide which route to take. Most of the times getting the one recommended by a prestigious guide is enough. Sometimes other routes are even faster or allow us to see beautiful views of the countryside while driving.  Eventually, we can find unexpected roadblocks in the selected route and alternatives are mandatory.  A good knowledge of a particular territory implies that we are able to use different routes. When we know many routes (and the connections between them), traveling along this particular region has no mystery to~us.

In classical mechanics, for example, most of the times the recommended route is taking the \emph{Newtonian} one. In other occasions, because of the specific characteristics of the trip, it is better to take the \emph{Lagrangian}, the \emph{Hamiltonian}, or the \emph{Hamilton--Jacobi} routes \cite{o.goldstein2014book}.  Quantum mechanics is not different. Many times  practical problems are solved with the formalisms associated to the so-called \emph{standard} route, also known as the orthodox or Copenhagen\footnote{The term \emph{Copenhagen interpretation} refers to a set of rules for interpreting quantum phenomena devised by Born,  Bohr, Heisenberg and others \cite{o.Born1926,o.Heisenber1925c,o.Heisenber1925b}. Note, however, that some people argue that the fathers of the \emph{orthodox} interpretation contradict each other on several important issues \cite{o.valentini2009Solvay}.} route. The \emph{standard} route itself has many subroutes.
For example, the quantum harmonic oscillator problem is cleanly and easily studied with the raising and lowering operators of the (Heisenberg) matrix formulation, while many other problems are better addressed directly with the (Schr\"{o}dinger) wave function formalism\cite{o.cohen1978book}.
Another relevant route is the Feynman path integral formulation which is rarely the easiest way to approach a nonrelativistic quantum problem, but which has innumerable and very successful applications in quantum statistics and quantum field theory \cite{o.feynmann1965book}. Certainly, having a good knowledge of all possible routes (and their connections) in the quantum territory is very helpful when facing a particular quantum problem.  However, there are routes that do not appear usually in the guides. One of these unexplained routes is Bohmian mechanics\footnote{We have chosen the name \emph{Bohmian mechanics} when referring to the work of Louis de Broglie and David Bohm because it is perhaps one of the most widespread names nowadays~\cite{o.Bohmian1996book,o.oriols2011book,o.durr2009book,o.durr2012book}. We are not completely satisfied with this choice because it seems to imply that Bohmian mechanics is not exactly the same as quantum mechanics.
We would prefer a title like \emph{applied quantum mechanics with trajectories}, but this possibility would be misleading, since quantum hydrodynamics, Feynman paths integrals and others would fit under that title.}.

The Bohmian formalism was proposed by Louis de Broglie \cite{o.dB_AnnPhys,o.debroglie1927b} even before the Copenhagen explanation of quantum phenomena was established. Bohmian mechanics provides an explanation of quantum phenomena in terms of point-like particles guided by waves\footnote{Quoting Bell\cite{o.bell2004book}: \emph{While the founding fathers agonized over the question `particle' or `wave', de Broglie in 1925 proposed the obvious answer `particle' and `wave'. Is it not clear from the smallness of the scintillation on the screen that we have to do with a particle? And is it not clear, from the diffraction and interference patterns, that the motion of the particle is directed by a wave? De Broglie showed in detail how the motion of a particle, passing through just one of two holes in screen, could be influenced by waves propagating through both holes. And so influenced that the particle does not go where the waves cancel out, but is attracted to where they cooperate. This idea seems to me so natural and simple, to resolve the wave-particle dilemma in such a clear and ordinary way, that it is a great mystery to me that it was so generally ignored.} }. One object cannot be a wave and a particle simultaneously, but two can. In the fifties, David Bohm \cite{o.Bohm1952a,o.Bohm1952b,o.Bohm1953b} clarified the meaning and applications of this explanation of quantum phenomena showing, for example, how  the measurement process can be explained as another type of interaction, without any \emph{ad-hoc} rule for it.

In the previous paragraph we made an analogy between routes and formalisms. The formalism of a theory is a set of mathematical tools used to explain and make predictions for a series of phenomena. Apart from its formalism, each theory also includes an interpretation, which describes how the elements of the formalism are related to the natural objects. The \emph{scientific method} developed in the 17th century provided a clear difference between the roles of the formalism (related to physical discussions and empirical evidences) and the interpretation (related to more metaphysical-like discussions). By its own construction, the formalism of Bohmian mechanics does exactly reproduce all experimental results dealing with (nonrelativistic) quantum phenomena \cite{o.oriols2011book,o.durr2009book,o.bell2004book,o.Holand1993,bohm-hiley-bk,o.durr2012book}. There are many scientists who defend that, after ensuring that a theory reproduces the empirical data of a laboratory, metaphysical discussions on the \emph{meaning} of the formalism
become unnecessary. Others, however, argue that such discussions provide a \emph{deeper} understanding on how the theory works and, ultimately, how nature is built.  Historically, the Bohmian theory has been involved in many metaphysical disputes about the role of the waves and the particles when trying to provide a hierarchy between different quantum theories. As far as one looks for a formalism that reproduces experimental results, all quantum theories (standard, Bohmian, many-worlds, etc.) are perfectly valid. The relevant discussion in this review is the practical usefulness of the Bohmian formalism in our everyday research, not its ontological implications\footnote{We think that, for general purposes, a better theory is that one with a better ontology. However, for the practical applications that we discuss in this review, the relevant point is the usefulness of each formalism in solving practical problems, not its ontological coherence.}. We will review the efforts done in the literature in
different research fields to solve practical quantum problems using the Bohmian route: what has already been done and also what can be done.

For practical computations, the knowledge of several routes and their connection is always helpful when traveling trough the quantum territory. Hence a pertinent question is ``\emph{Why is the pilot wave picture [Bohmian mechanics] ignored in textbooks?''} \cite{o.bell2004book}. The answer seems to be that many people believe that Bohmian mechanics is not useful in practical applications.
For example, Steven Weinberg wrote in a private exchange with Sheldon Goldstein \cite{o.Weinberg}:
\textit{``In any case, the basic reason for not paying attention to the Bohm approach is not some sort of ideological rigidity, but much simpler --- it is just that we are all too busy with our own work to spend time on something that doesn't seem likely to help us make progress with our real problems.''}
Researchers with the opinion that Bohmian mechanics has a limited utility argue that, apart from computing the wave function, Bohmian mechanics requires tracking a set of trajectories that, at the end of the day, will exactly reproduce the time-evolution of the wave function, which was already known. Then, \emph{what is the utility of the extra effort for computing Bohmian trajectories?}
In fact, this argument is based on a poor understanding of the abilities of the Bohmian theory.
The best way of refuting this claim is to present (counter) examples showing that the Bohmian route can be indeed very useful in some scenarios.

\subsection{The structure of the review}

This review introduces a discussion on the practical usefulness of the Bohmian theory in our daily research activity focused on three main goals.
First, showing with explicit numerical examples the applicability of the Bohmian formalism in solving practical problems in many different fields.
The second goal is presenting a brief introduction on (nonrelativistic) Bohmian mechanics for newcomers.
The third goal is pointing out that \emph{Bohmian mechanics is much more than reproducing wave functions with trajectories!}  and emphasizing the capabilities of the Bohmian formulation of quantum mechanics to deal with nonunitary quantum evolutions such as many body interactions, measurements, decoherence, etc.

In order to fulfil these goals we have chosen an unusual structure for this review.
One could expect first a discussion on the postulates of Bohmian mechanics followed by a development of the mathematical formalism for computing trajectories to finally present some examples of its usefulness.
However, we have selected the inverse order.
The reason is because we are interested in convincing the reader about the utility of Bohmian mechanics with \emph{practical examples}.
Thus, we first directly guide the reader to such examples in the largest part of this review, consisting of \sandsref{sec:application1}{sec:application2}.
Later on, the reader can get a deeper knowledge of the mathematical formalism, its various subroutes, or interpretative issues in \sref{sec:formalism} or somewhere else \cite{o.oriols2011book,o.durr2009book,o.Holand1993,o.bell2004book,o.durr2012book,bohm-hiley-bk}.

We have divided the examples shown in this review in two sets.
First, in \sref{sec:application1}, we show some examples of Bohmian solutions whose development or predictions are valid for some very specific research fields, and then, in \sref{sec:application2}, we discuss Bohmian solutions which can be applied in many different fields of research.

The first examples of \sref{sec:application1} emphasize how visualizing some quantum problems in terms of trajectories guided by waves can be very enlightening, even after computing the wave function. See, for example, \sref{sec:atoms} where it is shown, through Bohmian trajectories, that an accurate description of the adiabatic transport of cold atoms in a triple well potential could require relativistic corrections. Some practical Bohmian approximations to the many-body problem in electron-nuclei coupled dynamics are presented in \sref{sec:electron-nuclei}. \Sref{sec:light-matter} is devoted to the investigation, through Bohmian trajectories, of the interaction of intense light fields with matter. Furthermore, in \sref{sec:electron-transport}, quantum electron transport with Bohmian trajectories is presented with a discussion on how the measurement of high-frequency electrical currents can be modeled with Bohmian trajectories.
All this review is focused only on nonrelativistic quantum mechanics for massive particles.
However, the understanding of nonrelativistic quantum phenomena in terms of trajectories does also show a path to rethink some problems of relativistic quantum mechanics, quantum cosmology, or even classical optics. Some of these last examples are mentioned in \Sref{sec:beyond}.

In \sref{sec:application2} we group topics of research that are transversal to many research fields. For example, the works in \sandsref{sec:scattering}{sec:elastic} show how scattering and collisions (which commonly appear in almost all research fields dealing with quantum problems) are understood with the use of Bohmian trajectories. A Bohmian approximation to the many-body problem is presented in \sref{sec:manybody} with the use of conditional Bohmian wave functions.
\Sref{sec:measurements} discusses the Bohmian formalism for quantum measurements and its application to recent experimental progress on the measurement of local velocities, as well as the old tunneling time problem.
\Sref{sec:chaos} discusses how quantum chaos can be illuminated with Bohmian mechanics, while some Bohmian ideas on how the classical world emerges from a quantum one are presented in \sref{sec:quantum-classical}.

After these two sets of examples on Bohmian applications, in \sref{sec:formalism}, we discuss the original routes, i.e. the formalism, opened by the Bohmian theory. For example, the trajectories can be computed from the Schr\"odinger  or from the Hamilton--Jacobi equations. See the mathematical differences in \sandsref{sec:Analytical}{sec:Hamilton-Jacobi}. We also discuss the complex action formalism in \sref{sec:complex action}. Afterwards, in \sref{sec:conditional}, we present the  so-called conditional wave function formalism, which has many potential applications for practical computations since it provides a \emph{natural} bridge between the high-dimensional (computationally inaccessible) configuration space and the physical (ordinary) space.
We provide two additional subsections discussing how expectation values can be extracted from Bohmian mechanics.  One possibility,
briefly reviewed in \sref{sec:pointer}, is getting the expectation values directly by averaging the (Bohmian) position of a pointer.
Another possibility, shown in \sref{sec:operators}, is using the \emph{standard} operators.
We end this section with a summary of the formalism  in \sref{sec:postulates}.

In any case, in spite of the unusual structure of this review, Sects. \ref{sec:application1}, \ref{sec:application2}, and \ref{sec:formalism} (and their subsections) have been written independently and can be read in any order. Back and forth from applications to formalism.
Finally, \sref{sec:conclusions} contains the final remarks of this review.

\subsection{Only the tip of the iceberg has been investigated}

We would like to stress that, in fact, most of the works mentioned above deal only with the (unitary) time-evolution of the Schr\"odinger equation.
In that case, Bohmian mechanics coincides with quantum hydrodynamics in many aspects.
The hydrodynamic route, which was initially developed by Madelung \cite{o.Madelung}, also deals with the concept of local velocity fields but does not explicitly assume individual trajectories.
Therefore, many parts of \sref{sec:application1} and in \sref{sec:application2} can also be understood as a review on quantum hydrodynamics.
On the other hand, other parts of the review discuss the nonunitary evolutions of a quantum subsystem. Then, the Bohmian explanation provides a completely different route to the hydrodynamic one. Let us further develop this last point, which remains mainly unexplored by the scientific community.

As we have already indicated, many practical quantum problems require the knowledge of the unitary evolution of a quantum system. The standard route solves this type of problems by looking for the solution of the Schr\"odinger equation. When the Bohmian route is selected, these problems are solved, for example,  by computing the trajectories that reproduce the previous (unitary) evolution of the wave function (or similar techniques mentioned above). The hydrodynamic route follows an identical path for this type of problems. Although the Bohmian contribution can be of great relevance, these practical quantum problems imply (somehow) a short trip through the quantum territory.

Other practical problems, however, imply taking a much longer trip. For example, let us imagine that we want to study a system formed by a few interacting atoms. The exact solution of the wave function is computationally inaccessible, what is known as the many-body problem.
Since almost all quantum problems of interest deal with many degrees of freedom, \emph{can the Bohmian route help in the many-body problem of the Schr\"odinger equation?} The answer is yes. Further discussion on this topic can be found in \sref{sec:manybody} and \sref{sec:conditional}. Success on such direction would produce an important impact on many different research fields. In any case, it seems obvious that the success of this approach will require a significant effort to connect Bohmian and standard many-body approximations.

Another (but somehow connected) type of problems implying a long trip is the study of open quantum systems, whose solutions are inaccessible from the Schr\"odinger equation. An open quantum system can be viewed as a distinguished part, a quantum subsystem, of a larger closed quantum system. The other subsystem is the environment, the measuring apparatus, etc.\footnote{The roles of measurement, decoherence or environment are interchangeable in our present qualitative discussion. They all refer to the interaction of the system with the rest of the universe.} The quantum subsystem does no longer follow a unitary evolution. Some times even the superposition principle is no longer valid in such a subspace. For example, let us imagine that we are interested in predicting the total current measured in an electronic quantum-based device. Following the standard route, whenever the interaction between the electrons of the quantum device and those of the measuring apparatus are relevant, a second law (different from the Schr\"odinger equation)
is used, i.e., the so-called \textit{collapse} of the wave function \cite{o.Born1926}. This second law requires a new nonunitary operator (different from the Hamiltonian present in the first law) to \emph{encapsulate} all the interactions of the quantum systems with the rest of the particles (including the ammeter, the cables, the environment, etc). The addition of this postulate brings many questions with it. \emph{Which is the operator that determines the (nonunitary) evolution of the wave function when measuring the total current?} \emph{Is this measurement process ``continuous'' or ``instantaneous''? Does it cause a ``weak'' or a ``strong'' perturbation on the wave function?}
The answers to these questions are certainly not simple. Over the years, physicists have identified the operators, by developing instincts on which are the effects of measurements on the wave function. Let us fully clarify that we are not interested on metaphysical discussions about quantum measurement.
We are only interested in practical computations (of the total current in the previous example) and how useful the different available formalisms are.
The standard quantum formalism has an extraordinary ability to provide very accurate and successful predictions on many types of measurements,
but for some particular problems where the role of the apparatus is not so obvious, insights provided by other routes can be very useful.

In the Bohmian formalism the measurement process is treated just as any other type of interaction. All degrees of freedom of the quantum system, the measuring apparatus and the  environment are present in the many-particle wave function and in the many-particle trajectory. The Schr\"odinger equation (in this larger configuration space) determines the time evolution of the many particle wave function of \emph{everything}. By construction, the Bohmian as well as the other more traditional approaches produce the same statistical predictions of any type of measurements \cite{o.durr2004equilibrium,o.oriols2011book,o.durr2009book,o.durr2012book}.
However, they may seem to follow two very different mathematical paths when dealing with quantum phenomena that imply open quantum systems with nonunitary evolutions such as measurements, decoherence, etc.
The standard separation between what is defined as a quantum system and what is the apparatus is not needed within the Bohmian formalism.\footnote{See for instance an enlarged discussion on Bohmian measurements in \sref{sec:pointer} and its practical application to the computation of THz electrical currents in \sref{sec:electron-transport}.}

Whether or not the Bohmian  route can be useful in quantum problems dealing with nonunitary and nonlinear evolutions remains almost unexplored in the literature. The recent experimental and theoretical interest on ultrafast dynamics, nanometric manipulation of quantum particles, weak measurements, etc., suggest that the intrinsec characteristics of the Bohmian route (with a microscopic description on how the system interacts with the measuring apparatus) can be very useful. We also discuss along this review the quite novel concept of the Bohmian conditional wave function. There is, therefore, much more to explore along the Bohmian route.

\section{Applications to forefront research fields}
\label{sec:application1}

In this section and in order to show the usefulness of the Bohmian formalism, we will discuss its use to address some particular problems in different forefront research fields in physics ranging from matter-wave transport via tunneling to nanoelectronics. More transversal problems such as those involved typically in chemistry, e.g., reactive scattering and elastic collisions, nonlinear dynamics, and conceptually fundamental problems in physics, e.g., quantum measurement, decoherence, and the quantum-to-classical transition, will be discussed in detail in Sec.~\ref{sec:application2}. We will also present at the end of this section a brief discussion on the application of Bohmian mechanics to scenarios beyond those governed by the spinless nonrelativistic Schr\"odinger equation.

The central problem in simulating a many-body quantum system comes from the fact that its wave function lives in a $3N$-configurational space, being $N$ the number of particles.
Therefore, even the simulation of the dynamics of the simplest atom, namely the hydrogen atom, requires the numerical integration of the Schr\"odinger equation in six dimensions, three for the electron and three for the proton. At present, our best computers are able to numerically solve the Schr\"odinger equation at most in five dimensions \cite{parker_uv_2009}, which means that even the numerical integration of a two-particle three-dimensional system is out of our computational scope. Thus, to numerically solve many-body quantum systems two approaches are feasible: (i) to reduce the number of dimensions being simulated, e.g., to deal with one or two-spatial dimensions, to recall some symmetries of the problem, to apply the Born--Oppenheimer approximation, etc.; or (ii) to use coupled-conditional wave functions each of them describing one workable partition of the whole quantum system. In the following, we will show how Bohmian quantum mechanics can be used in several forefront fields of research. With this
aim, we will consider next either a single particle system or a many-body quantum system where any of the two previous approaches has been applied.

\subsection{Ultracold atoms: matter wave transport}
\label{sec:atoms}

Ultracold neutral atoms are atoms whose temperature is typically below some tenths of microkelvin, such that their quantum-mechanical properties become relevant, i.e., their dynamics is governed by the (nonrelativistic) Schr\"o\-din\-ger equation. Magneto-optical traps are commonly used to reach such low temperatures while further manipulation can be achieved by means of the dipole force of focused far-detuned laser beams.
In this context, single neutral atoms can be stored, for instance, in optical dipole traps created with microlens arrays \cite{schlosser_scalable_2011}. Alternatively, stationary light fields can be used to create single-trap-occupancy optical lattices from a Bose-Einstein condensate via the Mott insulator transition \cite{greiner_quantum_2002}.
Interest in the physics of ultracold atoms comes from both academic and practical perspectives.
Ultracold atoms have become a fundamental system to test the principles
of quantum mechanics and condensed-matter physics \cite{lewenstein_ultracold_2012}, e.g., for understanding quantum phase transitions, bosonic superfluidity, many-body spin dynamics, Efimov states, quantum magnetism, etc. However, they also constitute the building blocks of future devices for quantum engineering technologies like quantum metrology, quantum simulation, and quantum computation.

\begin{figure}
\centerline{ \resizebox{0.99\columnwidth}{!}{\includegraphics{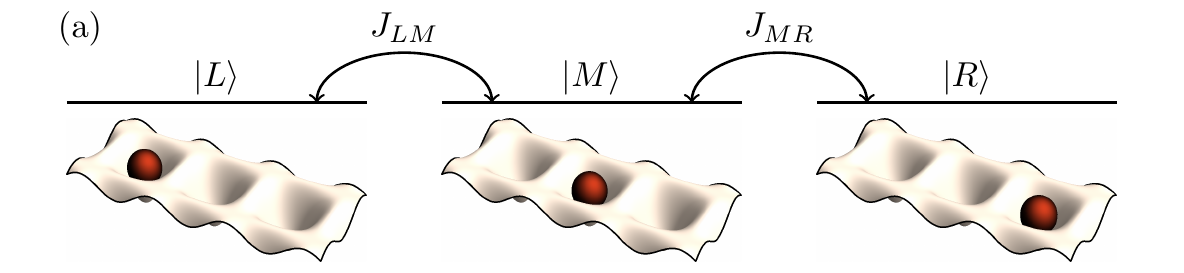} } }
\centerline{ \resizebox{0.99\columnwidth}{!}{\includegraphics{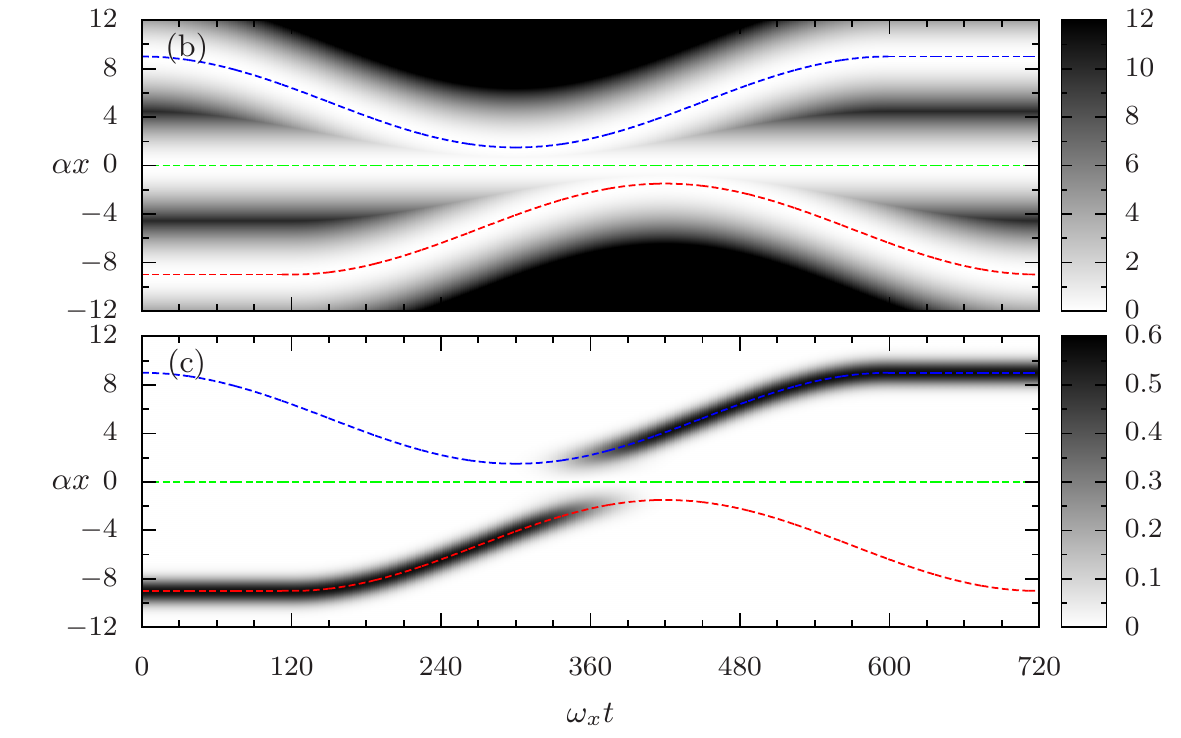} } }
\caption{(a) Sketch of three-level atom-optical system for a single atom in a triple-well potential. For the definition of states and couplings see the text. (b) Temporal evolution of the trapping potential with the dashed color lines indicating the trap centers. (c) Density probability $\left| \psi (x,t) \right|^2$ showing the spatial adiabatic passage of a single neutral atom between the outermost traps of the triple-well potential.}
\label{fig:SAP}
\end{figure}

In ultracold atom physics, Bohmian mechanics has been applied to investigate the adiabatic transport of a single atom between the outermost traps of a system formed by three identical traps \cite{benseny_need_2012}, see Fig.~\ref{fig:SAP}(a). In the adiabatic regime, where the system's parameters, such as the tunneling rates, are smoothly varied in time, there is one energy eigenstate of the system, the so-called spatial dark state, that reads $\ket{D(\theta)} = \cos \theta \ket{L}- \sin \theta \ket{R}$, being $\ket{L}$ and $\ket{R}$ the localized ground vibrational states for left and right traps, respectively, and $\tan \theta = J_{LM}/ J_{MR}$ with $J_{LM}$ and $J_{MR}$ being the tunneling rates between left-middle and middle-right traps, respectively. Note that direct tunneling between the outermost traps is neglected. Let us assume that an individual atom is initially located in the ground vibrational state of the left trap and that one has the ability to move the trap centers such that the tunneling rates can be temporally varied at will. Thus, if one approaches and separates first the middle and right traps and, with a certain time delay, the middle and left traps are also approached and separated, as indicated in Fig.~\ref{fig:SAP}(b), it becomes possible to adiabatically transfer the atom from the left to the right trap following $\ket{D(\theta)}$ and without populating the middle trap, see Fig.~\ref{fig:SAP}(c). This robust and efficient transport process is called spatial adiabatic passage \cite{eckert_three-level_2004} and is the atom-optics analog of the well-known quantum optical Stimulated Raman adiabatic passage (STIRAP) technique \cite{bergmann_coherent_1998}. It was argued in Ref.~\cite{rab_spatial_2008} that, as in the adiabatic limit the dark state $\ket{D(\theta)}$ does not involve the middle trap, transport takes place directly from the left to the right trap, which is known as the transport-without-transit paradox:
\begin{quote}
{\it ``Classically it is impossible to have transport without transit, i.e., if the points
1, 2, and 3 lie sequentially along a path then an object moving from 1 to 3
must, at some time, be located at 2. For a quantum particle in a three-well
system it is possible to transport the particle between wells 1 and 3 such that
the probability of finding it at any time in the classical accessible state in well
2 is negligible.''}
\end{quote}

\begin{figure}
\centerline{ \resizebox{0.99\columnwidth}{!}{\includegraphics{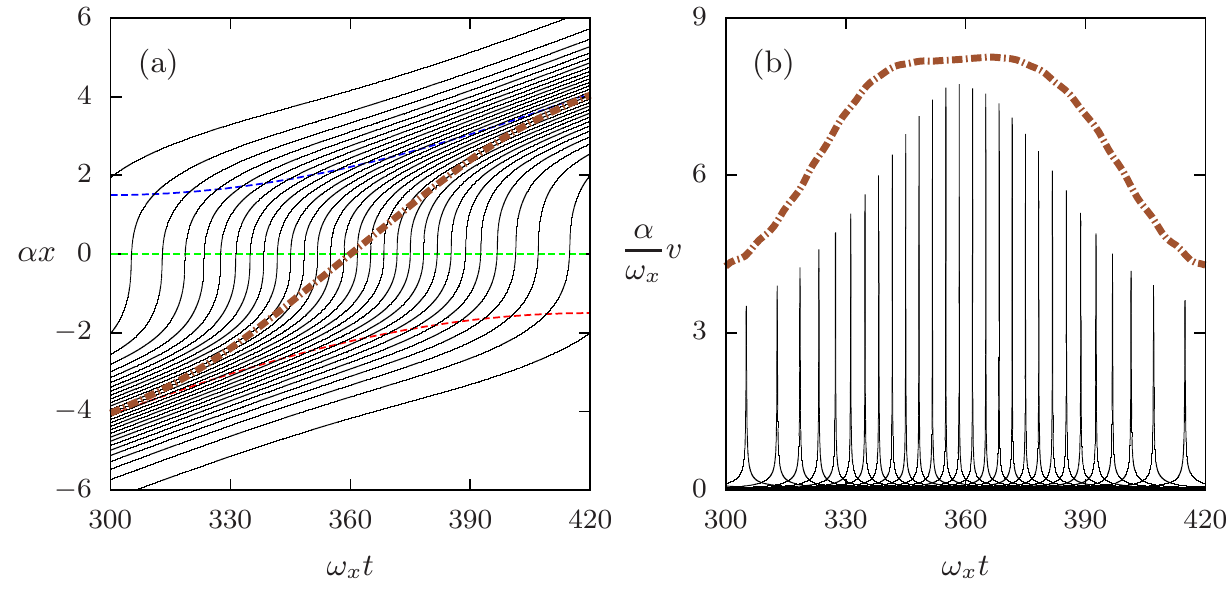} } }
\caption{(a) Positions and (b) velocites of the Bohmian trajectories for the central region of the spatial adiabatic passage sequence shown in Fig.~\ref{fig:SAP2}.
Thick dashed lines represent the mean value of (a) the position, and (b) the velocity scaled up by two orders of magnitude.}
\label{fig:SAP2}
\end{figure}

Clearly, quantum transport without transit is in contradiction with the continuity
equation that can be derived from the Schr\"odinger equation. The unraveling of this
equation in terms of Bohmian trajectories provides
a very clear physical picture of continuity, see \eref{eq.continuityanalytic} in \sref{sec:Analytical}.
Thus,
Fig.~\ref{fig:SAP2}(a) displays, at the central region, the corresponding Bohmian trajectories for the spatial adiabatic passage process under discussion, while (b) shows their corresponding velocities as a function of time.
The slope of the trajectories around the position of the middle trap center indicates
that the trajectories accelerate when crossing the central trap. The fact that the
wave function presents at all times a quasinode in the central region, implies that
the trajectories must cross it at a high velocity since the probability density around
it is always very small, see Fig.~\ref{fig:SAP2}(b). Note that at variance with other quantum
systems where the high Bohmian velocities are found in energetically-forbidden
regions, in this case this happens around the
trapping potential minima of the central well. Note also that the velocity of an
individual trajectory is not a quantum observable since one should average over
all the trajectories. In any case, the trajectories clearly show that the continuity
equation is perfectly fulfilled.

Moreover, by slowing down the total spatial adiabatic sequence, the transport process
will become more adiabatic and the instantaneous state of the atom will
remain closer to the ideal dark state. Thus, the middle trap will be less populated,
resulting in an increase of the velocity of the trajectories in the central
region. Therefore, a very counterintuitive phenomenon appears: by increasing the time for the transport of the entire
wave packet, the peak velocity that each of the trajectories reaches increases.
There is no apparent bound to the trajectory velocities in the middle region
as the limit of perfect adiabaticity is approached, and at some point Bohmian trajectories might
surpass the speed of light. As discussed by Leavens and Mayato \cite{leavens_are_1998} in their investigations
of the tunnel effect, superluminal tunneling times are {\it ``an artifact
of using the nonrelativistic Schr\"odinger equation''}, and that, with a correct relativistic
description, i.e., by using Dirac equation, Bohmian velocities cannot surpass the
speed of light, see also \cite{rel.Tausk2010}.
Therefore, it can be concluded that the appearance of superluminal trajectories would
mean that our system is no longer correctly described by Schr\"odinger's equation,
and relativistic corrections would be needed to properly describe its
dynamics. It is surprising that Schr\"odinger's equation ceases to be valid and one should consider such corrections in situations
where the process is performed very slowly. Note that an infinite transport
time is not needed for Schr\"odinger's equation to fail, because faster-than-light
trajectories would start appearing for finite ti\-mes, albeit very long compared to
the time scales considered here. Ultimately, the origin of the transport-without-transit
paradox is the incorrect use of the (nonrelativistic) Schr\"o\-din\-ger equation
in the adiabatic limit. 

It is also worth noting that very recently Huneke {\it  et al.} \cite{Huneke2013} have investigated, in an open triple quantum dot system, steady-state electronic transport via spatial adiabatic passage showing that noise in the resulting current correlates with the population in the middle dot. Thus, it could be possible to experimentally investigate the main signature of spatial adiabatic passage, i.e. the vanishing population of the middle dot, without the back action that would produce a direct measurement of the population in the middle dot.

Spatial adiabatic passage for two identical atoms in a triple-well potential has been also discussed in terms of Bohmian trajectories \cite{benseny_atomtronics_2010}, proposing efficient and robust methods to coherently transport an empty site, i.e., a hole, which, eventually, could be
used to prepare defect-free trap domains, to perform quantum computations or to design atomtronic devices.
In fact, taking into account the bosonic or fermionic statistics of these atoms and making use of both the collisional interaction and
the exchange interaction, hole transport schemes for the implementation of a coherent single hole diode and a coherent single hole
transistor have been discussed \cite{benseny_atomtronics_2010}.

To sum up, in ultracold atom physics, Bohmian trajectories have been used to get physical insight into the adiabatic transport of single atoms, Bose--Einstein condensates, and holes in triple well potentials \cite{benseny_need_2012,benseny_atomtronics_2010}. One can foresee that future research in this field would be focused on applying Bohmian algorithms for mesoscopic systems \cite{o.oriols2007prl} to the dynamics of a few cold atoms.

\subsection{Nonadiabatic molecular dynamics}
\label{sec:electron-nuclei}

The dynamics of chemical processes can nowadays be treated on a relatively routine basis.
The molecular dynamics method provides a description of the microscopic motion of individual atoms driven by classical forces computed from semiempirical force fields.
Despite the success of molecular dynamics to describe systems ranging from simple liquids and solids to polymers and biological systems,
force fields have a number of serious limitations \cite{AIMD}.
To surpass these problems, one of the most important developments in molecular dynamics, is the so-called ab-initio molecular dynamics method, which combines nuclear dynamics with forces
obtained from electronic structure calculations. The Born-Oppenheimer potential energy surfaces (BOPESs) are the central concept for understanding ab-initio molecular dynamics.
BOPESs have been mapped out with higher and higher accuracy for larger and larger molecular systems with accurate first principles electronic structure methods (such as density functional theory or time-dependent density functional theory).
Under the assumption that electrons adjust adiabatically to the slower motion of the nuclei, nuclear dynamics simulations have been carried out on top
of single BOPESs, both assuming classical equations of motion or with more accurate quantum mechanical propagation schemes for
small systems, sometimes with spectacular success in reproducing experiments \cite{Nature}.

Many challenging chemical processes, however, cannot be properly described with a single potential energy surface. The assumption that electrons adjust instantaneously to the
motion of the nuclei becomes meaningless whenever electron and nuclear motions occur on comparable time-scales \cite{Perspective}.
Electronic (nonadiabatic) transitions between potential energy surfaces play, indeed, a pivotal role in numerous chemical processes, such as electron transfer in
electrochemical reactions, ion-molecule reactions, or in proton-coupled electron transfer \cite{Schiffer}.
Similarly, electronic transitions between different BOPESs are essential to asses the performance of single-molecule electronic devices \cite{MolecularNonadiabatic}.
To study these nonadiabatic processes it is necessary to go beyond the quasi-static view of the electron-nuclear interaction.
Mixed quantum-classical approaches, where electrons are treated quantum mechanically and the nuclei are described with classical mechanics, have become particularly appealing because of the
localized nature of the nuclei in many relevant scenarios.
The interaction between classical and quantum degrees of freedom is usually addressed assuming a self-consistent field, i.e. nuclei evolve on top of a single effective potential energy surface defined as a weighted average of the involved adiabatic BOPEs. Entanglement is hardly described by these (Ehrenfest-like) approaches because the
back-reaction between classical and quantum subsystems is described under mean-field assumptions \cite{Ehrenfest1,Ehrenfest2,Ehrenfest3,Ehrenfest4}.
Multi-configuration schemes, such as Tully's surface hopping, are in general required to account for bifurcation paths with entaglement \cite{TSH1,TSH2}.
Although the undeniable success of these mixed approaches to describe many nonadiabatic phenomena, some limitations arise when quantum nuclear effects such as tunneling \cite{tunneling},
decoherence \cite{decoherence} or interferences \cite{interferences} occur.
Only the so-called quantum wave packet methods \cite{MCTDH,BurghardtCederbaum,Martinez} provide a complete description of nuclear quantum effects, although their computational cost becomes rapidly unaffordable with the size of the system.

Bohmian mechanics offers a trajectory-based scheme to describe quantum nuclear effects, and represents in this way an alternative to the quantum wave packet 
methods \cite{MCTDH,BurghardtCederbaum,Martinez}. 
Since the pioneer work of Wyatt in 1999 \cite{wyatt:prl:1999}, several schemes based on Bohmian mechanics
have been proposed to describe molecular dynamics beyond the adiabatic regime.
Based on a \textit{diabatic} representation of the molecular wave function, wave packets representing the nuclear motion are discretized into a set of Bohmian fluid elements.
These trajectories are followed in time by integrating coupled equations of motion which are formulated and solved
in the Lagrangian picture of fluid motion according to the Hamilton-Jacobi equations (see section \ref{sec:Hamilton-Jacobi})  \cite{Wyatt2,Wyatt3}.
For model two-state collision problems, even with a small number of fluid elements, the method accurately predicts complex oscillatory behavior of the wave packets.

\begin{figure}
\resizebox{0.99\columnwidth}{!}{\includegraphics{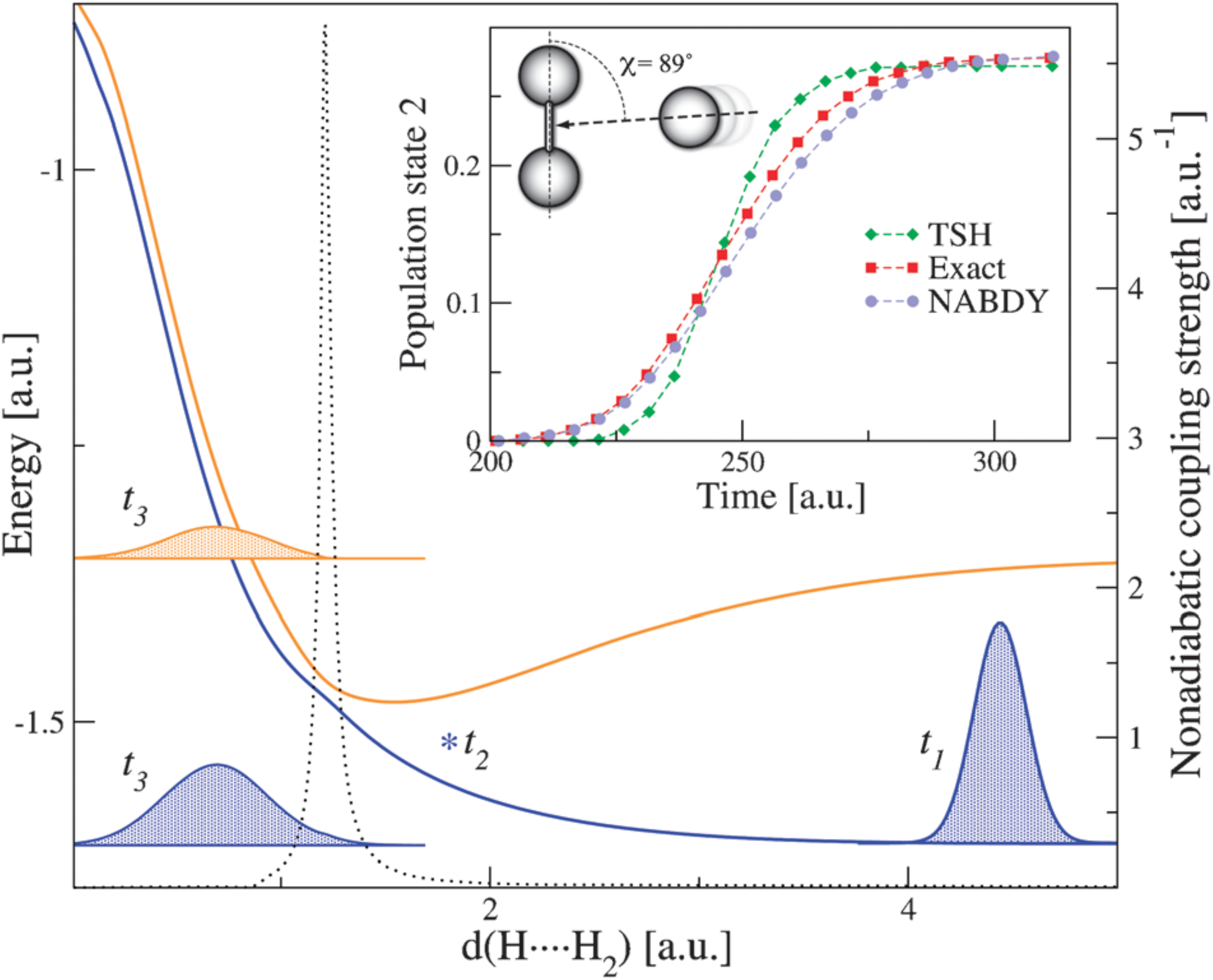}}
\caption{NABDY applied to the collision of $H$ with $H_2$ ($\chi = 89^o$ and $d(H–H) = 1.4$ a.u., see figure in the inset). An initial Gaussian
wave packet is prepared on the ground state ($t1 = 0$ a.u.) with
an initial momentum $k = 75$ a.u. In the figure, it is shown the probability density of
the nuclear wave packet obtained with 352 trajectories at the initial time ($t_1$) and
after the region of coupling ($t_3 = 300$ a.u.). The displacement of the
resulting wave packets in the vertical direction is arbitrary. Blue: wave packet on state 1;
orange: wave packet on state 2; black dotted line: nonadiabatic coupling
strength. The inset shows the time evolution of the population transfer
obtained using the different schemes (TSH: 3112 trajectories).}
\label{non-adiabatic}
\end{figure}

Alternatively, based on an \textit{adiabatic} decomposition of the electron-nuclear wave function, Tavernelli and co-work\-ers have recently presented an interesting approach which is suited for the
calculation of all electronic structure properties required for the propagation of the quantum trajectories \cite{Tavernelli1}. The nuclear equations of motion are
formulated in terms of the Hamilton-Jacobi equations (see section \ref{sec:Hamilton-Jacobi}) while density functional theory and time-dependent density functional theory are used to solve the electronic
structure at each time step.
As an example of the potential of this method to deal with electron-nuclear coupled dynamics, in \cite{Tavernelli1} the authors perform on-the-fly Bohmian dynamics of the collision of $H$ with $H_2$
using time-dependent density functional theory with the local-density approximation functional for the description of the BOPESs and the nonadiabatic coupling vectors \cite{Tavernelli2}.
In Fig. \ref{non-adiabatic}, results obtained for the colliding $H$ atom along the collision path (displayed in the inset) are shown.
Due to the strong nonadiabatic coupling, a partial population of the excited state is obtained by simple collision without the need of an external radiation field.
The agreement between nonadiabatic Bohmian dynamics (referred as NABDY in the figure) and the exact propagation for the amount of population transferred to the upper surface (inset Fig. \ref{non-adiabatic})
is very good, while in the case of Tully's surface hopping (TSH in the figure) the transfer occurs at a slightly faster rate.

Bohmian approaches to electron-nuclear coupled dynamics have been also derived without relying on a basis-set (diabatic or adiabatic) representation of the full molecular wave function.
In \cite{Meier1,Meier2},
electrons are described by waves that parametrically depend, via the total potential energy of the system, on nuclei trajectories.
The electronic waves are used to calculate Bohmian trajectories for the electrons which are required to calculate the force acting on the nuclear variables
described by approximations on the quantum Hamilton-Jacobi equations (see section \ref{sec:Hamilton-Jacobi}).
Even in the classical limit \cite{Prezhdo}, these approaches offer a solution to the trajectory branching problem
by creating a new type of quantum back-reaction on the classical subsystem. In the quantum-classical Ehrenfest approximation, which
is the most common approach, a single average classical trajectory is generated \cite{Ehrenfest1,Ehrenfest2,Ehrenfest3,Ehrenfest4}.
In contrast, in \cite{Prezhdo} an ensemble of quantum-classical Bohmian trajectories is created for a single initial quantum-mechanical wave function.
The Bohmian quantum-classical method is uniquely defined and gives results that are similar to surface hopping \cite{PrezhdoBook}.

Christov also presented an ab-initio method to solve quantum many-body problems of molecular dynamics whe\-re both electronic
and nuclear degrees of freedom are represented by ensembles of Bohmian trajectories. In \cite{Christov}, the guiding waves are solutions of a set of approximated Schr\"odinger equations evaluated along 
electronic and nuclear trajectories.
The quantum nonlocality is incorporated into the model through effective potentials which are efficiently calculated by Monte Carlo integration. Unlike other many-body methods based on density functional
calculations of the electronic structure, this approach uses explicit Coulomb potentials instead of parametrized exchange-correlation potentials.
The calculation of quantum potentials, which is a major bottleneck for those methods based on the Hamilton-Jacobi equations, is also avoided in \cite{Christov}.

Despite the potential of Bohmian approaches to deal with nonadiabatic processes, equations of motion retaining the quantum flavor of the nuclei have only been applied to model systems of very 
small molecules.
Their extension to systems made of more than a few atoms remains questionable due to the instabilities associated with the calculation of the quantum potential, and/or
the lack of a proper procedure to couple them with well established electronic structure methods.
Certainly, Bohmian approaches based on the propagation of the exact Hamilton-Jacobi equation cannot avoid the bothersome computation of the quantum potential. Only under certain approximations 
this problem can be relieved \cite{Meier1,Meier2,Prezhdo}. Unfortunately, since the quantum potential does carry crucial information about the quantum nature of the nuclei, these approaches often fail to 
capture quantum nuclear effects such as tunneling, interference or the splitting of the nuclear probability density.
There exist alternative quantum trajectory-based approaches that do avoid the calculation of the quantum potential, e.g.
the complex action formalism \cite{tannor:JCP-1:2012,tannor:JCP-2:2012}, the so-called quantum mechanics without wave functions \cite{BPoirier2012}, or the 
recently proposed extension of the conditional wave function scheme \cite{arXive_GA}. 
These approaches are still immature and their suitability to be coupled to well established electronic structure methods has not been yet demonstrated.  
Surpassing these drawbacks may be just a matter of time, and would result in a prominent computational tool to deal with general nonadiabatic phenomena.

\subsection{Intense light-matter interaction}
\label{sec:light-matter}

Since the invention of the laser there has been a pressing need to obtain light sources with increasingly higher intensities.
This has been possible with the development of techniques such as Q-switching, mode-locking and chirped pulse amplification.
For relatively low laser intensities, Einstein's photoelectric effect is enough to describe the main features of photoionization. As the light intensity increases, a plethora of inherently quantum phenomena appear~\cite{plaja_attosecond_2013}.
These effects include the ionization of atoms in multiphoton transitions with energies well above the ionization threshold~\cite{milosevic_above-threshold_2006} and the emission of high-order harmonics of the incident light~\cite{winterfeldt_colloquium:_2008} which allows for the generation of ultrashort pulses, the emission of ultrahigh frequency light or the imaging dynamics of chemical processes, see for instance Refs.~\cite{altucci_single_2011,popmintchev_bright_2012,vozzi_generalized_2011}.
All these processes require a quantum description of the interaction between light and matter.

In order to overcome the computational limit imposed by the exponential growth of the configuration space with the system dimensions, it is a usual approximation to consider only electron dynamics, while fixing the atomic (much more massive) nuclei positions, and to restrict the dynamics to one or two dimensions.
This allows to compute the dynamics of a one-dimensional lithium atom (with three electrons) with a desktop computer~\cite{ruiz_lithium_2005}.
Even though this is a drastic approximation, it retains the main physics of the photoionization dynamics.
The study of the full dynamics of more complex atoms is very computationally demanding, and only helium has been studied so far~\cite{parker_single-ionization_2007,parker_uv_2009}.

\begin{figure}
\centerline{ \resizebox{0.99\columnwidth}{!}{\includegraphics{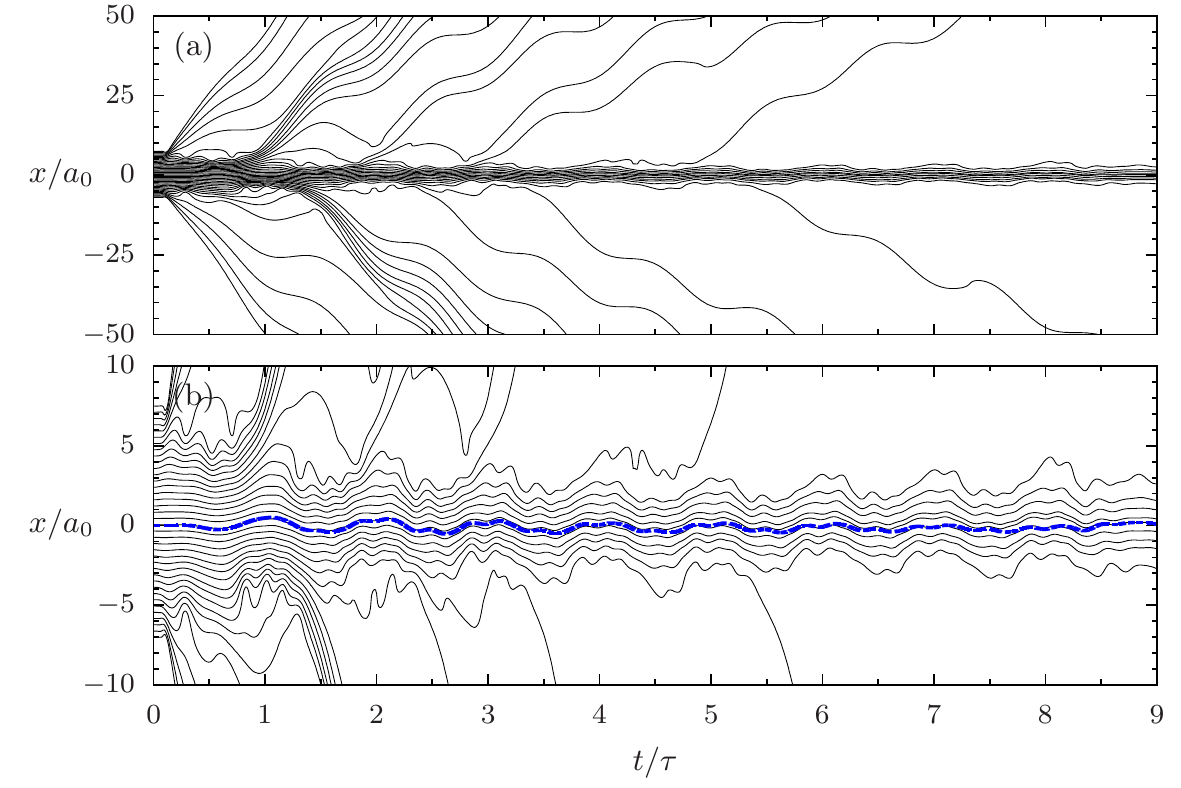} } }
\caption{(a) Bohmian trajectories associated with the one-dimensional electron dynamics of a hydrogen atom interacting with a light field of 300 nm wavelength and an intensity of $1.26 \times 10^{24}$ W/cm$^2$
(b) Enlarged view at the region around the atomic nucleus. The blue dashed curve represents the mean value of the electron position.
$\tau = 1$ fs is the light period and $a_0$ is the Bohr radius.
}
\label{fig:LMI_1D}
\end{figure}

The one or two-dimensional hydrogen atom has been extensively studied in the literature, also with Bohmian trajectories, as a prototype system to investigate photoionization.
\Fref{fig:LMI_1D} shows the time evolution of some trajectories in the one-dimensional hydrogen model.
Some of the outer trajectories escape from the nucleus, and thus should be associated with ionization.
The ones around the nucleus correspond to the internal electron dynamics, and as we will see later, are more relevant for the main features of harmonic generation.
This model has been used for instance to perform straightforward calculations of the above-threshold ionization and harmonic-generation emission spectra~\cite{lai_above-threshold_2009,lai_bohmian_2010,benseny_hydrogen_2012}, as well as for an insight into the role of the quantum potential in photoionization~\cite{lai_quantum_2009},
or even to study the chaotic behavior of classical and Bohmian trajectories~\cite{faisal_Broglie--Bohm_1998}.
Moreover, a self-consistent method~\cite{botheron_self-consistent_2010} based on the quantum Hamilton--Jacobi formalism (see \sref{sec:Hamilton-Jacobi}) has been proposed to study hydrogen photoionization.

\hyphenation{mol-e-cu-le}
\hyphenation{di-men-sion-al}

A quantum Monte Carlo method based on Bohmian trajectories to simulate the dynamics of multielectronic atoms in ultrastrong fields has been developed by Christov~\cite{christov_time_2010}.
This approximate method is related to the conditional wave function formalism, see  \sref{sec:conditional}, and reduces the problem of solving the $N$-body Schr\"odinger equation to solving a set of $N$ coupled pseudo-Schr\"odinger equations.
Each of these sets yields the dynamics of a single trajectory (for all the particles), which can be repeated for different initial positions to recover the full dynamics.
This time-dependent quantum Monte Carlo method provides a polynomial scaling for the integration time with the number of particles~\cite{christov_polynomial-time-scaling_2009} and can be applied to both finding the ground state of an atom~\cite{christov_time-dependent_2007} and studying its dynamics under an ultraintense laser pulse~\cite{christov_correlated_2006,christov_time-dependent_2007-1}.
While most of its applications have been focused in one-dimensional helium, it has also been applied to three-dimensional helium~\cite{christov_correlated_2011}.
This method also allows for the use of an effective potential which can model the nonlocal interaction between electrons, introducing correlations in their quantum state~\cite{christov_dynamic_2008}, and thus, is an adequate tool to study the role of nonlocality in multielectron states~\cite{christov_exploring_2012}.

Bohmian trajectories have also been used to study the dynamics of high-order harmonic generation.
The harmonic generation spectrum can be calculated from the Fourier transform of the electric dipole induced in an atom inside an oscillating field, which corresponds to the mean position of the electron inside a hydrogen atom.
It has been shown that trajectories from different parts of the electron wave packet contribute to different parts of the harmonic spectrum~\cite{song_investigation_2012}.
On the one hand, those trajectories starting far from the nucleus which ionize and oscillate with the field frequency provide a better representation of the low-frequency part of the spectrum.
On the other hand, the plateau and the cut-off characteristics of the high-order harmonic spectrum are better represented by the inner trajectories which start closer to the atomic nucleus and have much richer dynamics very similar to the mean electron position, as can be seen in \fref{fig:LMI_1D}(b).
This has been confirmed by more in-depth studies with long-range and short-range potentials but also have assessed that outer trajectories affect nonlocally the central ones~\cite{wu_local_2013,wu_bohmian-trajectory_2013}.

\begin{figure}
\centerline{ \resizebox{0.99\columnwidth}{!}{\includegraphics{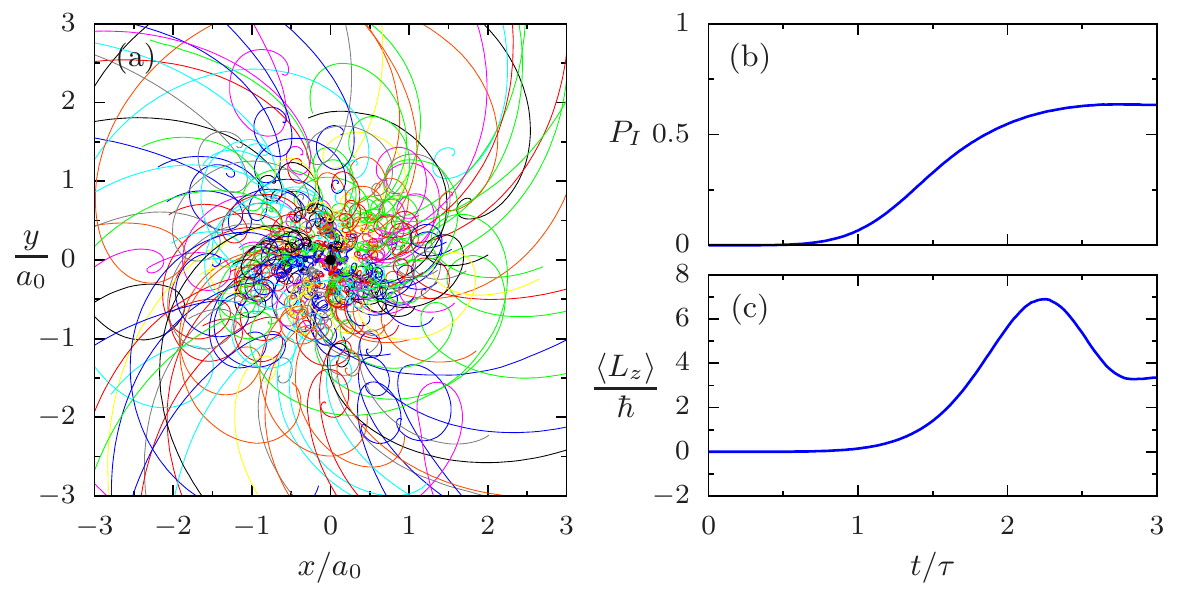} } }
\caption{Interaction of a hydrogen atom with a left-circularly polarized ($s=1$) Laguerre--Gaussian beam with $\ell = 1$.
The light pulse has a peak intensity of $6.7 \times 10^{24}$ W/cm$^2$, a beam waist of 4.79 $\mu$m, a 45 nm wavelength, and is three cycles long.
(a) Projection on the transverse plane of the electron Bohmian trajectories.
The black dot at the centre corresponds to the electron position mean value (at all times).
Time evolution of (b) the ionization probability and (c) the expectation value of the electron angular momentum along the light propagation axis.
$\tau = 150$ as is the light period and $a_0$ is the Bohr radius.
}
\label{fig:LMI_OAM}
\end{figure}

It is widely known that light carries $s \hbar$ angular momentum per photon due to its polarization ($s=\pm 1$ for left/right circular polarizations), but light can also carry orbital angular momentum due to its transverse profile.
For instance, Laguerre--Gaussian beams have an azimuthal phase dependence $\exp(i \ell \phi)$ which endows them with an angular momentum of $\ell \hbar$ per photon.
The interaction of such beams with an atom has particular selection rules which allow for a transfer of angular momentum to the electron state of more than $\hbar$ per photon~\cite{picon_photoionization_2010}.
A detailed study of the dynamics of a hydrogen atom interacting with such light pulses was carried out in Refs.~\cite{picon_photoionization_2010,picon_transferring_2010,benseny_hydrogen_2012}.
Due to the spatial profile of the light, no reduction of the system dimensionality could be taken, and the Schr\"odinger equation was integrated in three-dimensions for different polarizations, relative positions between the atom and the pulse, and pulse lengths.
In this particular case, Bohmian trajectories were used to illustrate how electrons absorb angular momentum due to the light polarization and due to its orbital angular momentum\cite{picon_transferring_2010,benseny_hydrogen_2012}.
For instance, in the case where the spin and orbital and angular momentum of the incident light point in the same direction, the trajectories associated to the electron clearly rotate around the light vortex while the electron mean position remains at rest at the origin, see \fref{fig:LMI_OAM}(a).
This exchange results in an increased ionization and a large transfer of angular momentum to the electron, cf. \frefs{fig:LMI_OAM}(b-c).

Another interesting case which has been studied with Bohmian trajectories is the interaction of a molecule with an electromagnetic field.
As in the atom case, one would expect that the ionization is largest when the electric field of the incident light is maximum, since the tunnel barrier is thinner as the electric field is stronger.
However, the interaction of a H$_2^+$ ionic molecule with an intense infrared laser pulse presents multiple bursts of electron ionization within a half-cycle of a laser field oscillation~\cite{takemoto_multiple_2010}.
The tunneling dynamics of the electron between the localized states around each hydrogen nucleus is much faster than the electric field changes, and thus it modulates the ionization bursts, as can be seen from the Bohmian trajectories of the electron~\cite{takemoto_visualization_2011}.
Furthermore, the trajectories can be used to build a two-level model for the relative phase of the two localized states which allows to predict the Bohmian velocity of the trajectories inside the molecule and obtain the subcycle ionization structure~\cite{takemoto_visualization_2011}.

Bohmian trajectories have also been used in the study of the ionization dynamics of a one-dimensional H$_2$ molecule~\cite{sawada_analysis_2013}.
The main aim of this work was to distinguish between two different classes of ionizations depending if the dynamics of the two electrons are correlated or uncorrelated.
A trajectory interpretation easily allows to distinguish between these two different ionizations, while the wave function alone does not help in elucidating {\it ``from which part of the wave packet the ejected electron originates.''}

In conclusion, Bohmian mechanics is a novel tool to study a wide range of situations in the field of strong light-matter interaction.
Bohmian trajectories have been used to both perform calculations and get insights on the dynamics.
The application of the time-dependent quantum Monte Carlo method~\cite{christov_time_2010,christov_polynomial-time-scaling_2009,christov_time-dependent_2007,christov_correlated_2006,christov_time-dependent_2007-1,christov_correlated_2011,christov_dynamic_2008,christov_exploring_2012} could allow in the future to study more complex atoms and molecules.

\subsection{Nanoelectronics: from DC to the THz regime}
\label{sec:electron-transport}

The seek of faster and smaller devices is inevitably driving the electronics industry to develop electron devices made of solid-state structures that rapidly approach  
the quantum regime \footnote{The 2012 edition of the International Technology Roadmap for Semiconductors can be found online at http://www.itrs.net. Its objective is to ensure cost-effective advancements in the performance of the integrated circuit 
and the products that employ such devices, thereby continuing the health and success of this industry.}.
Under these circumstances, electron motion can no longer be described by classical mechanics because it obeys quantum-mechanical laws.
After Landauer's seminal work \cite{landauer} in 1957, relating the electrical resistance of a conductor to its scattering (tunneling) properties, 
significant effort has been devoted to improve our ability to predict the performance of such quantum electron devices. 
In the stationary regime (DC), conductance quantization, quantum Hall effects or Friedel oscillations are, just to mention a few, 
quantum phenomena that emerge when confining charged particles in nanostructures exposed to electrostatic (or electromagnetic) driving fields.   
Scattering matrices, Green functions, quantum master equations or density functional theory among many other formalisms have been used to model quantum electron transport \cite{o.DiVentra2008book}. 
As will be shown in this section, 
Bohmian mechanics has also been successfully used by the scientific community to improve our understanding of electron transport.

For simple model systems, Bohmian trajectories derived from the probability density and probability current density (see section \ref{sec:Analytical}) 
have been used to reveal simple pictures of particle flow in quantum structures. A particularly appealing example is that of quantum vortices occuring when electron transport 
takes place across nodal points of the wave function. Due to its hydrodynamic analogy, quantum vortices are effects that can be well understood in terms of trajectories. 
The vicinity of a nodal point constitutes a forbidden region for the set of Bohmian trajectories associated with the net transport from source to drain. 
Therefore, the net current passing from source to drain cannot penetrate the vortex regions but skirts around them as if they were impurities. 
In \cite{vortex1}, 
a simple description of this effect was given, showing how quantum vortices around wave function nodes originate from the crossings of the underlying classical ray paths in quantum dot structures.  
A quantitative description of trajectories in a particular quantum wire transmission problem with vortices was also given in \cite{vortex2}. 

In order to predict the preformance of more realistic electron devices, one has to deal with several degrees of freedom. In this regard, approximations on the grounds of Bohmian mechanics 
have been developed to deal with the many-body problem. In \cite{quantumFerry}, 
the authors showed that by using an appropriate effective potential, obtained by convolving the electrostatic potential with a Gaussian, one can replicate certain quantum behavior by using classical physics.
Significantly, in contrast to the Bohm potential method, one is not required to actually solve Schr\"odinger's (or the Hamilton-Jacobi) equation in all situations using this method. 
This effective potential approach has already been successfully incorporated into a particle-based ensemble Monte Carlo simulation of a silicon MOSFET \cite{quantumFerry}.
Many-body effects also include dissipation, which plays a crucial role at room temperature. This is an aspect that was studied in terms of Bohmian trajectories in \cite{dissipation},
where the inelastic scattering was modelled by a spatially varying imaginary potential. This approach is closely related with the complex terms appearing in the conditional formulation of Bohmian mechanics 
described in section \ref{sec:conditional}, and provides new insight into the effects of electron-phonon scattering and decoherence.

Towards a broader electron device simulation tool, a generalization of the classical ensemble Monte Carlo device simulation technique was proposed to simultaneously deal with quantum-mechanical 
phase-coherence effects and scattering interactions in quantum-based devices \cite{MCBOHM,o.oriols1999sst,o.oriols2001apl,o.oriols2002apl,o.oriols2003ted,o.oriols2004apl}. 
The proposed method restricts the quantum treatment of transport to the regions of the device where the potential profile significantly changes in distances
of the order of the de Broglie wavelength of the carriers. Bohm trajectories associated with time-dependent Gaussian wave packets are used to simulate the electron transport in the quantum window. 
Outside this window, the classical ensemble Monte Carlo simulation technique is used. A self-consistent one-dimensional simulator for resonant tunneling diodes was developed to demonstrate the 
feasibility of this proposal.

The computational capabilities to describe dynamic properties of electron transport are still far from the degree of maturity of the equivalent ones for DC transport. 
Electron transport beyond the stationary regime (AC) constitutes an extremely valuable source of information to gain insight into relevant dynamical quantum phenomena such as 
the AC conductance quantization \cite{AC_quantum,o.oriols2005prb}, the quantum measurement back-reaction \cite{Backaction1,Backaction2}, high-moments of the electrical current \cite{higher_moments1,higher_moments2}, 
classical-to-quantum transitions \cite{quant_class1,quant_class2}, Leggett inequalities \cite{Legget1,Legget2,Legget3}, etc. 
Moreover, the prediction of the dynamic (AC, transients and noise) performance of electron devices is of crucial importance to certify 
the usefulness of emerging devices at a practical level. In principle there is no fundamental limitation to correctly model the high-frequency electrical current and its fluctuations, 
although one has to model such properties with far more care than DC. 

The measurement in a quantum system plays a crucial role in the predictions of the fluctuations of the electrical current around its DC value \cite{beyond_DC}.
Electronic devices work properly only below a certain cut-off frequency. In this regard, ammeters are not able to measure the entire spectra of 
the electrical current but only the power spectral density of the noise below this cut-off. 
Such power spectral density is related to the correlation function, which is the ensemble value of an event defined as measuring the current at two different times. The perturbation of the active region due to its interaction with the ammeter is the ultimate reason why modeling the measurement process plays a fundamental role 
in determining the noise. 

Second quantization offers a route to circumvent the multi-time measurement problem, simplifying its effect on the system by introducing the Fock-space as an alternative
basis for the electronic system \cite{2_quant1,2_quant2,2_quant3}. 
Alternatively, Bohmian mechanics provides a conceptually easier recipe. 
To deal with multi-time measured systems, one has to add the degrees of freedom of the measuring apparatus to those of the system of interest and solve the Schr\"odinger equation for the combined 
system (see \sref{sec:pointer}). 
At the computational level this scheme could result in a huge additional complexity. 
Some preliminary attempts to tackle this problem can be found in \cite{time_resolved}, where an effective potential for the system-apparatus interaction is considered. 
The authors were able to relate the total current measured on an ammeter to the Bohmian trajectories of the electronic system. The acceleration of the center of mass of the pointer was demonstrated 
to be directly proportional to the total (particle plus displacement) current, and a qualitative estimation of the back-reaction of the measuring apparatus on the electronic system was discussed. 
A discussion on how the ammeter produces the channeling of the many-particle wave function is missing in the simple model of \cite{time_resolved}. 
However, as concluded in \cite{time_resolved}, whenever the main branching of the system (into transmitted and reflected parts) comes mainly from the active region itself, not from the ammeter, 
the effects of the measurement on the active system can be neglected up to very high frequencies.  

Improved solutions to the many-body problem are also required in order to go beyond the DC regime. 
Approximations to the electron transport problem must be able to reproduce charge neutrality and quantify displacement currents \cite{beyond_DC}. 
These requirements constitute an additional source of complexity with respect to the stationary regime because in the latter the value of the displacement current is zero when time-averaged, 
and overall charge neutrality is trivially fulfilled when fluctuations are disregarded. 
On one hand, the total (conduction plus displacement) current satisfies a current conservation law, a necessary condition to assume that the current 
measured by an ammeter (far from the simulation box) is equal to the current that we compute on the simulation region \cite{beyond_DC,time_resolved}. On the other hand, 
positive and negative deviations from charge neutrality inside electronic devices approach zero after periods of time larger than the dielectric relaxation time 
\cite{2_quant1,albaredaPRB2}.   

Both the computation of the displacement current and the imposition of overall charge neutrality require the Poisson equation and electron dynamics to be solved in a 
self-consistent way \cite{beyond_DC}. The convenience of Bohmian mechanics to face this particular problem has been extensively studied. 
In \cite{albaredaPRB1}, a many-particle Hamiltonian for a set of particles with Coulomb interaction inside an open system was solved without any perturbative or mean-field approximation by means of a 
conditional trajectory algorithm (see section \ref{sec:conditional}) \cite{o.oriols2007prl}. In order to guarantee overall-charge-neutrality, a set of boundary conditions for the above mentioned Hamiltonian was 
derived to include the Coulomb interaction between particles inside and outside of the active region \cite{albaredaPRB2,JCE_boundaries1,JCE_boundaries2}. 
In the high-frequency domain the assessment of current conservation has been achieved through a generalization of the Ramo-Shockley-Pellegrini theorems 
\cite{RSP1,RSP2,RSP3,RSP4,RSP5} for Bohmian mechanics \cite{albaredaFNL,alarcon2009computation}. 
Over the last ten years, as a result of the above mentioned works, Oriols and coworkers have developed a trajectory-based quantum Monte Carlo simulator based on Bohmian mechanics specially designed for 
the description of electron transport in nanoscale devices, both for DC and beyond DC regimes (see Refs. \cite{MCBOHM,o.oriols1999sst,o.oriols2001apl,o.oriols2002apl,o.oriols2003ted,o.oriols2004apl,o.oriols2007see} 
and a recent review in \cite{time_resolved}). The simulator includes also a
package based on the semiclassical limit of Bohmian mechanics \cite{AbdelilahAPL,albaredaJAP,albaredaJSTAT,AbdelilahFNL,AlbaredaMC} and has been rececently generalized to 
include spin-dependent electron transport \cite{o.alarcon2009pps}.
Such simulator is named Bohmian Interacting Transport for nonequiLibrium eLEctronic Structures (BITLLES).

As an example of the predicting capabilities of BITLLES, the authors have investigated the main characteristics of a resonant-tunneling diode (RTD), a diode with a resonant-tunneling structure in 
which electrons can tunnel through some resonant states at certain energy levels. 
Characteristic to the current-voltage relationship of a tunneling diode is the presence of one or more negative differential resistance regions, which enables many unique applications. 
Resonant tunneling is of general interest in many applications of quantum mechanics (see \cite{RTD} and references therein); the particular
case of RTDs is very intriguing, not only for their peculiar properties, but also for their potential applications in both analogue
\cite{analogue} and digital \cite{digital} electronics. Nevertheless, technology solutions to integrate RTDs in electronic circuits are still under investigation.
\begin{figure}
\resizebox{\columnwidth}{!}{\includegraphics{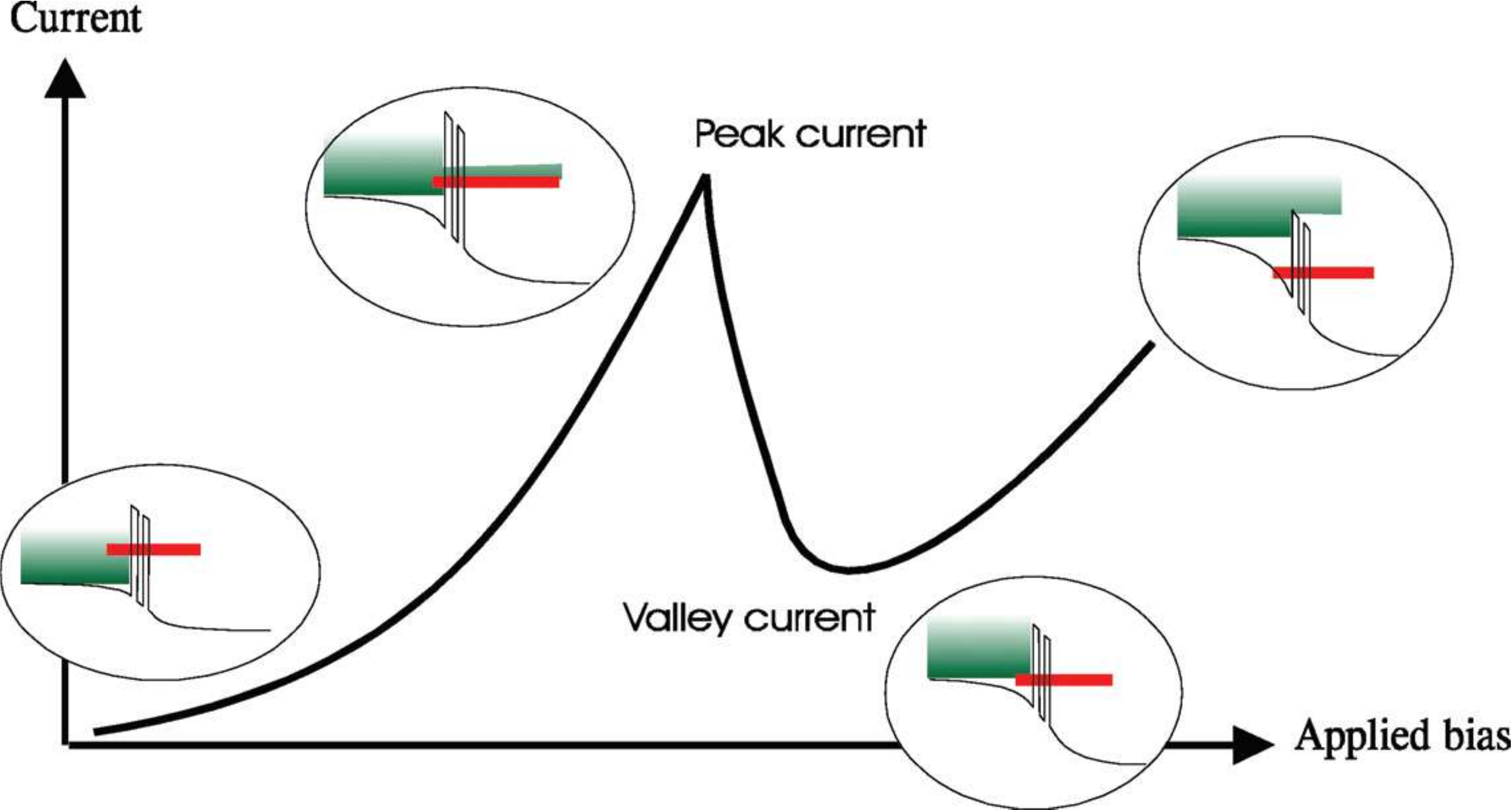}}
\caption
{Schematic representation of the current-voltage characteristic of a typical RTD.
The resonant energy inside the quantum well acts like an energetic filter
that lets the electrons from the source to arrive at the drain. See also Color
Insets.}
\label{RTD}
\end{figure} 

\begin{figure}
\resizebox{\columnwidth}{!}{\includegraphics{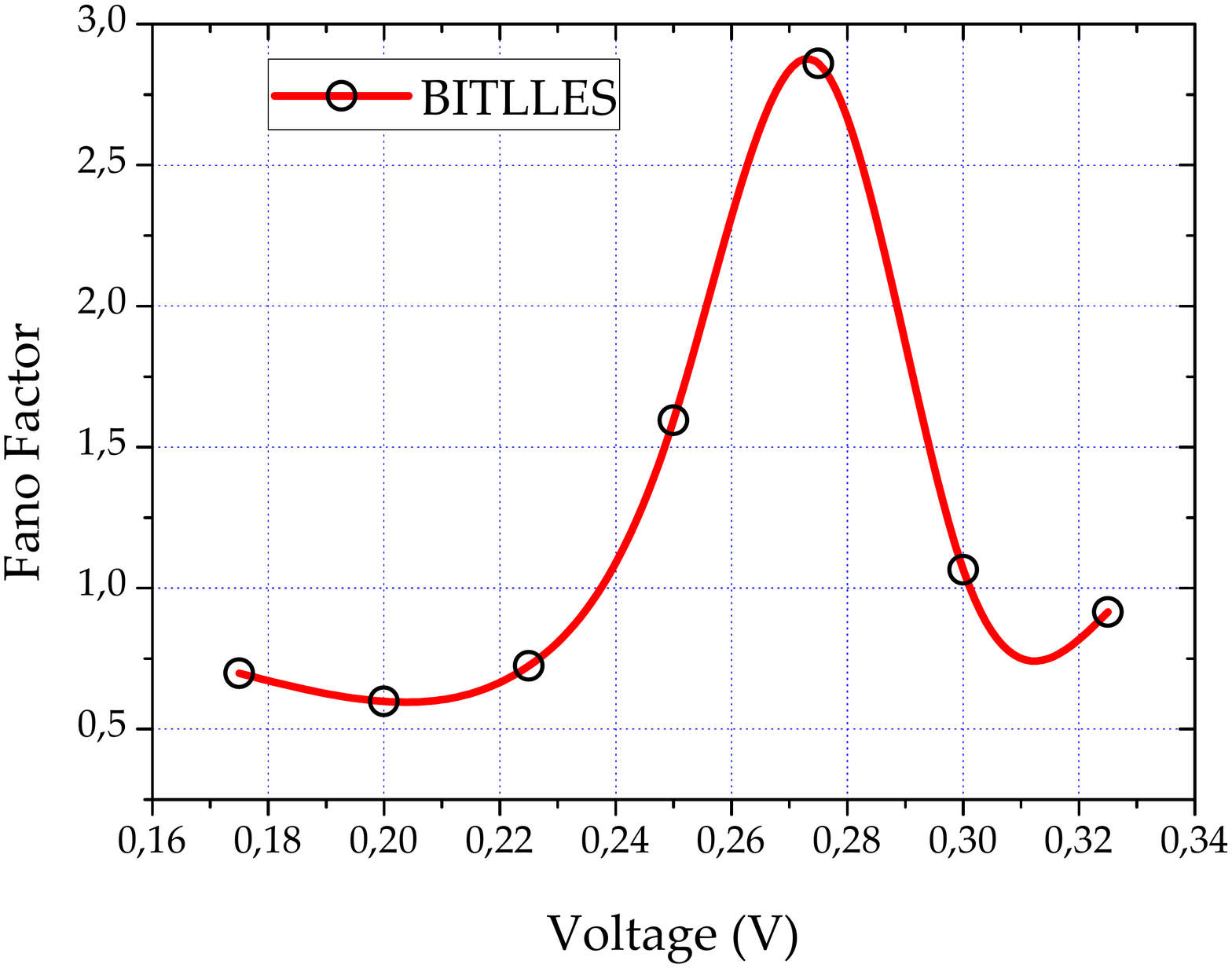}}
\caption
{Current noise power spectrum referred to Poissonian shot noise at different biases. }
\label{Fano_factor}
\end{figure}

From a computational viewpoint, the single-particle theory for mesoscopic structures is valid for capturing the basic behavior of RTDs, but not appropriate to describe the totality of
the typical behavior of these devices \cite{o.oriols1996ssc,o.oriols2001apl,o.oriols2002apl,o.oriols2003ted,o.oriols2004apl}. Also in the most idealized case of RTDs, the inclusion of the Coulomb correlation between electrons
is enough to spoil the results of the single-particle theory. Many-body theories and simulations, confirmed by experimental measurements, show, for example, different current patterns 
\cite{albaredaPRB1,RTDMB1,RTDMB2} or a very enhanced noise spectrum in the negative differential conductance region \cite{RTDnoise1,RTDnoise2}.
Of particular interest is the behavior of the intrinsic current fluctuations, where the correlations between  
electron trapped in the resonant state during a dwell time and those remaining in the left reservoir play a crucial role \cite{fabioIEEE}. 
This correlation occurs essentially because the trapped electrons perturb the potential energy felt by the electrons in the reservoir. 
This phenomenon can be analysed through the value taken by the Fano factor, which can be viewed as a measure of the noise-to-signal ratio. 
This quantity depends directly on the correlation function and thus a proper modeling of the measuring process is required. In addition, 
in the limit of noninteracting electrons or simple mean-field approximations, the Fano factor reduces to the partition noise, a wrong result for finite applied bias. 
Only if the dynamical (self-consistent) Coulomb correlations are taken into account, the Fano factor recovers the correct behavior, showing both the sub- and super-Poissonian behaviors 
that characterize the low-frequency noise of a RTD (see Fig. \ref{Fano_factor}). 

Bohmian mechanics has been used in the simulation of electron transport for more than ten years, its utility ranging from that of a pure interpretative tool to that of a powerful
approximation to the many-body problem at high-frequency regimes. Approximations to decoherence and dissipation are however agenda items still to be addressed with Bohmian mechanics. 
In particular, combining trajectory-based approximations to the electron-nuclear coupled motion together with the above described Bohmian approaches to electron transport would result into a powerful and 
versatile simulation tool to describe molecular devices.

\subsection{Beyond spinless nonrelativistic scenarios}
\label{sec:beyond}

All  examples of applications of Bohmian mechanics that we have discussed so far deal with the nonrelativistic spatio-temporal dynamics of quantum systems formed by spinless particles, where the spin is only taken into account by the antisymmetrization (symmetrization) of the particle's wave function for identical fermions (bosons). However, and contrary to what is sometimes stated, Bohmian mechanics allows for an accurate description of the dynamics of spin particles and can be extended to other domains such as relativity, quantum field theory, quantum cosmology or even to classical optics.

There are basically two alternatives to include the spin degree of freedom within the Bohmian formulation.
Firstly, and resembling the standard quantum mechanics procedure, spin can be accounted for by replacing the usual scalar wave function with a spinor-valued function whose dynamics is given by the appropriate generalization of the Schr\"odinger equation \cite{spin.Bell1971}. Secondly, one could include particle's spin into the dynamics following a full Bohmian approach by adding three Euler angles $\left\{ \alpha , \beta, \gamma \right\}$ to the wave function for each of the spin particles of the system such that the amplitude and phase of the wave function do not only depend on the position of the particles, but also on these angles \cite{spin.Dewdney1986}.  See Chapter 10 in Ref.~\cite{o.Holand1993} for a straightforward derivation of the corresponding equations of motion for the positions and the Euler angles within the Bohmian formulation.

Quantum mechanics is a nonrelativistic covariant theory and, as a consequence, neither its standard nor its Bohmian formulation is compatible with relativity.
In fact, for a quantum system of $N$ particles, the velocity of each particle is determined by the $N$-particle wave function around the actual configuration point of the system such that the motion of each particle depends on the instantaneous positions of all the other particles, no matter how distant they are. Thus, nonlocality is the main concern in developing a satisfactory relativistic quantum theory. As it has been recently shown by D\"urr \textit{et al}., \cite{rel.Durr2013}, one promising approach consists in extracting, from the wave function, a privileged foliation of space-time into space-like hypersurfaces to define the Bohmian dynamics. A similar approach was previously discussed with the space-time foliation extracted from the actual configuration space \cite{rel.Horton2004}.
In addition, ``synchronization'' of particle trajectories has been also considered as a resource to obtain a Lorentz covariant Bohmian formulation in references\cite{rel.Berndl1996,rel.Horton2001,rel.Dewdney2002,o.nikolic2005,o.nikolic2009,o.nikolic2011}, which, at variance with \cite{rel.Horton2004}, do not completely agree with the standard predictions of quantum mechanics.  See \cite{rel.Tumulka2007} for an overview of all these different approaches. In particular, Nikoli\'c \cite{o.nikolic2005,o.nikolic2009,o.nikolic2011} makes use of two elements: it generalizes the space probability density of standard quantum mechanics to a space-time probability density, and introduces a many-time wave function for many-particle systems. With these two ingredients, a relativistic covariant formulation of quantum mechanics for both spinless and spin particles is derived. The relativistic covariant character of this extension is explicitly shown for the relativistic version of Bohmian mechanics.
Worth noting, the relativistic Dirac equation has been also analyzed in terms of Bohmian trajectories \cite{rel.Tausk2010} showing that the probability that an electron reaches the speed of light at any time is equal to zero.

Quantum field theory is particularly useful for those physical systems where the number of particles is not fixed. In particular, an accurate quantum description of the measurement process or the interaction with the environment assumes that particles can be created or destroyed. In this regard, different models that account for the particle creation and annihilation in a Bohmian way which reproduce the standard quantum predictions have been proposed \cite{qft.dur2004,qft.col2007,qft.str2011}. Alternatively, the Bohmian formulation of Ni\-ko\-li\'c's relativistic covariant quantum theory is also particularly interesting \cite{o.nikolic2005}. Thus, for instance, when the conditional wave function associated with a quantum measurement does not longer depend on one of the space-time coordinates, then the corresponding particle has zero four-velocity (with respect to its own four-dimensional space-time Minkowski coordinate), i.e., such a particle has no longer an associated trajectory but instead it is represented by a dot in space-time. Trajectories in space-time may have beginning (creation) and ending (annihilation) points, which correspond to positions where their four-velocities vanish. This mechanism allows effectively for the nonconservation of the particle's number in quantum systems.  Particle's positions are usually the ``hidden variables'' in Bohmian mechanics but this is not mandatory \cite{qft.str2010}. Fields (or even strings) could be also taken as the hidden variables. In fact, Bohmian quantum field theories have been developed to account for the free quantized electromagnetic field \cite{o.Bohm1952a,o.Bohm1952b}, bosonic quantum fields \cite{o.Holand1993,takabayasi:ProgTheorPhys:1952,qft.Holland1988}, fermionic quantum fields \cite{o.Holand1993,qft.Holland1988,qft.Valentini1996}, and quantum electrodynamics \cite{qft.Valentini1996,qft.Struyve2007}.

Bohmian mechanics has been also used as a realistic causal model for quantum cosmology \cite{qc.Callender1994,qc.Pinto2005,qc.Shojai2007} to address several open problems such as the still universe resulting from the fact that the Hamiltonian of classical general relativity equals zero, the so-called problem of time. Thus, even for a stationary wave function, the Bohmian formulation can provide a time evolution through the Bohmian trajectories.  On the other hand, the quantum force naturally appearing in Bohmian mechanics has been discussed as a mechanism to avoid singularities due to gravity \cite{qc.deBarros1998,qc.Shojai1998,qc.Pinto2005b}. 

In a completely different physical scenario, Kocsis \textit{et al}., \cite{kocsis:Science:2011} have experimentally reported the statistically average paths taken by single-photons in a Young double-slit experiment via the weak measurement technique. It was shown that these average trajectories match indeed with the corresponding Bohmian trajectories. This very relevant experiment will be further commented in \sref{sec:measurements}. In fact, the connection between Bohmian trajectories for massive particles and optical trajectories for light beams has been investigated in detail by Orefice \textit{et al}., \cite{o.orefice2009} beyond the geometrical optics approximation. In particular, it has been explicitly demonstrated that the Helmholtz equation of a classical optics wave allows, without any approximation, for a Hamiltonian set of ray-tracing equations that take into account interference and diffraction. The trajectories associated with these rays are shown to strongly depend on the beam amplitude distribution through the so-called ``wave potential'' term that is the source of nonlocality and that it is typically omitted in the geometric optics approximation. This wave potential is shown to be equivalent to the quantum potential of the Bohmian theory.

\section{Applications to general problems}
\label{sec:application2}

In the previous section we have presented some examples on how the Bohmian theory provides predictions that become of great utility in understanding some state-of-the-art problems in forefront research fields, ranging from atomtronics  or nanoelectronics  to light-matter interaction or cosmology.
However, there are many other problems which have been successfully tackled with Bohmian mechanics that cannot be ascribed to a unique research field, but to many of them.
For example, the elastic collisions or scattering which are reviewed below are present in almost all research fields dealing with quantum phenomena.
In this section we show Bohmian solutions to some of these general problems.
We emphasize that the Bohmian formalism also provides practical solutions to this type of problems, such as the many-body or quantum measurements problems.

\subsection{Reactive scattering}
\label{sec:scattering}

Scattering processes play a fundamental role in determining physical
and chemical properties of materials.
Quantum scattering dynamics is thus ubiquitous to many different
problems in atomic and molecular physics, chemical physics, or
condensed matter physics ---among other fields of physics and the
interphase between physics and chemistry.
Within the standard quantum formulations it is
common to carry out analytical expansions in the energy domain in
order to study quantum scattering (e.g., partial wave analyses).
These approaches provide stationary asymptotic solutions, usually in
terms of plane waves with well-defined energies.
Since the development of the first efficient wave packet propagation
methods (with the advent of the also first efficient computers) by the
end of the 1960s\footnote{A solitary pioneering work
already appeared as early as 1959 by Mazur and Rubin dealing with
collinear scattering \cite{mazur:JCP:1959}.}
\cite{goldberg:AJP:1967,mccullough:JCP:1969,mccullough:JCP:1971-1},
analyses in the time-domain gained in popularity and relevance.
The advantage of these methods is that
they allow us to explore the scattering dynamics in the
configuration space, thus offering a pictorial representation of
what is going on all the way through, i.e., from the initial state
to the final asymptotic one.
Nevertheless, the evolution of the system probability density still
lacks the intuition or insight about the dynamics that one otherwise
obtains with trajectories.
This is precisely where Bohmian mechanics comes into play as a quantum
formulation that allows us to investigate scattering under a trajectory-based perspective.

Given the ample scope of scattering dynamics, we have established a
distinction between reactive and nonreactive scattering.
This criterion is equivalent to separate scattering problems with
exchange of energy and momentum during the scattering event, from those
with only exchange of momentum (this type of problems will be treated
in Sect.~\ref{sec:elastic}).
In the case of reactive scattering, the trip starts by the end of the
1960s and beginning of the 1970s, when McCullough and Wyatt published
a series of works where the collinear H+H$_2$ reaction was analyzed
within a quantum hydrodynamic-like formulation
\cite{mccullough:JCP:1969,mccullough:JCP:1971-1}.
At that moment Bohm's theory was not in fashion at all, nor anything
related with a formulation of quantum mechanics out of the Copenhagen cannon.
Although these authors did not employ what we nowadays know as Bohmian
mechanics, interestingly they proposed the use of the quantum
probability flux vector as a tool, as defined by
Eq.~(\ref{fluxvector}) (see Sect.~\ref{sec:Analytical}) in the case of a neutral particle.
They argued that, if $\psi(\vec{r},t)$ is expressed in its (polar) form
$R(\vec{r},t)\exp [iS(\vec{r},t)/\hbar]$, as in Eq.~(\ref{eqtrans}),
then the probability density and flux are given by $\rho = R^2$ and
$\vec{j}=\rho\nabla S/m$, respectively.
Accordingly, a quantity $\vec{v}=\nabla S/m$ can be defined as a local
velocity for the probability flow, so that $\vec{j}=\rho \vec{v}$,
which {\it ``emphasizes the similarity with fluid flow in classical
hydrodynamics''} \cite{mccullough:JCP:1971-1}.

By computing the quantum flux and representing it in terms of arrow
maps, McCullough and Wyatt found a dynamical explanation for the
{\it quantum bobsled effect}, formerly predicted by Marcus
\cite{marcus:JCP:1966}.
In the transit from reactants to products in the H+H$_2$ reaction,
an excess of energy leads a portion of the system probability density
out of the reaction path (just as a kind of centrifugal effect),
climbing up the potential energy surface that describes this reaction.
The reflection of the wave function with the hard wall of the potential
energy surface causes that it folds back onto itself, giving rise to a
series of ripples by interference.
Although the monitoring of the probability density offers a picture of
how the system spreads beyond the region that it should cover, it is
the flux (or the local velocity) the quantity that specifically shows
the direction of this flow and how it evolves in time.

McCullough and Wyatt also observed a remarkable dynamical vortical
behavior whenever a node of the probability density develops.
In these cases, the flux spins around the node, giving rise to a
vortical dynamics, the {\it quantum whirlpool effect}
\cite{mccullough:JCP:1971-1}.
Later on, this behavior was further analyzed by Hirschfelder and
coworkers in terms of quantum streamlines
\cite{hirsch:JCP:1974-2,hirsch:JCP:1977}, including applications
to reactive atom-diatom scattering \cite{hirsch:JCP:1976-1}.
They found that these vortices display an interesting property,
namely that the circulation around them is quantized, as already
noted in the early 1950s by Takabayasi \cite{takabayasi:ProgTheorPhys:1952}.
That is, if the circulation is defined by a line integral along a close
loop around the vortex, the result from this integral is a nonzero value,
namely $(2\pi\hbar/m)n$, which denotes the change in the phase of the
wave function $\Psi$ after completion of a number $n$ of full loops
($m$ denotes the system mass).
Notice that because of the complex-valuedness of the wave function,
its phase is always well defined except for an integer multiple of $2\pi$.
If no nodes are present, the phase of the wave function changes
smoothly from one point to another of the configuration space.
However, as soon as a node is present, it undergoes a change that
is a multiple of $2\pi$, as any complex function ---somehow it behaves
like the noncompact Riemann surface associated
with the complex variable function $f(z) = \ln z$, which displays a
$2\pi$ increase after completing a full loop around $z=0+i0$.
The presence of quantum vortices (or {\it whirlpools}) can be observed
in many different physical problems characterized by two or more
dimensions, caused by the coalescence on a certain region of the
configuration space of different parts of the wave function.
Notice that we have already found vortical dynamics in
Sect.~\ref{sec:electron-transport}, in the context of quantum transport
through constrictions in nanostructures.
On the other hand, it is also remarkable the fact that these dynamics
are closely related to the presence of chaos in Bohmian mechanics (see
Sect.~\ref{sec:chaos}).

\begin{figure}
 \begin{center}
 \resizebox{0.95\columnwidth}{!}{\includegraphics{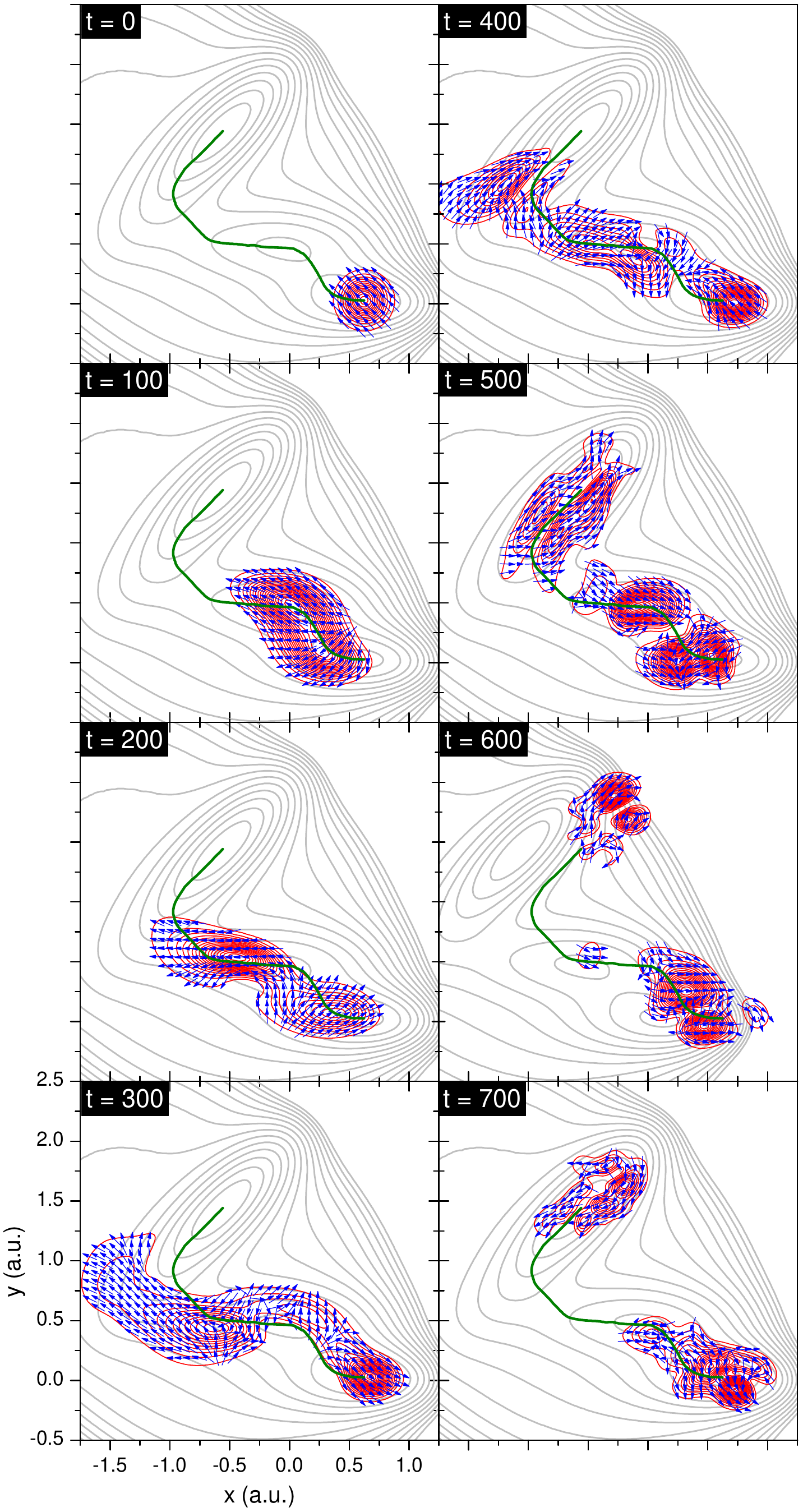}}
 \end{center}
 \caption{Snapshots of the time-evolution of the probability density
  (red contour lines) describing the passage from reactants to products
  in a prototypical chemical reaction \cite{sanz:cpl:2009}.
  The gray contour lines represent the corresponding potential energy
  surface, while the reaction path is denoted by the thicker green line.
  Arrows provide an insight on the hydrodynamics of the process (they are
  preferred to the detriment of Bohmian trajectories due to visual
  clarity).
  Positions and time are given in arbitrary units.}
 \label{fig9revABM}
\end{figure}

The behaviors observed by McCullough and Wyatt have also been observed
more recently when analyzing chemical reactivity
\cite{wyatt:JCP:1999,sanz:cpl:2009,sanz:cpl:2010E,sanz:CP:2011}.
In Fig.~\ref{fig9revABM} the dynamics of a prototypical chemical
reaction is displayed in terms of a series of snapshots.
As it can be noticed, at $t=300$ (in arbitrary units) part of the
probability density tries to surmount the leftmost part of the potential
energy surface, although it is relatively high in energy.
This is an example of the quantum bobsled effect mentioned above.
On the other hand, in the region of reactants it is possible to observe
the appearance of a series of whirlpools as a consequence of the
interference of the part of the wave function getting back to reactants
with the part of the initial wave packet that is still leaving the
region.
Beyond $t=600$, the reaction can be considered as almost finished, since
the dynamics has reached a certain equilibrium, with part of the wave
function being localized in the products region, while another part
(the nonreactive one) is in the reactants region.

In Fig.~\ref{fig9revABM} we notice that the transition from
reactants to products takes place through a local maximum, which
readily appeals to the notion of tunneling. This is actually a
central issue in many problems of reactive scattering and reaction
dynamics. Tunneling constitutes one of the most intriguing
properties exhibited by quantum systems, and therefore the reason
why it was one of the first quantum phenomena in being attacked from
a Bohmian perspective. In this regard, the first contribution that
we find in the literature is due to Hirschfelder and coworkers
\cite{hirsch:JCP:1974-1}, who analyzed the scattering produced by a
square two-dimensional barrier, finding the quantum analogs of the
frustrated total reflection of perpendicularly polarized light and
the longitudinal Goos-H\"anchen shift. Because of the
two-dimensionality of the problem, the authors found the appearance
of vortices, as indicated above. It is worth noticing that the
results obtained by Hirschfelder, not very well known out of the
chemical physics community, predate those reported by Dewdney and
Hiley about eight years later, starting directly from Bohm's
approach \cite{hiley:foundphys:1982}. In this work, the authors
reproduced by means of trajectories some of the results earlier
reported by Goldberg {\it et al.}~\cite{goldberg:AJP:1967} on
scattering off one-dimensional barriers and wells (which involve
tunneling).

In order to explain how tunneling takes place along the transversal
direction of propagation, one-dimensional simulations are also
reported in Ref.~\cite{hirsch:JCP:1974-1}. They show that, within
the Bohmian scheme, tunneling takes place because of particles {\it
``riding above the barrier.''} In other words, from a Bohmian
perspective there is nothing such as particles {\it traversing} the
barrier (as it is commonly taught in standard quantum mechanics
courses), but they {\it surmount} the barrier. This latter result
was numerically rediscovered about 25 years later by Lopreore and
Wyatt, when they proposed the first quantum trajectory method
\cite{wyatt:prl:1999}.

The interest and importance of reactive scattering, including tunneling
(mainly through Eckart barriers, which smartly describe the transition
from reactants to products) has given rise to a vast literature on
trajectory-based methods, from classical trajectories to wave packet
propagation schemes.
Within this framework, Bohmian mechanics has also been considered as
an alternative resource of numerical quantum-propagation methods.
They are the so-called quantum trajectory methods, summarized by Wyatt
in 2005 in a detailed monograph about the issue \cite{wyatt-bk}.

\subsection{Elastic collisions}
\label{sec:elastic}

In the case of elastic collisions (nonreactive scattering), the first
Bohmian outcomes that we find are on scattering off localized targets,
also produced by Hirschfelder and coworkers \cite{hirsch:JCP:1976-2}.
They studied the elastic collisions between two particles that interact
through a spherically symmetric square potential, i.e., $V(r) = V_0$
for $r<a$ and 0 everywhere else, where $V_0>0$ for a potential barrier
and $V_0<0$ for a potential well.
To some extent this system, which approximates the interaction of
neutrons with protons, for example, constitutes a three-dimensional
generalization of the former one-dimensional movies obtained by
Goldberg et al.~\cite{goldberg:AJP:1967}, or later on by Galbraith
et al.~\cite{abraham:AJP:1984} in two dimensions.
The motivation for this work somehow summarizes the leitmotif of other
works developed in the area of elastic scattering and diffraction (or
even other fields of physics and chemistry):
\begin{quote}
{\it ``The emphasis in scattering theory has been on obtaining the scattered
wave function in terms of the incident wave function.
What happens {\it during} the collision has become a black box.
By plotting the quantum mechanical streamlines and probability density
contours, we can see what is taking place inside the black box.
In this manner, we obtain additional details which should be helpful
in understanding collision dynamics.''}
\end{quote}
In particular, the analysis carried out by the authors exhibits an
interesting feature, namely the appearance of a rather complex vortical
dynamics around the target during the time of maximal interaction.
As will be seen, this type of dynamics is typical of any elastic
scattering process regardless of the system analyzed, and play a
fundamental role in the formation of resonances \cite{sanz:jcp:2005}.

More recently and independently, Efthymiopoulos and coworkers have also
tackled a similar issue by studying the diffraction of charged particles
by thin material targets \cite{efthymio:IJBC:2012,efthymio:AnnPhys:2012}.
The exhaustive analysis presented by these authors is in agreement with
the earlier findings by Hirsch\-fel\-der and coworkers.
Nonetheless, probably the most original aspect in these works is the
estimation made by the authors of arrival times and times of flight
using as a tool the Bohmian trajectories, a problem that is totally
ambiguous within the other more standard formulations of quantum
mechanics, where time is just regarded as an evolution parameter or a
label \cite{muga:PhysRep:2000,muga-bk-1:LNP:2002,muga-bk-2:LNP:2009}.
A more detailed discussion on the role of time in Bohmian mechanics
can be found in Sect.~\ref{sec:measurements}.

With an analogous purpose, Bohmian mechanics has also been applied to the
field of atom-surface scattering.
In this field various aspects have been analyzed since 2000 in order
to determine the relationship between surface diffraction and
classical rainbow features \cite{sanz:prb:2000,sanz:EPL:2001,sanz:JPCM:2002},
the role of vortical dynamics in adsorption process \cite{sanz:JPCM:2002,sanz:jcp:2004,sanz:prb:2004},
or the dynamical origin of selective adsorption resonances below the
onset of classical chaos \cite{sanz:jcp:2005}.
Differently with respect to the diffraction by localized targets, the
presence of an extended object provides very interesting results from a dynamical
or, more precisely, hydrodynamical point of view.
In the case of perfectly periodic surfaces, it is shown that the longer
the extension covered by the incoming wave packet representing the
atom (i.e., the higher its monochromaticity) the better the resolution
of the diffraction features, which in these cases correspond to Bragg
maxima.
Obviously, one could conclude this from the optics with gratings and
periodic arrays.
However, by inspecting the corresponding Bohmian trajectories one
immediately realizes that such a behavior is a direct consequence of
the redistribution of Bohmian momenta along various groups of
trajectories.
That is, as the extension parallel to the surface of the incoming
wave becomes larger, the information about the periodicity of the
surface becomes more precise.
This is in sharp contrast with a classical (or semiclassical) situation,
where only the knowledge of a single lattice is enough to characterize
the scattering process.
This is a trait of the quantum nonlocality as well as the fact that
the distribution of trajectories along Bragg angles also depends on
the number of cells covered by the incoming wave \cite{sanz:prb:2000}.

The previous effect can be regarded as a parallel one, in the sense
that it is related to the parallel direction to the surface.
If one observes the dynamical behavior of the wave packet as it
approaches the surface from a Bohmian viewpoint, there is an also
interesting perpendicular effect: the trajectories starting in positions
located around the rearmost parts of the initial wave function never
reach physically the surface, but bounce backwards at a certain
distance from it \cite{sanz:JPCM:2002,sanz:jcp:2004,sanz:prb:2004,sanz:jcp:2005}.
On the contrary, the trajectories with initial positions closer to
the surface are pushed against the surface and obliged to move parallel
to it until the wave starts getting diffracted and abandons the
surface.
This has been regarded as an effect similar to a quantum pressure
associated with the Bohmian noncrossing property --- Bohmian
trajectories cannot cross through the same spatial point at the
same time.
The possibility to define such kind of pressure was already proposed
by Takabayasi \cite{takabayasi:ProgTheorPhys:1952,takabayasi:ProgTheorPhys:1953}
in 1952 (the same year that Bohm proposed his hidden-variable model).

The appearance of a quantum pressure effect has important consequences
regarding surface trapping, involved in adsorption and desorption
processes.
First, because of this pressure, part of the wave or, equivalently,
the corresponding swarm of trajectories, is forced to keep moving
parallel to the surface.
This may cause its trapping within the attractive well near the surface
and therefore to confine the trajectories temporarily.
It is in this way how adsorption and in particular adsorption
resonances appear.
The second aspect to stress is the development of temporary or
transient vortical
dynamics while the wave function is near the surface.
Because part of the wave is still undergoing a motion towards the
surface while the other part is already being deflected, a web of nodes
emerges from their interference (this is precisely the idea behind the
analytical treatment developed by Efthymiopoulos et al. in
\cite{efthymio:IJBC:2012,efthymio:AnnPhys:2012}).
As can be shown numerically \cite{sanz:jcp:2004,sanz:prb:2004}, if
the circulation integral is performed along the part of a Bohmian
trajectory that encloses a node of the wave function, the result is
a multiple of the quantum flux $2\pi\hbar/m$, in agreement with the
result discussed in Sect.~\ref{sec:scattering}.

Perfectly periodic surfaces behave as reflection gratings, which are
a direct analog of the well-known transmission gratings formed by
arrangements of slits.
Probably the better known of this type of gratings is the two slit,
which was actually the first system analyzed in terms of Bohmian
trajectories, in 1979 by Philippidis et al.~\cite{dewdney:NuovoCimB:1979}.
These authors considered two Gaussian slits, for which the diffraction
problem has an analytical solution, and explained the dynamics exhibited
by the corresponding trajectories in terms of the topology displayed by
the quantum potential.
It was observed that the trajectories essentially move along regions
where this potential is flat, while avoid those where the potential
undergoes sharp, canyon-like variations, in analogy to the dynamical
behavior that a classical particle would also show.
Each time that a trajectory reaches one of those canyons, it feels a
strong force that drives it to the next plateau.

Apart from explaining the formation of the interference fringes in
Young's experiment, the trajectories also display a very interesting
behavior: at the central plateau two groups of trajectories coexist
without mixing, each one coming from a different slit.
Obviously this effect has important consequences at a fundamental
level, since it indicates that it is always possible to elucidate
the slit that the particle passed through without even observing it
directly.
This result, which a priori may seem to be particular of Bohmian
mechanics, is actually a distinctive trait of quantum mechanics,
although its detection requires the use of the quantum density current
\cite{sanz:JPA:2008}.
Indeed data recently obtained from a Young-type experiment with
light have confirmed that the feasibility of this phenomenon
\cite{kocsis:Science:2011}, which was theoretically observed for
light earlier on by several authors
\cite{prosser:ijtp:1976-1,herrmann:AJP:2002,sanz:AnnPhysPhoton:2010}.
A more thorough discussion about the measurement process required to
experimentally find these results can be found in
Sect.~\ref{sec:measurements}.

\begin{figure}
 \begin{center}
 \resizebox{1.00\columnwidth}{!}{\includegraphics{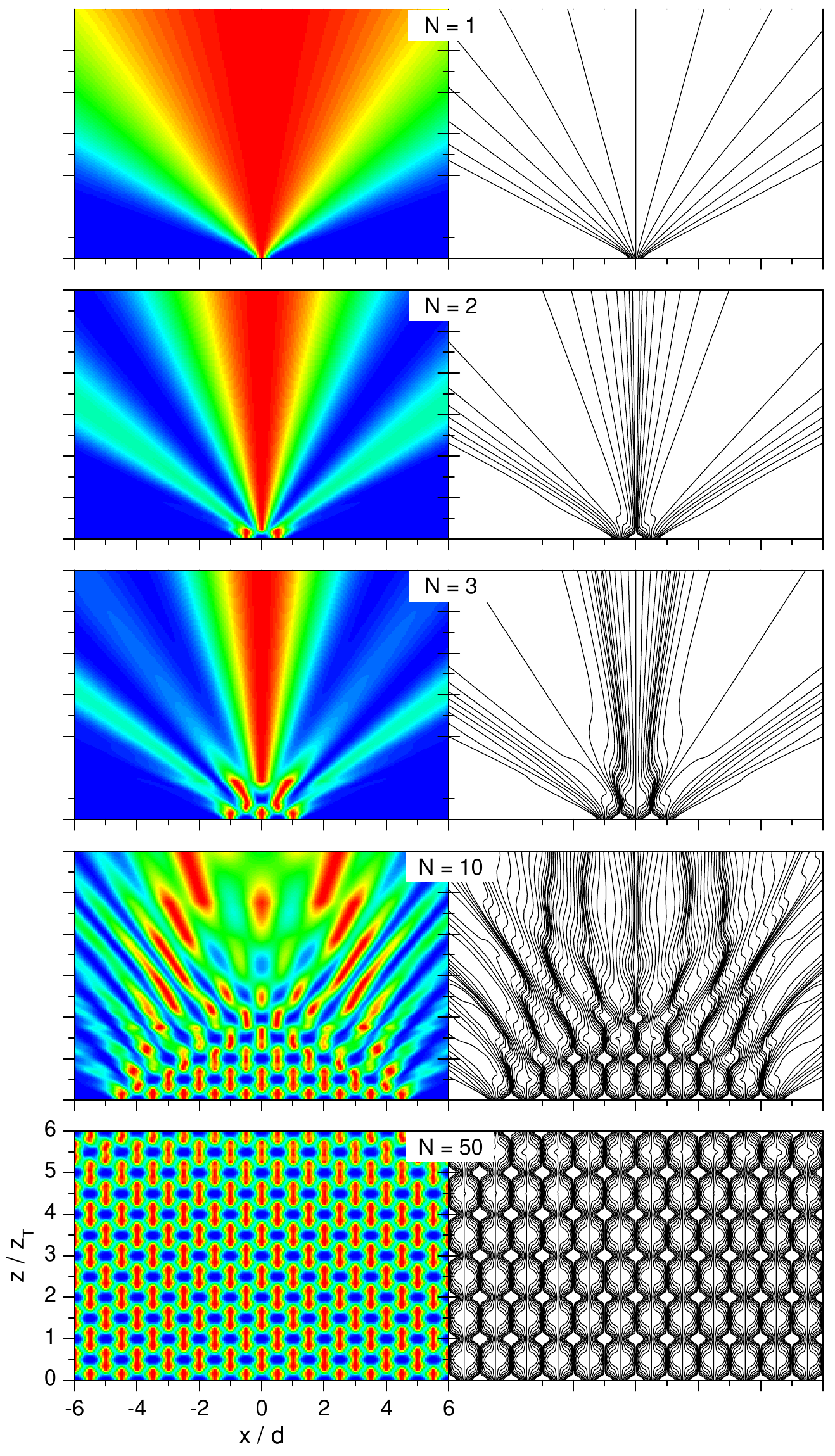}}
 \end{center}
 \caption{Emergence of the Talbot carpet in the near field as the
  number of slits ($N$) in a grating increases (left column) and
  associated Bohmian representation of the phenomenon (right column).
  The $z$-axis is given in terms of the Talbot distance ($z_T$) and
  the $x$-axis in terms of the grating period ($d$) (for particular
  numerical details involved in the simulation, see
  Ref.~\cite{sanz:JCP-Talbot:2007}).}
 \label{fig10revABM}
\end{figure}

More recently, grating diffraction has also been analyzed for
different types of matter waves and various slit arrangements
\cite{sanz:JPCM:2002,sanz:JCP-Talbot:2007}.
In particular, as the number of slits increases it can be noticed
the emergence of a well-ordered structure (see Fig.~\ref{fig10revABM}),
which in the case of a relatively large number of slits becomes a
kind of regular pattern.
This structure is known as Talbot carpet \cite{sanz:JCP-Talbot:2007},
a near field effect that consists of the repetition of the transmission
function of the grating at multiples of the so-called Talbot distance
--- for even integers the pattern is in phase with the grating, while
for odd integers there is half-way displacement.

\subsection{The many body problem}
\label{sec:manybody}

For a system of $N$ particles with a separable Hamiltonian, a many-particle wave function \emph{living} in the $3N$-dimensional configuration space can be constructed from single-particle wave functions. However, for nonseparable Hamiltonians, such a procedure is not possible. Then, the computational burden associated with deriving the $N$-particle wave function makes the exact solution inaccessible in most practical situations. This is known as the many-body problem. This problem was already acknowledged by Dirac\cite{o.dirac1929} in 1929:
\begin{quote}
\emph{``The general theory of quantum mechanics is now almost complete. The underlying physical laws necessary for the mathematical theory of a large part of physics and the whole of chemistry are thus completely known, and the difficulty is only that the exact application of these laws leads to equations much too complicated to be soluble.''}
\end{quote}

There has been a constant effort among the scientific community to provide solutions to the many-body problem. The quantum Monte Carlo solutions of the Schr\"{o}\-din\-ger's equation provide approximate solutions to exact many-particle Hamiltonians \cite{o.nightingale1999}.  The Hartree--Fock algorithm \cite{o.Hartree2,o.fock1930} approximates the many-particle wave function by a single Slater determinant of noninteracting single-particle wave functions. Alternatively, density functional theory (DFT) shows that the charge density can be used to compute any observable without the explicit knowledge of the many-particle wave function \cite{o.Kohn1, o.kohn1964}. Practical computations within DFT make use of the Kohn--Sham theorem \cite{o.kohn1965}, which defines a system of $N$ noninteracting single-particle wave functions that provide a system of equations to find the exact charge density of the interacting system. However, the complexity of the many-body system is still present in the so-called exchange-correlation functional, which is unknown and needs to be approximated. DFT has had a great success, mostly, in chemistry and material science \cite{o.Siesta}, both, dealing with equilibrium systems. Similar ideas can also be used for nonequilibrium time-dependent scenarios, through the \linebreak[4] Runge--Gross theorem \cite{o.Runge1984}, leading to the time-dependent density functional theory (TDDFT). In contrast to the stationary-state DFT, where accurate exchange functionals exist, approximations to the time-dependent exchange-correlation functionals are still in their infancy.  TDDFT has been reformulated in terms of the current density and extended into a stochastic time-dependent current density when the system is interacting with a
bath \cite{o.DiVentra2008book}.

The common strategy in all many-particle approximations is to obtain the observable result from mathematical entities defined in a real space, $\mathds{R}^{3}$, (single-particle wave functions for Hartree--Fock and charge density for DFT) rather than from the many-particle wave function, whose support is defined in the configuration space $\mathds{R}^{3N}$. The Bohmian formalism has also proposed several techniques to get good approximations to the many-body problem. For example, see the early works from Nerukh and  Frederick \cite{o.dmitry2000}, the works by Christov \cite{Christov} already mentioned in \sref{sec:electron-nuclei}, some discussion on the (Bohmian) quantum potential for many-particle systems\cite{o.dellesite2004}, the mixture of classical and quantum degrees of freedom \cite{o.donoso2001,o.drassolov2005} seen in \sref{sec:quantum-classical} or the use of conditional wave functions\cite{o.oriols2007prl}, just to cite a few of them. Many more works on this many-body problem, which become transversal to most research fields, are also mentioned along this review.

\begin{figure}
\resizebox{0.7\columnwidth}{!}{\includegraphics{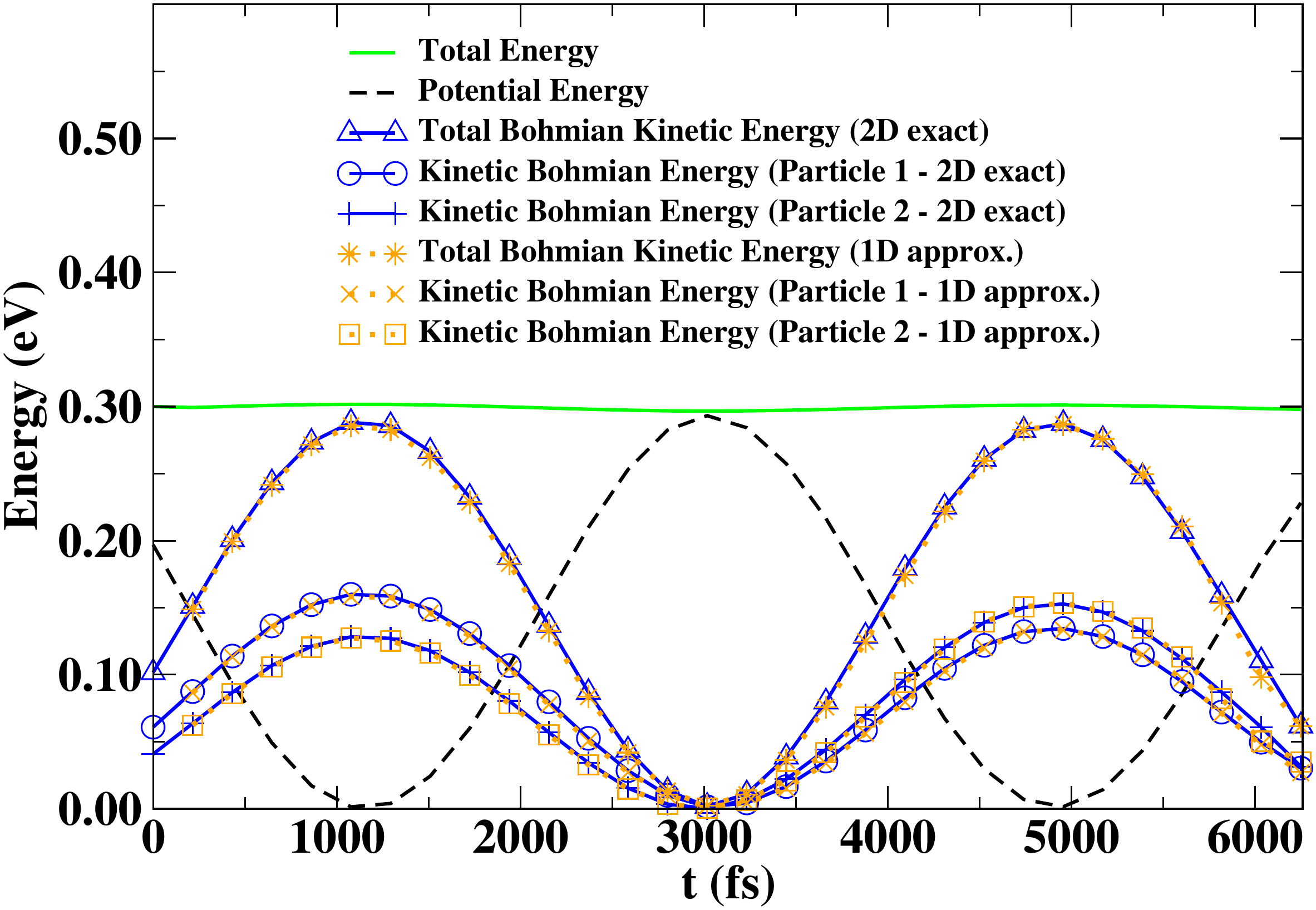}}
\caption{(Color online) Time evolution of individual (ensemble averaged) Bohmian kinetic energies of two identical electrons under a nonseparable potential computed from $2D$ exact and $1D$ approximate solutions. See Ref.  \cite{o.alarcon2013pcm} for details. }
\label{figconditional}
\end{figure}

In this subsection we present and discuss in more detail one of these many-particle Bohmian approaches. As shown in \sref{sec:conditional}, the Bohmian route offers a natural way of finding a single-particle wave function defined in $\mathds{R}^{3}$, while still capturing many-particle features of the system, the so-called \emph{conditional wave function}. It is built by substituting all degrees of freedom present in the many-particle wave function, except one, by its corresponding Bohmian trajectories.  This substitution produces a single-particle wave function with a complicated time dependence. In order to numerically illustrate the ability of the conditional (Bohmian) trajectories presented in \sref{sec:conditional} to treat many-particle systems, we study a simple two-elec\-tron system in $\mathds{R}^2$ under a nonseparable harmonic Hamiltonian with a potential energy:
\begin{align}
U(x_1,x_2) = F \cdot (x_1 - x_2)^2,
\end{align}
with $F = 10^{12}\,$eV/m$^2$ quantifying the strength of the many-body interaction. See Ref.~\cite{o.alarcon2013pcm} for details. Once the exact 2D wave function   $\Phi(x_1, x_2, t)$ is known, we can compute the exact 2D Bohmian trajectories straightforwardly. The initial wave function is a direct product, ${\psi}_{1}(x_1,0) \cdot {\psi}_{2}(x_2,0)$ of two Gaussian wave packets. In particular, we consider $E_{o1}=0.06$ eV, $x_{c1}=50$ nm and $\sigma_{x1}=25$ nm for the first wave packet, and $E_{o2}=0.04$ eV, $x_{c2}=-50$ nm and $\sigma_{x2}=25$ nm for the second. In \fref{figconditional}, we have plotted the ensemble (Bohmian) kinetic energy (from the expectation values defined in \sref{sec:operators}) as:
\begin{align}
\langle K^a_{Bohm}(t) \rangle =   \mathop {\lim }\limits_{M \to \infty } \frac{1}{M}\sum\limits_{\alpha = 1}^M \frac{1}{2}m^*v_a^2(x_a^\alpha(t),t),
\end{align}
where $m^*$ is the (free) electron mass.
We first compute the results directly from the 2D exact wave function. We emphasize that there is an interchange of kinetic energies between the first and second particles (see their kinetic energy in the first and second oscillations in \fref{figconditional}) indicating the nonadiabatic (many-particle) nature of the system. This effect clearly manifests that the Hamiltonian of that quantum system is nonseparable.

Alternatively, we can compute the Bomian trajectories without knowing the many-particle wave function, i.e. using the conditional wave function $\Psi_1(x_1,t)$. This wave function described by Eq.~(\ref{Conditional}) can be easily understood in this simple case. Here the wave function $\Psi_1( x_1,t)$ represents a 1D slice of the whole 2D wave function centered on a particular point of $x_2^\alpha(t)$, i.e. $\Psi_{1}(x_1,t) = \Phi(x_1,x_2^\alpha(t),t)$. The relevant point is that we want to compute $\Psi_{1}(x_1,t)$ without knowing $\Phi(x_1,x_2,t)$. Instead we look for a direct solution of \eref{Conditional_eq} with a proper approximation of $G_{a}$ and $J_{a}$. The reader is referred to section \ref{nonlinearnonunitary}. Here, we consider a zeroth order Taylor expansion around $x_a^\alpha(t)$ for the unknown potentials $G_{a}$ and $J_{a}$.
That is, we consider $G_{a}(x_{a},x^\alpha_{b}(t),t) \approx G_{a}(x^\alpha_a(t),x^\alpha_{b}(t),t)$ as a purely time-dependent potential, and identically \linebreak[4] $J_{a}(x_{a},x^\alpha_{b}(t),t) \approx J_{a}(x^\alpha_a(t),x^\alpha_{b}(t),t)$. This constitutes the simplest approximation that we can adopt. Then, we know that these purely time-dependent terms only introduce a (complex) purely time-dependent phase in the solution of \eref{Conditional_eq}, so we can write $\Psi_a(x_a,t)$ as:
\begin{eqnarray}
\Psi_a(x_a,t)= \tilde{\psi}_{a}(x_a,t) \exp (z^\alpha_{a}(t)),
\label{mpnocoulomb}
\end{eqnarray}
where the term $z^\alpha_{a}(t)$ is a (complex) purely time-dependent phase that has no effect on the trajectory $x^\alpha_a(t)$. Under the previous approximation, \eref{Conditional_eq} can be simplified into the following equation for the computation of $\tilde{\psi}_{a}(x_a,t)$:
\begin{align}
i\hbar\frac{\partial \tilde{\psi}_{a}(x_a,t)}{\partial t} = \Big[ -\frac{\hbar^2}{2m}\frac{\partial^2}{\partial {x^2_a}} + U_a (x_a, t) \Big] \tilde{\psi}_{a}(x_a,t),
\label{setpseudo}
\end{align}
where the potential energies are $U_1 (x_1, t)=F(x_1-x^\alpha_2(t))$ for $a=1$ and $U_2 (x_2,t)=F(x^\alpha_1(t)-x_2)$ for $a=2$.

For this particular scenario, even this rather simple approximation for the unknown terms works perfectly and the agreement between 2D exact results and the 1D approximation mentioned above is excellent (see \fref{figconditional}). The ensemble energies are computed in order to justify that the algorithm is accurate not only for an arbitrarily selected set of Bohmian trajectories, but for an ensemble of them.

An improvement over the simple approximation used here for  $G_{a}$ and $J_{a}$, when constructing the conditional (Bohmian) wave functions, is necessary in other types of interacting potentials to get the same degree of accuracy shown in \fref{figconditional}. One possible approach could be to follow the ideas introduced in Refs. \cite{o.travis2010fp,o.travis2014} where a full (infinite) set of equations for an exact description of the conditional wave functions is presented.

As we mentioned in the introduction, most of the computational quantum tools are developed for (isolated) systems that suffer linear and unitary evolutions. However, most of the quantum systems of interest are far from these idealized conditions. We deal with nonisolated systems that interact with the environment, the measuring apparatus, etc. Such quantum subsystems are not governed by the linear and unitary Schr\"odinger equation. The nonlinear and nonunitary Eq. (\ref{Conditional_eq}) provides an alternative, mainly unexplored, route to tackle these problems.  In principle, one could criticize this route because terms $G_{a}$ and $J_{a}$ are unknown and they introduce ambiguity on the attempt to deal with the conditional wave functions. However, a similar ambiguity is present in the standard route. For example, the \emph{standard} evolution of the wave function of a quantum system is controlled by the unitary (well-defined) Hamiltonian operator (while isolated) and by a nonunitary operator (when interacting with the environment). Over the years,  physicists have developed useful ways to anticipate what type of (nonunitary, projective, weak, continuous, etc.) operator is recommended for each particular (nonlinear and nonunitary) scenario. Additionally, as we have explained in the first paragraphs of this section, the powerful DFT and the TDDFT techniques rely on exchange-correlation functionals that are unknown and need educated guesses. \emph{Why similar instincts can not be developed for the terms  $G_{a}$ and $J_{a}$?}

\subsection{Quantum measurements}
\label{sec:measurements}

A measurement is a process in which a physical system of interest interacts with a second physical system, the apparatus, that is used to inquire information of the former. The word ``measurement'' can be misleading. It is better to use the word ``experiment'' because \emph{``when it is said that something is measured it is difficult not to think of the result as referring to some preexisting property of the object in question''}\cite{o.bell1990pw}. On the contrary, in an experiment, it is natural to think that everything (in the system and apparatus) can change during interactions.

The Bohmian explanation of measurement, which is discussed in references \cite{o.durr2004equilibrium,o.oriols2011book,o.durr2009book,o.durr2012book} and briefly elaborated in \sref{sec:pointer}, is based on avoiding the \emph{artificial} division between what we call the \emph{quantum system} and the measuring  \emph{apparatus}. In Bohmian mechanics, the $N$ particles that define the quantum system and also the $M$ particles of the apparatus have their own Bohmian trajectory and they all share a common many-particle wave function (in an enlarged $N+M$ configuration space). Bohmian mechanics treats quantum measurements as any other type of interaction between two quantum (sub)systems. The outputs of the measurement and their probabilities are obtained from the trajectories that conform the pointer of the apparatus. No need for additional operators (different from the Hamiltonian) or ad-hoc rules.

One of the healthy lessons that one learns from Bohmian mechanics is that most of the time that we talk about a \emph{quantum system}, we are indeed referring to a \emph{quantum subsystem}. Understanding the measurement process in the $N$ configuration space is much more complicated than understanding it in a $M+N$ configuration space. The nonunitary evolution suffered by the \emph{standard} wave function (the so-called collapse of the wave function\cite{o.Born1926}) in the $N$ space can be trivially understood as the evolution of a subquantum system in a larger configuration space. The concept of the conditional wave function explained in \sref{sec:conditional} provides the mathematical bridge between the unitary evolution in a large $N+M$ configuration space and the nonunitary (nonlinear) evolution in a smaller $N$ configuration space.

\subsubsection{Bohmian velocity}

Almost all textbooks on quantum mechanics do only explain the projective (ideal) measurement, where an eigenvalue of some particular operator is obtained as an output, while the quantum system is transformed into an eigenstate of such an operator\footnote{ See, for example, references \cite{o.durr2004equilibrium,o.durr2009book,o.durr2012book} or section 1.4.2 in reference \cite{o.oriols2011book} for the detailed mathematics showing how the Bohmian explanation exactly reproduces this type of measurement.}.  Recently, there has been an enormous interest and significant progress in the study of general quantum measurements. In particular, in pre- and postselection as well as in weak or, more generally, nonprojective measurements. Nothing strange. They are just different types of interaction between system and apparatus, or different treatments of the measured data.

An apparatus that provides a weak measurement is characterized first of all by having a very weak interaction with the quantum subsystem. The final state is only very slightly different from the one associated with the (free) evolution of the state without apparatus. As a consequence of such a small disturbance, the information transmitted from the quantum subsystem to the apparatus is also very small. A single experiment does not produce any useful information because of the weak coupling, therefore, the experiment has to be repeated many times on many identical (preselected) quantum subsystems to obtain reliable information. The information of the system is obtained from a statistical analysis of the data. One of the most striking developments in such studies was the discovery that measurements that are both weak and pre- and postselected provide the so called \emph{weak value} \cite{o.AVV1988}.

\begin{figure}
 \begin{center}
\resizebox{0.95\columnwidth}{!}{\includegraphics{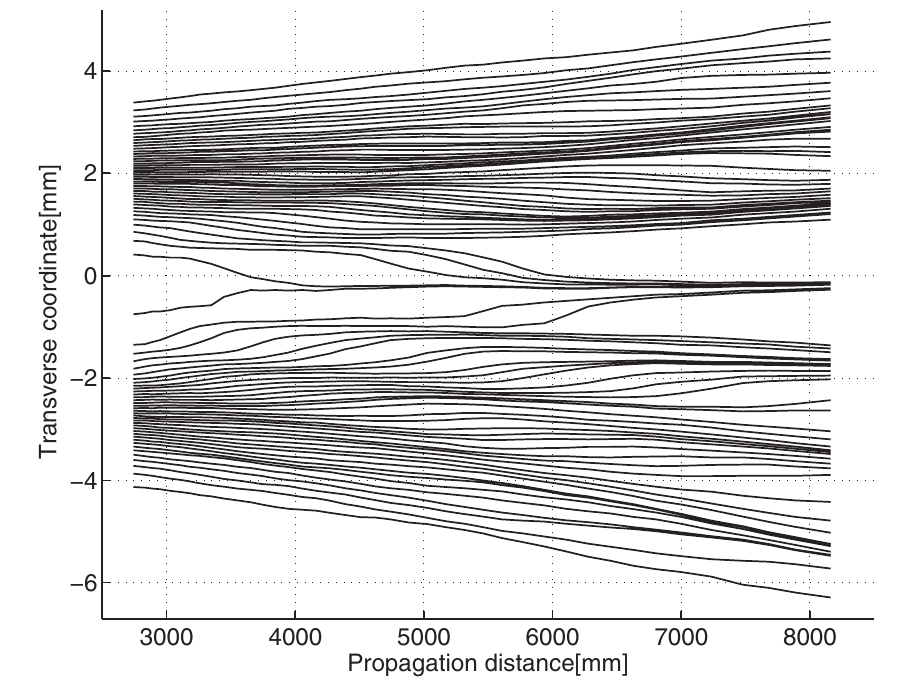}}
 \end{center}
\caption{The reconstructed average trajectories of an ensemble of single photons in the double-slit experiment. Here, 80 trajectories are shown. To reconstruct a set of trajectories,
the weak momentum values are obtained for each transverse $x$ positions at the initial plane. From this initial position and momentum information, the $x$ position on the subsequent imaging plane that each trajectory lands is calculated, and the measured weak momentum value $k_x$ at this point found. This process is repeated until the final imaging plane is reached and the trajectories are traced out.   Results from reference \cite{kocsis:Science:2011},
reprinted with permission from AAAS.}
\label{Kocsis}
\end{figure}

For example, we can preselect the system to some particular (initial) physical state and then make a weak measurement of the momentum (that provides a very small distortion of the system). Later, we make a strong (projective) measurement of the position and we postselect only those weak values of the momentum measurement that provide later a determined position. It has been demonstrated that such a procedure provides information of the local Bohmian velocity of a particle \cite{o.wiseman2007}.  Nowadays, there is an increasing (experimental and fundamental) interest in measuring local velocities of quantum particles.  Recently,  Kocsis \emph{et al.} \cite{kocsis:Science:2011} implemented the mentioned sche\-me to reconstruct the average trajectories for photons in the two-slit experiment. The beautiful experimentally-reconstructed trajectories (see \fref{Kocsis}) are indeed congruent with the iconic images of two-slit Bohmian trajectories \cite{dewdney:NuovoCimB:1979}.

It was recently pointed out by Braverman and Simon \cite{o.Braverman_prl2013} that such measurements, if performed on one particle from an entangled pair, should allow an empirical demonstration of the nonlocal character of Bohmian trajectories. Recently, Traversa et al. \cite{o.traversa_con2013} showed that the measurement of the Bohmian velocity using the concept of weak value developed by Aharanov \emph{et al.}\cite{o.AVV1988} is fully compatible with the more formal concept of positive-operator valued measure (POVM) \cite{Backaction2}. In any case, the measured Bohmian velocity is a \emph{weak value}, i.e. a value obtained after a large ensemble of experimental results. An individual result of the two-times measurement explained above does not provide such a (unperturbed) Bohmian velocity because of the perturbation done on the quantum subsystem by the measurement. It is important to emphasize that it is precisely the ensemble over a large number of experiments that is able to cancel out (in average) the effect of the \emph{true} perturbation done by the measuring apparatus. In some experiments, the measured velocity is larger than what corresponds to the Bohmian velocity of the  (unperturbed) quantum system, while smaller in others\cite{o.DGZ2009}. We emphasize that the trajectories in \fref{Kocsis} cannot be obtained from a single experiment and that the Bohmian velocity for (relativistic) photons is not properly defined (a similar experiment for nonrelativistic particles will be very welcome). In any case, those experimental (ensemble) trajectories for nonrelativistic systems would be numerically equal to the Bohmian (or hydrodynamic) trajectories computed from the wave function solution of the Schr\"odinger equation (when the perturbation due to the measurement process is not taken into account).

Recently, Lundeen \emph{et al.} \cite{o.lundeen_prl2012,o.lundeen_nature2011} showed that the wave function of a particle can be ``directly measured'' using weak measurements. Travis et al.\cite{o.Travis_con2013} showed that if the same weak technique is applied to an entangled system, the result is precisely the conditional wave function discussed in \sref{sec:conditional}.

\subsubsection{Tunneling and arrival times}

There are many experiments where a particle is sent towards a detector and we are interested in knowing \emph{When will the detector click? }or \emph{How much time will the particle spend in a particular region?} The first question is related to the \emph{arrival time} and the second to the \emph{dwell time} or sojourn time.

In most formulations of quantum mechanics, one cannot provide a clear answer to the measure of time, since time itself is not a quantum observable (there is no time operator), but a parameter \cite{pauli-hbk-1,pauli-hbk-2}. In other words, even if time can be somehow measured, as it is inferred from Pauli's theorem \cite{pauli-hbk-1,pauli-hbk-2}, it is not a proper observable because it is not possible to define an associated self-adjoint time-operator consistent with all axioms of quantum mechanics for a system with an energy spectrum bounded from below. Nonetheless, there have been efforts in the literature to introduce such a time-operator \cite{olkhovsky:NuovoCim:1974,olkhovsky:AMP:2009,wang:AnnPhys:2007} as well as various approaches to introduce a proper definition of time
in terms of quantum probability distributions for time observables \cite{muga:PhysRep:2000,muga-bk-1:LNP:2002,muga-bk-2:LNP:2009}.

In principle, it seems that Bohmian mechanics provides an unambiguous answer to the problem of time in quantum mechanics, since the time
connecting two different points along a Bohmian trajectory is a well-defined quantity \cite{leavens:SSC:1990-1,leavens:SSC:1990-2,leavens:PhysLettA:1993,leavens:FoundPhys:1995}. However, one cannot forget the lessons of the Bohmian measurement emphasized in this review: it is a mistake to think of measurable quantities as something intrinsic to the quantum subsystem and independent of the measuring apparatus (see \sref{sec:pointer}). Measuring is a quantum interaction between an apparatus and the system. In other words, there is a difference between the outcomes of measurements, and the actual trajectories of particles belonging to the quantum (sub)system. The utility of the different point of view provided by the Bohmian trajectories on the tunneling/arrival times and the discussion of the several difficulties that the Bohmian measurement involve can be found in Ref. \cite{o.vona2013,o.vona2014,muga:PhysRep:2000,muga-bk-1:LNP:2002,muga-bk-2:LNP:2009}.

In summary, the Bohmian explanation of a quantum measurement is, perhaps, the most attractive (and also ignored) feature of the Bohmian explanation of the quantum nature \cite{o.durr2004equilibrium,o.oriols2011book,o.durr2009book,o.durr2012book}. Apart from reproducing the unitary time-evolution of quantum systems with waves and particles, Bohmian mechanics provides its own formalism to explain all types of (ideal, nonideal, strong weak) of nonunitary evolutions of the measurements, without any additional ad-hoc rule. To be fair, from a computational point of view, Bohmian mechanics does not provide \emph{magical} receipts. The direct application of the Bohmian formalism for measurements needs solving a quantum many-body problem in a $N+M$ configuration space. See \sref{sec:pointer}. In order to make easier practical computations that involve only the quantum (sub)system, the use of operators (if needed) is also very welcomed. See \sref{sec:operators}. For (strong) projective measurements, anticipating which is the operator that we need in our particular problem can be easily found. However, in other practical scenarios finding the explicit operator is even harder than trying to follow the Bohmian route with an explicit modeling of the interaction between the quantum (sub)system and the apparatus. Some very preliminary example is given in \sref{sec:electron-transport}.

\subsection{Quantum chaos}
\label{sec:chaos}

\begin{quote}
{\it ``There is no quantum chaos, in the sense of exponential
sensitivity to initial conditions, but there are several novel
quantum phenomena which reflect the presence of classical chaos. The
study of these phenomena is quantum chaology.''}
\end{quote}
With these words Michael Berry\cite{berry:PhysScr:1989}  somehow summarized
the feeling regarding the idea of extending the notions of classical
chaos to quantum mechanics in the late 1980s and the 1990s.
Over the last decades there has been a very active research on the
question about how the properties of classical chaotic systems manifest
in their quantum counterparts.
From the 1990s on these studies merged with problems involving decoherence
and entanglement \cite{zurek:PRL:1994} due to its interest in modern
quantum information theory and related quantum technologies, in particular
in relation to how nonlinear system-environment couplings eventually affect
the system dynamics \cite{sanz:PRE:2012}.
Nevertheless, the notion of quantum chaos is still ambiguous, and in the
last instance implies the use of quantum-classical correspondence
argumentations and semiclassical methods to ``label'' a quantum system
as regular or chaotic \cite{gutzwiller:1990,reichl:1992}.
These criteria range from level-space statistics of energy
spectra, based on the random matrix theory \cite{bohigas:PRL:1984,mehta:1991},
to the analysis of the behavior of the wave function
\cite{zurek:PRL:1994,zaslavskii:ZhETF:1973,zaslavskii:SPJETP:1974,mcdonald:PRL:1979,heller:PRL:1984}.

Due to the particularities of its formulation, Bohmian mechanics
constitutes an interesting candidate to shed some light on the problem
of quantum chaos.
More specifically, we have a theoretical framework that allows us
to analyze quantum systems with the same tools used in classical
mechanics.
The usefulness of Bohmian mechanics to investigate this kind of
questions was formerly suggested by Bohm and Hiley \cite{bohm-hiley-bk}
in the case of a single particle confined in a two-dimensional box.
About the same time, D\"urr {\it et al.} \cite{duerr:JStatPhys:1992}
and Holland \cite{o.Holand1993} also suggested that the concept of chaos
from classical physics can be extended or generalized to quantum
systems by means of Bohmian trajectories in a natural way.
In this theory quantum chaos arises solely from the dynamical law,
what occurs in a manner far simpler than in the classical case
\cite{duerr:JStatPhys:1992}.
Moreover, considering the complexity introduced in the guidance
equation by the wave function as it evolves and displays a more
intricate interference pattern, a fundamental feature intrinsic to
Bohmian mechanics is the high sensitivity of quantum motion on the
initial conditions \cite{o.Holand1993}.

\subsubsection{Global indicators of chaotic dynamics}

In general, unlike classical motion, in Bohmian mechanics it is
more difficult to find constants of motion \cite{o.Holand1993} ---e.g.,
the momentum or the energy---, and therefore to determine whether a
system is regular or chaotic using this criterion.
This is due to a more generalized conception of motion than in
classical mechanics, where the global structure of trajectories is
highly organized due to the requirement of forming a single-valued
trajectory field or congruence.
In spite of this, such structures (and so quantum motions) may become
rather complex even in the case of relatively simple external potential
functions, just by choosing initial wave functions consisting of
certain combinations of eigenfunctions
\cite{frisk:PLA:1997,parmenter:PLA:1995,parmenter:PLA:1996,makowski:PLA:2000,makowski:ActaPhysPolB:2001}.
This context-dependence makes quantum trajectories to display features
very different of those shown by their classical counterparts,
which cannot be reproduced under any mathematical limit
(see discussion on the quantum-classical transition in
Sect.~\ref{sec:quantum-classical}).
Here we find an example of the relationship between quantum
contextuality and chaos.
In this regard, we find that the problem of a particle in a stadium
potential, for example, is classically chaotic, the nodal patterns of
the quantum eigenfunctions display irregularities according to the
classical dynamics \cite{gutzwiller:1990,mcdonald:PRL:1979}.
From a Bohmian point of view, however, all particles associated with
those eigenfunctions are at rest, since they are real functions of the
spatial coordinates\footnote{Notice that the current density associated
with a real function is zero and therefore the Bohmian velocity is also
zero (see Sect.~\ref{sec:Analytical})}.
On the contrary, we have relatively simple potentials, such as square
or circular billiards \cite{bonfin:PRE:1998,bonfin:PLA:2000}, which are
regular systems, but that can exhibit quantum motions if we choose
some linear combination of their eigenfunctions.

Given the possibility to define trajectories in Bohmian mechanics, the
next reasonable step consists therefore in employing the tools of
classical mechanics for their analysis.
In this regard, it is known that a quantitative characterization of the
degree of chaos requires the calculation of global indicators, e.g.,
Lyapunov exponents or entropies.
The {\it Lyapunov exponent} of a system, $\Lambda$, describes the asymptotic
rate at which the distance between two initially nearby trajectories
evolves with time \cite{gutzwiller:1990}.
Alternatively, chaotic dynamics can also be characterized by measuring
the rate of information exchanged between different parts of the
dynamical system as it evolves \cite{baker-bk:1996}.
This is directly related to the notion of entropy, in particular the
so-called {\it Kolmogorov-Sinai} or {\it KS entropy} \cite{gutzwiller:1990}.
In brief, this type of metric entropy is related to the occupancy rate of
the phase space by a dynamical system or, in other words, the rate of loss
of information in predicting the future evolution of the system by
analyzing the behavior displayed by its trajectories.
Thus, for chaotic systems this entropy is positive, while for regular ones
it gradually decreases.
In 1995 Schwengelbeck and Faisal
\cite{faisal:PLA-1:1995,faisal:PLA-2:1995} proposed one of the first
quantitative measures of the chaoticity of a quantum system based on
the application of the KS entropy combined with the measure of Lyapunov
exponents.
More specifically, they established that in analogously to the
classical definition of chaos, quantum dynamics are chaotic if for a
given region of the phase space the flow of quantum trajectories has
positive Kolmogorov-Sinai entropy.
Nice examples of the application of this idea to describe the
transition from order to chaos quantum-mechanically in terms of
Bohmian trajectories can be found in the literature, as the hydrogen
atom acted by an external electromagnetic field
\cite{iacomelli:PLA:1996} or the H\'enon-Heiles system
\cite{chattaraj:PLA:1996,chattaraj:CurrentSci:1996,chattaraj:CurrentSci:1998}.
On the other hand, in the particular case of the coherent state
representation an alternative definition of quantum instability has
been given by Polavieja and Child
\cite{polavieja:PRA:1996,polavieja:PRE:1997,polavieja:PLA:1996}.

The possibility to establish a one-to-one comparison between classical
and quantum systems is very important, because it could happen that the
existence of classical chaos does not imply that its quantum
counterpart will also display it in Bohmian terms.
This was formerly observed by Schwengelbeck and Faisal
\cite{faisal:PLA-1:1995}, but also by Parmenter and Valentine
\cite{parmenter:PLA:1995,parmenter:PLA:1996}.
By also analyzing the time-evolution of Lyapunov exponents, these
authors found a series of requirements that quantum systems must
fulfil {\it at least} to display a chaotic Bohmian dynamics:
\begin{enumerate}
 \item The system should have two degrees of freedom.
 \item The wave function must be a superposition of three stationary states.
 \item One pair of these stationary states should have mutually
  incommensurate eigenenergies.
\end{enumerate}
Apart from the external potential $V$, quantum dynamics are thus
strongly influenced by the wave function, as mentioned above.
This is the reason why the requirements (2) and (3) are
necessary, and also why quantum chaos can appear in systems for
which classical chaos is not observed.
In particular, Parmenter and Valentine observed this behavior in
the two-dimensional anharmonic oscillator.
Later on Makowski and coworkers also developed a series of studies
exploring the relationship between chaoticity and two degree-of-freedom
wave functions built up of eigenstates and depending on some parameters
\cite{makowski:PLA:1998,makowski:PLA:2000,makowski:ActaPhysPolB:2001,makowski:ActaPhysPolB:2002}.

In summary, one of the remarkable contributions of the Bohmian analysis
to quantum chaos is that quantum instabilities derive from the complexity
of the quantum potential rather than from external classical-like ones.
Relatively simple potentials can then give rise to quantum chaotic motions.
But, how does this take place? What is the mechanism behind this behavior?

\subsubsection{Role of vortical dynamics}

To our knowledge, the first detailed analytical explanation of the
appearance of Bohmian chaos was provided by Frisk \cite{frisk:PLA:1997}
in 1997, who established a much stronger parallelism with classical
Hamiltonian systems.
Without going into mathematical details, it can be said that the most
important conclusion from Frisk's work is the link between
the presence of vortices (nodes of the wave function) and the
observation of a quantum chaotic behavior (from a Bohmian viewpoint).
This is a very interesting result.
In classical mechanics chaotic dynamics are related to the features
characterizing the flow or, in Hamiltonian terms, the potential
function.
In quantum mechanics, however, we find that even if such potential
functions are relatively simple and ``harmless'', the nonlinearity
of the Bohmian guidance equation may lead to chaotic dynamics driven
by the behavior displayed by the wave function along time.
More specifically, this behavior is connected to the presence of
vortices and how Bohmian trajectories may wander around them.
Early examples of this fact were observed, for example, in squared and
circular boxes \cite{frisk:PLA:1997,sprung:PLA:1999} as well as in
stadium-like billiards \cite{sprung:PLA:1999}).

The answer provided by Frisk was in the right direction, yet
incomplete.
In a series of works published later on by Wisniacki and coworkers
\cite{wisniacki:EPL:2005,wisniacki:JPA:2007} it was shown that the
origin of Bohmian chaos is in the evolution in time of those vortices.
An analogous conclusion was also found by Efthymiopoulos and coworkers
\cite{efthymio:JPA:2007,efthymio:CMDynAstr:2008,efthymio:PRE:2009,efthymio:JPA:2012,efthymio:IJBC:2012},
who showed that chaos is due to the presence of moving quantum vortices
forming nodal point-X-point complexes.
In particular, in reference \cite{efthymio:IJBC:2012} a theoretical analysis of
the dependence of Lyapunov exponents of Bohmian trajectories on the
size and speed of the quantum vortices is presented, which explains their
earlier numerical findings \cite{efthymio:JPA:2007,efthymio:CMDynAstr:2008}.

\subsubsection{Relaxation and quantum equilibrium}
\label{sec:relax}

The interest in Bohmian chaos extends in a natural way to the problem of
the quantum equilibrium hypothesis (a discussion on this issue in relation
to the {\it second postulate} can also be found in Sect.~\ref{sec:postulates}).
In analogy to classical systems, it is also argued that this type of
chaos is the cause behind the dynamical origin of quantum relaxation
\cite{valentini:PRSA:2005,valentini:PRSA:2012,schlegel:PLA:2008,bennett:JPA:2010,struyve:NJP:2010}.
It is in this way that Bohmian mechanics provides us with an
explanation of the Born rule $\rho = |\psi|^2$, since it predicts
that, under some conditions, the quantum trajectories lead to an
asymptotic (in time) approach toward this rule even if it was
initially allowed that $\rho_{\rm initial} \ne |\psi_{\rm
initial}|^2$. Nonetheless, it should be noted that not all choices
of $\rho_{\rm initial}$ warrant quantum relaxation, as shown by
Efthymiopoulos {\it et al.} \cite{efthymio:JPA:2012} by means of
an argument used to explain the suppression of quantum
relaxation in the two-slit experiment apply (which also applies to
many other cases, e.g., \cite{sanz:prb:2000}). In particular, a
necessary condition to observe quantum relaxation is that
trajectories should exhibit chaotic dynamics
\cite{valentini:PRSA:2005,efthymio:JPA:2012}, yet it is not
sufficient \cite{efthymio:JPA:2012}. On the other hand, it is also
important to stress that this is analogous to the usual Monte Carlo
sampling used in molecular dynamics simulations
\cite{brumer-elran:JCP:2013}, which also makes use of a ratio
$\rho/W$, where $\rho$ is a convenient distribution of classical
trajectories and $W$ is a given distribution function, e.g., a
Wigner distribution within the classical Wigner method ---in
Valentini's approach, $\rho$ is an arbitrary probability density and
$W$ is the probability distribution described by $|\psi|^2$. In this
regard, notice that Bohmian mechanics allows us to put quantum
mechanics at the same level of classical statistical mechanics.
Because of the probability conservation along trajectories
(or, more formally, tubes of probability in the limit of a vanishing
cross-section \cite{sanz:AnnPhys:2013}), the ratio is usually
evaluated at $t=0$.

Apart from the fundamental aspect related to the quantum equilibrium
hypothesis, in the literature it is also possible to find different
works with a more practical orientation, where Bohmian mechanics is
used as a tool to explore and analyze the relaxation dynamics of
quantum systems
\cite{maddox:JCP:2001,maddox:PRE:2002,holloway:JCP:2001,burghardt:JCP-1:2001,burghardt:JCP-2:2001,burghardt:IJQC:2004,burghardt:JCP:2004,burghardt:JPCA:2007,garashchuk:JCP:2013}.
Similarly, the appealing feature of dealing with ensembles has also
been considered to develop semiclassical-like statistical numerical
approaches \cite{bittner:JCP:2003,makri:JCP:2003,makri:JPCA-1:2004,makri:JPCA-2:2004}.
In spite of all these efforts, there is still much to be done and
developed, particularly in relation to large, complex systems
(see Sect.~\ref{sec:electron-nuclei}) as well as the system-apparatus
coupling in measurement processes (for related discussions, see
Sects.~\ref{sec:electron-transport}, \ref{sec:measurements},
\ref{sec:pointer}, and \ref{sec:operators}).

\subsection{Quantum-to-classical transition and decoherence}
\label{sec:quantum-classical}

Bohmian mechanics has a very appealing feature regarding the
quantum-classical transition: the quantum potential.
This additional term to the Hamilton--Jacobi equation
contains information about the topological curvature of
the wave function in the configuration space and is proportional to
$\hbar^2/m$ (see Sect.~\ref{sec:Hamilton-Jacobi}).
According to standard textbook criteria
for the classical limit, classical behaviors are expected in the
$\hbar \to 0$ limit (or, equivalently, as the mass or some
relevant quantum numbers of the system become increasing larger).
Therefore, in Bohmian mechanics everything should be very simple:
the condition $\hbar = 0$ should lead to classical mechanics. The
reality, however, is far more complicated, as it was already pointed
by Rosen
\cite{rosen:AJP-1:1964,rosen:AJP-2:1964,rosen:AJP:1965,rosen:FoundPhys:1985}
in the mid-1960s.
This is the traditional classical limit.
Now, $\hbar$ is just a constant and therefore one could also think the
classical limit in a more physical manner, in terms of larger and larger
masses.
Here these two limits will be examined, as well as the more recent route
in terms of {\it entanglement} and {\it decoherence}.

\begin{figure}[t]
 \begin{center}
  \resizebox{1.00\columnwidth}{!}{\includegraphics{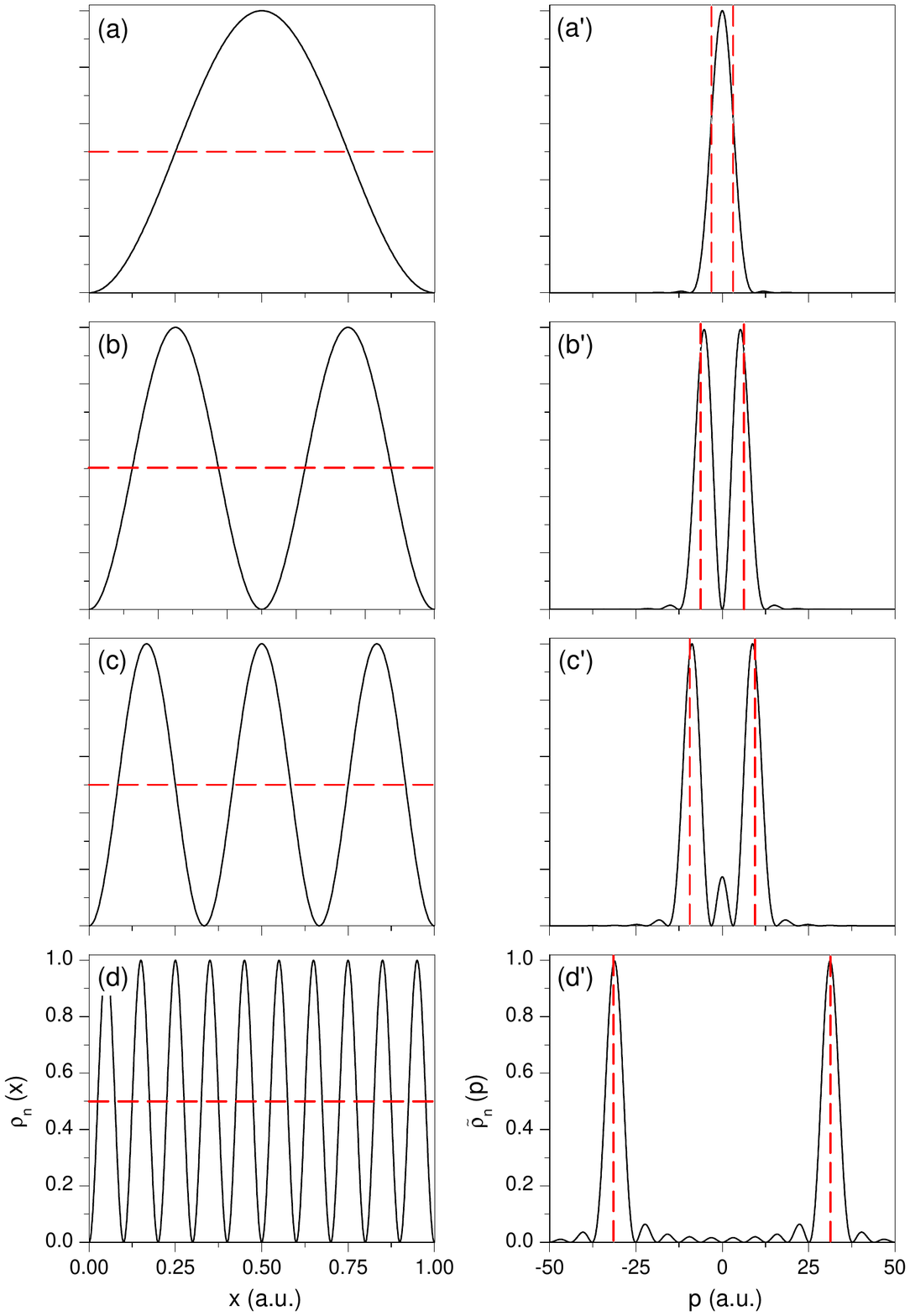}}
 \end{center}
 \caption{Probability density for the position (left) and momentum
  (right) eigenfunctions of the particle in a one-dimensional infinite
  well for: (a) $n=1$, (b) $n=2$, (c) $n=3$, and (d) $n=10$; the
  classical counterparts are denoted with red dashed lines.
  For visual clarity, the maximum of all densities has been set to
  unity.
  Arbitrary units (a.u.) have been considered.}
 \label{fig13revABM}
\end{figure}

\subsubsection{``Traditional'' classical limits}

Newtonian mechanics smoothly arises from special relativity as the
speed of the system becomes smaller and smaller than the speed of
light. This is possible because both theories are built upon the
same conception of systems as localized (point-like) objects in
space and time. In the limit $v \ll c$, we gradually recover the
Newtonian description in terms of speed-independent masses and
separation of space and time. In quantum mechanics, it is often
considered that the limit $\hbar \to 0$ (or equivalently $\hbar
\ll S_{\rm cl}$, where $S_{\rm cl}$ denotes a related classical
action) should also behave the same way. As pointed out by Michael
Berry \cite{berry:LesHouches:1991}, the limit $\hbar \to 0$ is
singular (and ``pathological'').
To illustrate this fact in a simple manner, we are going to consider
a system that has already been used in a similar fashion in different
contexts \cite{sanz:SSR:2004,drezet:ch:2014}, namely the particle in
a one-dimensional square infinite well of size $L$.
The eigenfunctions are of the form
\begin{align}
\phi_n (x) = \sqrt{\frac{2}{L}} \sin \left(\frac{p_n x}{\hbar}\right),
\end{align}
where $p_n = \pi n\hbar /L$, with $n = 1, 2, \ldots$
Therefore, the corresponding probability densities are
\begin{align}
\rho_n (x) = \frac{2}{L} \sin^2 \left(\frac{p_n x}{\hbar}\right).
\end{align}
Similarly, in the momentum space we also have time-in\-de\-pen\-dent
eigenfunctions, $\tilde{\phi}_n(p)$, from which we obtain probability
densities
\begin{align}
\tilde{\rho}_n (p) \propto \left\{ \begin{array}{lcc}
 \displaystyle \frac{p_n^2}{(p^2 - p_n^2)^2} \cos^2 \left(\frac{pL}{2\hbar}\right) , & \quad & \textrm{odd $n$} , \\
 \displaystyle \frac{p^2}{(p^2 - p_n^2)^2} \sin^2 \left(\frac{pL}{2\hbar}\right) , & \quad & \textrm{even $n$} .
 \end{array} \right. \quad 
\end{align}
These probability densities in the configuration and momentum
representations are displayed in Fig.~\ref{fig13revABM} (left and
right columns, respectively) for different values of $n$; in each
case, the red dashed line indicates the classical distribution (a
continuous function for positions and two $\delta$-functions at
$\pm p_n$ for momenta).
As it can be noticed, the probability density in the configuration
space is a strongly oscillatory function in the classical limit
($n \to \infty$), whose average (but not the distribution itself)
coincides with the classical one.
On the other hand, in the momentum space one approaches two
distributions centered around the two classical momentum values
allowed, $\pm p_n$, at a given energy ($E_n = p_n^2/2m$).
From a Bohmian perspective, the particles associated with $\phi_n$
are motionless.
The effective potential acting on them is constant (consisting only of
the quantum potential, since $V=0$ in the allowed region) and equal to
the total energy $E_n$.
On the other hand, the kinetic energy is zero, since
$p_B = \partial S/\partial x = 0$.
Therefore, in this case, no classical limit can be reached, since there
is no way that Bohmian particles can move with a well-defined momentum
$p_{\rm cl} = \pm \ p_n$.

\subsubsection{``Physical'' classical limit ($m\to\infty$)}

Now, what happens if instead of this limit we assume a more realistic
one, as it is the case with an increasing mass?
The quantum potential should also vanish and therefore the classical
Hamilton-Jacobi equation should start ruling the system dynamics.
To illustrate this case, let us consider a weakly corrugated surface,
such as the (110) copper surface, and normal incident conditions
(the atoms impinge on the surface perpendicularly from above, with
$z>0$, assuming the latter is on the $XY$ plane) with an incident
energy $E_i = 21$~meV.
The rare gas of incident atoms is described by an extended wave packet that
covers several unit cells above the surface; because of this extension,
it reaches the surface with almost no spreading \cite{sanz:JPA:2008}.
The propagation starts in the classical asymptotic region, where the
atom-surface interaction potential is negligible ($z > 12$~\AA). From this region, the wave packet is launched against the surface. After reaching the surface, the wave packet bounces backwards and is allowed to go far beyond the classical asymptotic region.
As for the particles, let us consider the sequence He, Ne, Ar, and
He$^*$, the latter being a fictitious atom with mass
$m_{\rm He^*}=500 \ m_{\rm He}$.
The question is, how is the far-field diffraction pattern modified by
the increasing mass of the incident particle?
If in the far field the quantum potential vanishes ($Q \to 0$),
the answer that one would expect is that this pattern
approaches the one displayed by a classical distribution.

\begin{figure}[t]
 \begin{center}
  \resizebox{1.00\columnwidth}{!}{\includegraphics{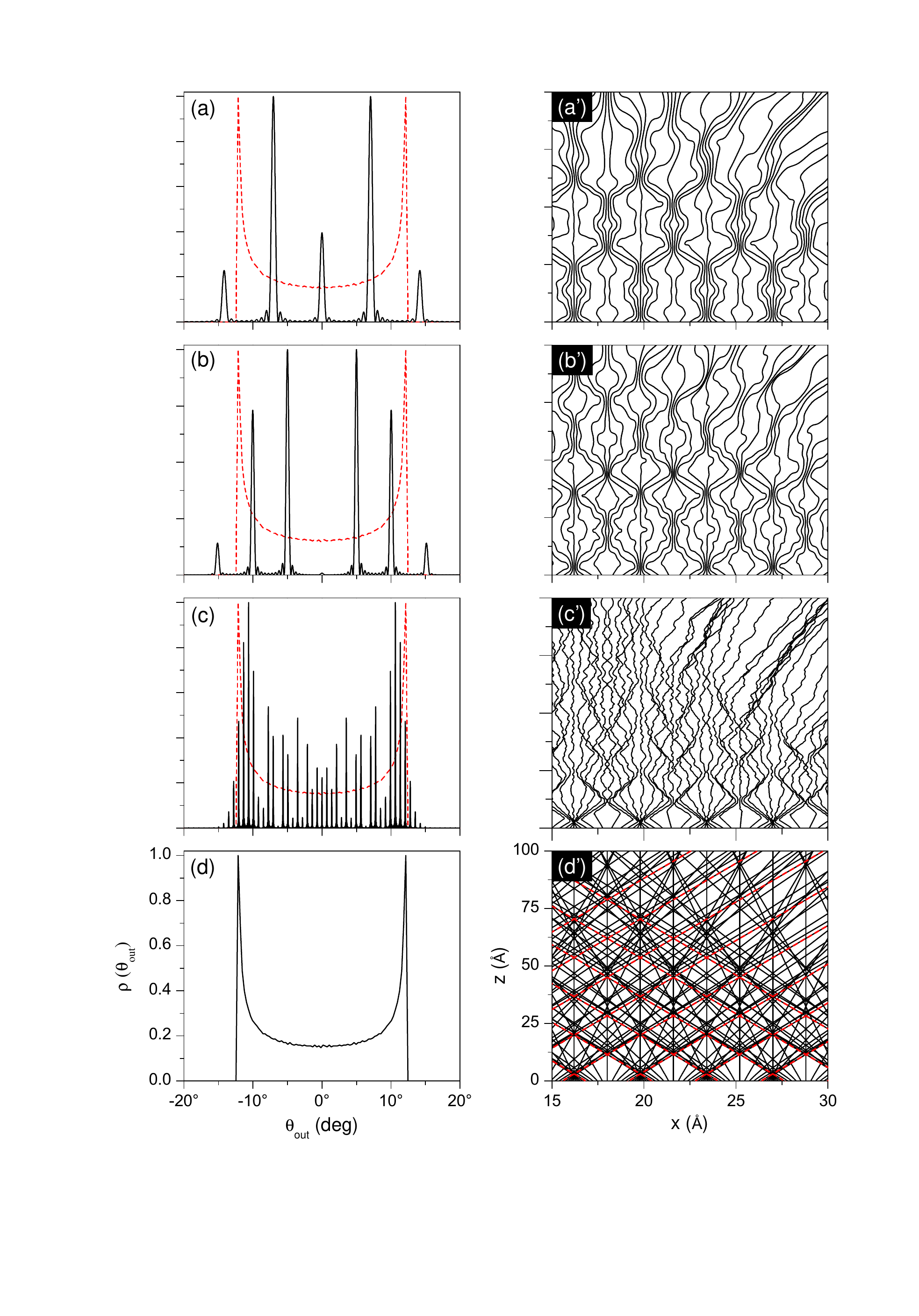}}
 \end{center}
 \caption{Angular intensity distribution (left) and Bohmian
  trajectories for X-Cu(110) scattering at normal incidence and
  $E_i=21$~meV, with X being: (a) Ne, (b) Ar, (c) He$^*$, and
  (d) He$_{\rm cl}$.
  For visual clarity, only the outgoing part of the trajectories near
  the surface has been displayed.
  In panel (d'), the dashed red lines denote classical trajectories
  deflected in the direction of rainbow angles.
  For computational details, see Refs.~\cite{sanz:SSR:2004,sanz-bk-2}.}
 \label{fig14revABM}
\end{figure}

For angles at which classical trajectories become maximally deflected
---typically when they impinge on inflection points of the
equipotential energy surface at $E_i=21$~meV---, the intensity goes to
infinity \cite{sanz:SSR:2004,sanz-bk-2}.
The effect is known as rainbow effect \cite{child-bk:1974} in analogy
to the optical rainbow, with the maximum angle of inflection being the
rainbow angle.
In X-Cu(110) scattering there are only two rainbow features, which play
the role of turning points for the angular distribution ---the intensity
for any other angle is confined between them.
This is in sharp contrast with the quantum case, where one observes a
series of diffraction intensity maxima at the corresponding Bragg
angles, as seen in Fig.~\ref{fig14revABM}(a).
However, as the mass of the incident particles increases,
not only more and more Bragg diffraction peaks
can be observed (their number scales with the mass of the incident
particle as $n \propto \sqrt{m}$), but their intensity is such that
the whole pattern, on average, approaches
that of the classical distribution \cite{sanz:SSR:2004}, as seen
in Fig.~\ref{fig14revABM}(c).
That is, if $m$ increases gradually, the intensity distribution
resembles the classical distribution {\it on average}, but the fine
structure still consists of an increasingly large series of
interference maxima, just as in the problem of the infinite well.
From a Bohmian perspective (see right column of Fig.~\ref{fig14revABM}),
what we observe as $m$ increases is that trajectories start mimicking
the behavior of classical trajectories.
One could be tempted to think that we are actually observing the
emergence of classical trajectories (note that for He$^*$ the prefactor
in the quantum potential is 500 times smaller than in the case of He).
However, because of the intricate interference structure of the wave
function, the quantum potential also exhibits a rather complex
topology \cite{sanz-bk-1}, such that the factor depending
on the curvature of the wave function does not vanish.
Consequently, one of the distinctive traits of Bohmian mechanics, the
noncrossing rule \cite{o.oriols1996pra}, still remains valid.

\begin{figure}[t]
 \begin{center}
  \resizebox{1.00\columnwidth}{!}{\includegraphics{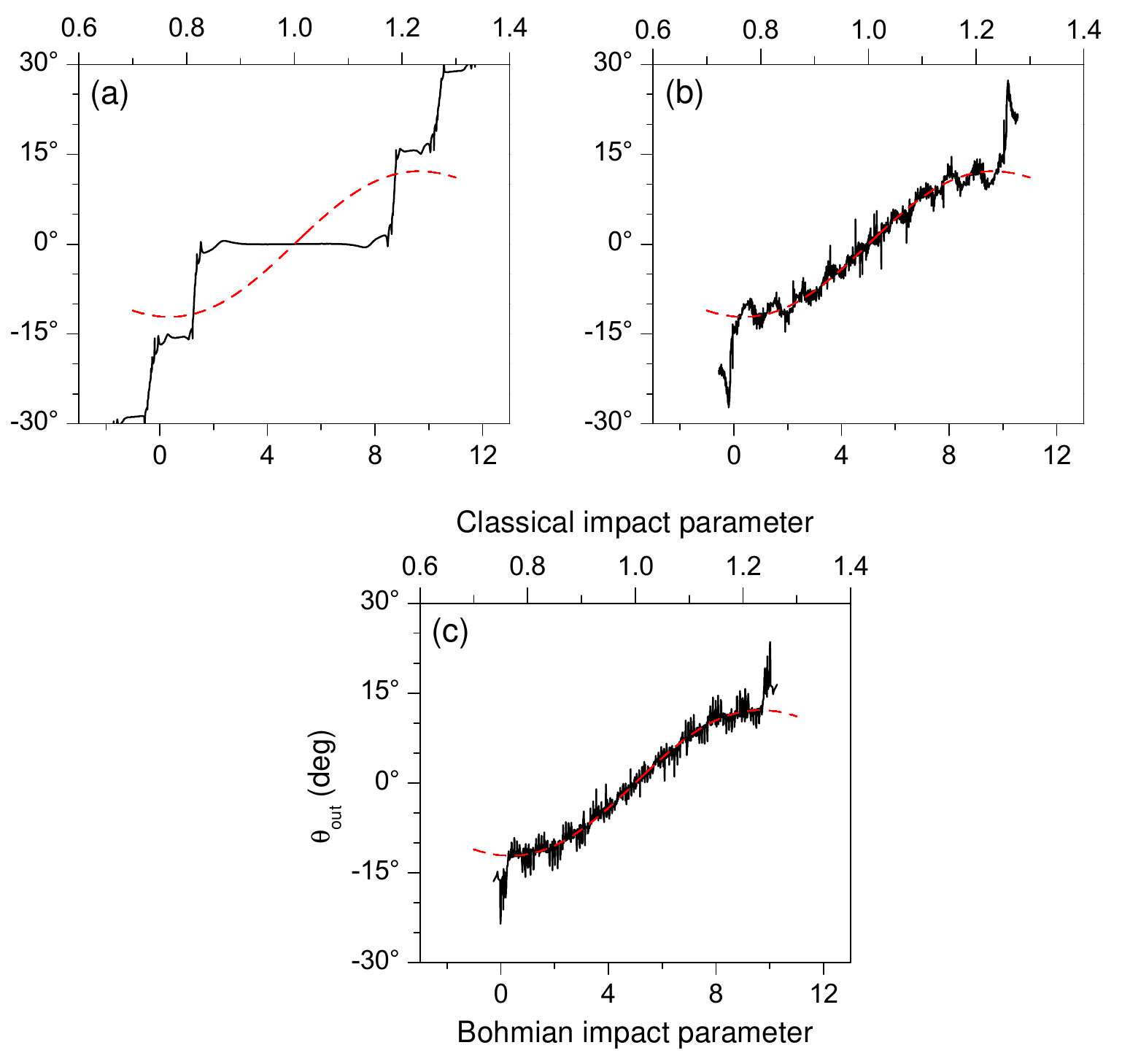}}
 \end{center}
 \caption{Quantum deflection functions obtained from the Bohmian
  trajectories of particles with masses: (a) $m_{\rm He}$,
  (b) $m_{\rm He^*} = 500 m_{\rm He}$, and
  (c) $m_{\rm He^*} = 1000 m_{\rm He}$.
  In each panel, the classical deflection function of He is displayed
  with thick red line (only the part between the two rainbow angles is displayed).}
 \label{fig15revABM}
\end{figure}

The trajectories show us the behavior of the system at a local level,
but what about a more global level?
We have observed that intensity patterns approach (on average) the
classical ones.
What about the trajectories?
To illustrate this point, let us consider the deflection function.
In classical mechanics, this is just a representation of the final
or deflection angle displayed by the trajectory (with respect to the
normal to the surface) versus its impact parameter (its position
within the distance covered by a unit cell).
By varying the particle position from 0 to 1 along the unit cell, we
obtain a map of the accessible final angles, which for normal incidence
and Cu(110) is essentially a sinusoidal function.
The maximum and minimum of this curve are the two rainbow angles.
We can proceed in the same way with the Bohmian trajectories, although
instead of covering a single unit cell it is necessary to cover the
whole extension of the incoming wave.
When we normalize this extension to the length covered by a single unit
cell, the result for He is shown in Fig.~\ref{fig15revABM}(a).
The staircase structure observed in the quantum deflection function is
related to the appearance of Bragg angles, each step corresponding to
a different diffraction order.
As the mass of the incoming particle increases, we find that this
staircase structure becomes more complex, with more steps, that make
it to resemble the classical deflection function [see panels (b) and (c)].
There are, however, three differences.
First, even if the trajectories cannot cross, they try to mimic the
classical behavior within each unit cell, which gives rise to a series
of oscillations along the deflection function.
Second, only a half of the classical deflection function can be
reproduced.
Third, the quantum deflection function can only reproduce globally the
classical one, but not within each cell.

\subsubsection{Role of entanglement and decoherence}

The above two examples show that the standard textbook argumentation
around the classical limit is rather vague and confusing.
Even if the very idea of quantum-classical correspondence is a
priori interesting in itself ---actually, it allowed Bohr to lay down the
foundations of modern quantum mechanics (to be distinguished from
Planck's and Einstein's former theory of quanta)---, we find it to
be incomplete to explain the appearance of the classical world, as
also argued by different authors in the literature.
Apart from Rosen, more recently the idea of identifying quantum
motions that are analogous to classical ones have also been
considered by Makowski and coworkers \cite{makowski:PRA-1:2002,makowski:PRA-2:2002}.
In any case, it seems that the criterion of a vanishing quantum
potential with $\hbar$ (or $m$, or large quantum numbers) is not a
general criterion of classicality
\cite{holland-cushing-bk:1996,bolivar:CanJPhys:2003,bolivar-bk:2004}.
Even in the case that semiclassical wave functions propagate or
spread along classical trajectories (e.g., scar-like quantum states
\cite{heller:PRL:1984}), Bohmian trajectories may look highly nonclassical
\cite{matzkin:PLA:2007,matzkin:SHPMP:2008,matzkin:JPCS:2009,matzkin:FoundPhys:2009},
thus breaking the traditional notion of quantum-classical correspondence.

To properly address the question of the classical limit or, to be more
precise, the quantum-to-classical transition, first it is worth noting
that Bohmian mechanics is confined to the configuration space (as it is
the wave function, unless other representation is chosen).
However, classical mechanics takes place in phase space, where coordinates
and momenta are independent variables, and therefore at each time it is
possible to observe that the same spatial point can be crossed by
trajectories with equal but opposite momenta.
This is a strong condition that should be satisfied in a proper
classical limit; reaching the classical Hamilton-Jacobi equation is
just a weak condition, as seen above, since it does not warrant the
observation of twofold momenta.
So, what can we do in this situation? Is there any way to get rid
of the fast oscillations that appear in the so-called classical limit
(independently of how this limit is reached)?
The answer to such a question seems to be linked to entanglement
\cite{giulini-bk,schlosshauer-bk:2007}, the inseparability among
different degrees of freedom that eventually leads to an apparent
loss of coherence or decoherence in a particular property of the
degree of freedom (or system) of interest.
Different authors have treated the problem of entanglement within the
Bohmian framework in various contexts
\cite{dewdney:FoundPhys:1988,dewdney:PLA:1990,hiley:FoundPhys:1999,golshani:JPA:2001,ghose:PRAMANA:2002,marchildon:JModOpt:2003,marchildon:JPA:2003,struyve:JPA:2003}
---actually Bohmian trajectories have also been used to analyze no-go
theorems and the appearance of nonlocality in quantum mechanics
\cite{durt:PRA:2002}.
The main idea underlying all these works is the following.
If one observes the dynamics of the full-dimensional system (system of
interest plus environment), the corresponding Bohmian trajectories
satisfy the usual noncrossing rule \cite{o.oriols1996pra}.
Now, these trajectories contain information about both the system and
the environment.
In order to examine the system dynamics, one has to select only the
respective components of those full-dimensional trajectories.
This is equivalent to observing the dynamics in a subspace, namely the
system subspace.
It is in this subspace where the system trajectories display
crossings, just because they are not (many-particle) Bohmian trajectories
in the high-dimensional space, but trajectories onto a particular subspace
---actually this leads us immediately to the notion of conditional wave
function developed in Sect.~\ref{sec:conditional}, which opens new paths
to formally establish a bridge between these various subspaces.
Here the role of the environment consists of relaxing the system noncrossing
property by allowing its (reduced) trajectories to reach regions of
the configuration subspace which are unaccessible when the system is
isolated.
This thus explains the phenomenon of decoherence
\cite{na:PLA:2002,na:PhysScr:2003}.
Probably this is one of the most important aspects that make Bohmian
mechanics worth exploring and using, since it provides an unambiguous
prescription to monitor the flow and exchange of quantum coherence
between system and environment.
Any other quantum approach only offers a rather abstract and unclear
picture of this phenomenon, even though it is the key question in the
theory of open quantum systems \cite{breuer-bk:2002}.

Taking into account how the system is influenced by the environment
within the Bohmian framework, it is now quite clear how the fast
oscillatory interferences previously mentioned will be washed out.
The issue was formerly discussed by Allori {\it et al.}\
\cite{allori:JOptB:2002}, who claimed that Bohmian mechanics
constitutes, precisely, the correct way to recover the classical
limit.
This limit should be essentially analogous to determine when Bohmian
trajectories look Newtonian, but in a rather different fashion to other
attempts based on the standard conception of correspondence of simply
varying a certain control parameter.
This was what they called the seven steps towards the classical world.
The main idea behind the approach followed in this work consists of
formulating a given quantum problem within the Bohmian representation
of quantum mechanics in the most complete way, i.e., including all
possible degrees of freedom.
Some of these degrees of freedom will describe the system of interest,
while the remaining will be treated like an environment.
Even without introducing any particular assumption or limit (as the
authors do), it is clear that the trajectories displayed by the system
---remember that they are projections of the real full-dimensional
Bohmian trajectories onto the system configuration subspace--- will be
able to cross and therefore to show that certainly a given spatial
point of the system subspace can be characterized by two (or more, if
the system is described by two or more degrees of freedom) values of
the momentum.

A current trend in different areas of physics and chemistry is the
study of large and complex systems (i.e., systems characterized by a
large number of degrees of freedom) as well as the implementation of
the corresponding numerical tools to achieve such a goal.
As mentioned at the end of Sect.~\ref{sec:chaos}, Bohmian mechanics
has been considered as one of the feasible routes to explore.
Various approaches have been developed in this regard, particularly
mixed Bohmian-classical methods
\cite{Meier1,Meier2,gindensperger:JCP:2002-1,gindensperger:JCP:2002-2,gindensperger:AdvQuantChem:2004,meier:JCP:2004}
and analogous extensions making use of the hydrodynamic approach
\cite{burghardt:JCP-1:2001,burghardt:JCP-2:2001,burghardt:IJQC:2004,burghardt:JCP:2004,burghardt:JPCA:2007}.
The conditional wave function (see Sect.~\ref{sec:conditional}) is
another alternative of interest worth exploring.
Nonetheless, there is still a lot to be done in this area, mainly in
connection to the system-apparatus interactions, which unavoidably
lead us to questions such as how entanglement manifests in ordinary
life, or how decoherence processes can be optimally controlled.
These are problems where Bohmian mechanics has a lot to say.

\section{Formalism}
\label{sec:formalism}

In this section we describe the different formalisms that are used to make Bohmian predictions, in general, and in most of the reviewed works of \sandsref{sec:application1}{sec:application2}, 
in particular. As emphasized in \sref{sec:intro}, the formalisms are just the mathematical tools that explain how predictions are obtained. 
One theory can have many valid formalisms (or subroutes). For example, the matrix or the wave formulations of the standard quantum theory. 
From a physical point of view, the only requirement for a valid formalism is that, by construction, it exactly reproduces experiments. 
For some particular quantum problems, it is better to compute trajectories by solving the Schr\"odinger equation, as explained in \sref{sec:Analytical}, while for others the computation using 
the Hamilton-Jacobi equation (or the quantum potential) described in \sref{sec:Hamilton-Jacobi} is preferred. Identically, expectation values can be computed from the operators as 
explained in \sref{sec:operators} or from the pointer
trajectories as detailed in \sref{sec:pointer}. For practical purposes, all Bohmian subroutes are valid\footnote{From a metaphysical point of view, different formalisms are associated with slightly different interpretations of the Bohmian theory. For example some researchers defend that \sref{sec:Analytical} is a \emph{better} formalism because it is a first order (velocity) formalism which is the correct ontologic understanding of Bohmian mechanics.  On the contrary, others defend the second order explanation done in \sref{sec:Hamilton-Jacobi} because their ontological understanding of the Bohmian theory is based on the quantum potential (acceleration). Most of Bohmian researchers will only accept an ontological explanation of the measurement in terms of pointer positions of \sref{sec:pointer} (\emph{Naive realism about operators} \cite{daumer1996naive,o.durr2004equilibrium,o.durr2012book}). However, in many practical situations, the subroute with operators in \sref{sec:operators} is very useful.
Once more, for our practical interest, any ontological hierarchy of the different formalisms is irrelevant.}.

\subsection{Trajectories from the Schr\"odinger equation }
\label{sec:Analytical}

In Bohmian mechanics, a quantum system is described by both a wave function and a particle position which describes a well-defined trajectory guided by the wave function.
There are two main approaches to compute the dynamics of a system in Bohmian mechanics:
analytic and synthetic algorithms to which we will dedicate the following two sections, respectively.
This distinction comes from the analytic-synthetic dichotomy in philosophy.

In synthetic algorithms, Bohmian trajectories play a key part in the algorithm to perform the computations, i.e., as the points where the wave function is evaluated.
Thus, these algorithms require an extra step in formulating them, the quantum Hamilton--Jacobi equation, which we will introduce in \sref{sec:Hamilton-Jacobi}.

The basis of analytic approaches, however, consists in computing first the wave function and then obtaining the Bohmian trajectories from it.
In a sense, the trajectories do not contribute to the structure of the algorithm, but are simply obtained by the equations in the formalism.
While these algorithms do not add, in principle, any computational advantage, e.g., the trajectory computation is an additional step to integrating the Schr\"odinger equation, they can be easily implemented to obtain the trajectory dynamics which can be very useful to gain insights into the dynamics, see for instance the works in Sects.~\ref{sec:atoms}, \ref{sec:light-matter}, and \ref{sec:chaos}.
On the other hand, they are at the foundation of the conditional wave function algorithms which will be discussed in \sref{sec:conditional}.

As in standard quantum mechanics,
the time evolution of the wave function is given by the Schr\"odinger equation.
For a system composed of $N$ spinless charged particles under the effect of an electromagnetic field it takes the form:
\begin{align}
\label{eq.schrodingervectorpotential}
i \hbar \frac{\partial \psi(\vec{r},t)} {\partial t} &= \sum_k \frac {1} {2 m_k} \left[- i \hbar\nabla_k - q_k \vec{A}_k(\vec{r}_k,t)\right]^2  \psi(\vec{r},t) \nonumber \\ &+ V(\vec{r},t) \psi(\vec{r},t).
\end{align}
where $m_k$ and $q_k$ are, respectively, the mass and charge of the $k$-th particle, and $\nabla_k$ is the del operator with respect to the coordinates $\vec{r}_k$.
The vector potential $\vec{A}_k(\vec{r}_k,t)$ represents the electromagnetic field (in the $k$-th particle's subspace) and $V(\vec{r},t)$ describes both the interactions between the particles in the system and the effects of an external potential.
The usual Schr\"odinger equation for neutral particles (or in the absence of an electromagnetic field) can be recovered by taking $\vec{A}_k \rightarrow 0$ in \eref{eq.schrodingervectorpotential}.
Thus, in the remainder of this section (and in the next one), this limit can be used at any time to obtain the expressions corresponding to the neutral particles case.

The solutions of the Schr\"odinger equation obey a continuity equation.
From \eref{eq.schrodingervectorpotential} it is easy to see that
\begin{align}
\label{eq.continuityanalytic}
\frac{\partial\rho (\vec{r},t)}{\partial t} + \nabla \cdot \vec{j}(\vec{r},t) = 0,
\end{align}
where we have defined a probability density and its associated current as
\begin{align}
\rho(\vec{r},t) &= |\psi(\vec{r},t)|^2, \\
\vec{j}(\vec{r},t)&= \sum_k \vec{j}_k(\vec{r},t) = \sum_k \left[\frac{\hbar}{m_k} \mathop{\textrm{Im}}\left( \psi^*(\vec{r},t) \nabla_k \psi(\vec{r},t)\right)\right. \nonumber \\
&\qquad\qquad\qquad\qquad- \left.\frac{q_k}{m_k} |\psi(\vec{r},t)|^2 \vec{A}_k(\vec{r}_k,t)\right] .
\label{fluxvector}
\end{align}

These densities lead to the definition of a velocity field,
\begin{align}
\label{eq.joverrho}
\vec{v}(\vec{r},t) =& \frac{\vec{j}(\vec{r},t)}{\rho(\vec{r},t)} \nonumber \\
=& \sum_k \left[\frac{\hbar}{m_k} \mathop{\rm Im}\left(\frac{\nabla_k \psi(\vec{r},t)}{\psi(\vec{r},t)}\right) - \frac{q_k}{m_k} \vec{A}_k(\vec{r}_k,t)\right],
\end{align}
which provides a guidance law for a $N$-particle trajectory $\vec{r}^\alpha(t) = (\vec{r}_1^\alpha(t), \ldots, \vec{r}_N^\alpha(t))$:
\begin{align}
\label{eq.guidancelaw}
\frac{d \vec{r}^\alpha(t)}{dt} = \vec{v}(\vec{r}^\alpha(t),t).
\end{align}
The initial positions $\{\vec{r}^\alpha(0)\}$ of the trajectories $\{\vec{r}^\alpha(t)\}$ are distributed according to the quantum equilibrium hypothesis, i.e. following the
probability density at time $t=0$, $\rho(\vec{r},0)$, 
as in \eref{om.sum_0f_particlesNDo}, see \sref{sec:postulates}.
Then, the continuity equation (\ref{eq.continuityanalytic}) ensures that the trajectories will be distributed following $\rho(\vec{r},t)$ at all later times.

It is important to note that the velocity field associated with each particle is defined on the entire configuration space and not only on the subspace of the particular particle.
Specifically,
\begin{align}
\frac{d \vec{r}_k^\alpha(t)}{dt} &= \vec{v}_k(\vec{r},t)|_{\vec{r}=\vec{r}^\alpha(t)}, \\
\vec{v}_k(\vec{r},t) &= \frac{\vec{j}_k(\vec{r},t)}{\rho(\vec{r},t)}
\nonumber \\ &= \frac{\hbar}{m_k} \mathop{\rm Im} \left(\frac{\nabla_k \psi(\vec{r},t)}{\psi(\vec{r},t)} \right) - \frac{q_k}{m_k} \vec{A}_k(\vec{r}_k,t).
\label{eqtrans}
\end{align}
Thus the trajectory of each particle in the system experiences nonlocal effects through the positions of the rest of the particles.

In summary, in the so-called analytic methods, the Schr\"odinger equation is first integrated to later calculate the trajectories by integrating the velocity field obtained from the wave function.
We want to point out that any wave equation that has a continuity equation associated with it allows for this kind of trajectory treatment.
Thus, this method can be also applied to obtain Bohmian equations of motion of, for example, relativistic systems or particles with spin (the interested reader can find a discussion about extensions of Bohmian mechanics in Sec. \ref{sec:beyond}).

\subsection{Trajectories from the Hamilton-Jacobi equation}
\label{sec:Hamilton-Jacobi}

A formulation of quantum mechanics analogous to the classical Hamilton--Jacobi can be obtained starting from the Schr\"odinger equation.
This ``synthetic'' approach to Bohmian mechanics was the one used by David Bohm in his formulation of the theory in the early fifties~\cite{o.Bohm1952a,o.Bohm1952b}.
This formalism for the quantum theory allows to obtain the trajectories without computing first the wave function, is the source of a lot of hydrodynamic algorithms (see, for instance, \cite{wyatt-bk}), and sets the basis for extensions such as Bohmian mechanics with complex action presented in \sref{sec:complex action}.

Working equations are obtained by expressing the wave function in polar form,
\begin{align}
\psi(\vec{r},t) = R(\vec{r},t) e^{i S(\vec{r},t) / \hbar},
\end{align}
and then introducing it into the Schr\"odinger equation, i.e. \eref{eq.schrodingervectorpotential}, to obtain
\begin{align}
\label{eq.math.DerivationBohm}
i \hbar \partial_t R - R \partial_t S  =
& \sum_k \left[
-\frac{\hbar^2}{2m_k} \nabla^2_k R
- \frac{i\hbar}{2m_k} R \nabla^2_k S
\right. \nonumber  \\ & \left.
- \frac{i\hbar}{m_k} \nabla_k S \cdot \nabla_k R
+ \frac{1}{2m_k} R \left(\nabla_k S\right)^2
\right. \nonumber  \\  & \left.
+ \frac{i\hbar q_k}{m_k} \vec{A}_k \cdot \nabla_k R
- \frac{q_k}{m_k} R \vec{A}_k \cdot \nabla_k S
\right. \nonumber  \\  & \left.
+ \frac{i \hbar q_k}{2m_k} R \nabla_k \cdot \vec{A}_k
+ \frac{q_k^2}{2m_k} R \vec{A}_k^2
\right] + V R.
\end{align}

On the one hand, the imaginary part of \eref{eq.math.DerivationBohm} leads to
\begin{align}
\frac{\partial R^2} {\partial t} + \nabla \cdot \sum_k  R^2 \frac{\nabla_k S - q_k \vec{A}_k}{m_k} = 0,
\end{align}
which is again the continuity equation (\ref{eq.continuityanalytic}) with
\begin{align}
\label{eq.math.rhoRR}
\rho(\vec{r},t) &= R(\vec{r},t)^2 , \\
\label{eq.math.jRnS}
\vec{j}(\vec{r},t)
&=  \sum_k R^2(\vec{r},t) \frac{\nabla_k S(\vec{r},t) - q_k \vec{A}_k(\vec{r}_k,t)}{m_k} .
\end{align}

On the other hand, taking the real part of \eref{eq.math.DerivationBohm} and defining the \textit{quantum potential} as
\begin{align}
\label{eq.math.QuantumPotential}
Q(\vec{r},t) = - \sum_k \frac{\hbar^2}{2 m_k} \frac{\nabla_k^2 R(\vec{r},t)}{R(\vec{r},t)},
\end{align}
we arrive at 
\begin{align} 
\label{eq.math.QHJ}
\frac{\partial S(\vec{r},t)}{\partial t} &+ \sum_k \frac{\left[\nabla_k S(\vec{r},t) - q_k \vec{A}_k(\vec{r}_k,t)\right]^2}{2m_k} \nonumber \\ &+ V(\vec{r},t) + Q(\vec{r},t) = 0,
\end{align}
which is known as the quantum Hamilton--Jacobi equation, analogous to its classical counterpart but with $Q$ as an additional potential term.
\Eref{eq.math.QHJ} can be then used to describe an ensemble of trajectories (labeled $\alpha$) defined by:
\begin{align}
\label{eq.math.guidance}
\frac{d \vec{r}^\alpha(t)}{dt} &=  \left. \sum_k \frac{\nabla_k S(\vec{r},t) - q_k \vec{A}_k(\vec{r}_k,t)}{m_k} \right|_{\vec{r}=\vec{r}^\alpha(t)} ,
\end{align}
and initially sampled according to the quantum equilibrium hypothesis, i.e. following the probability density at time $t=0$, $\rho(\vec{r},0)$, as in \eref{om.sum_0f_particlesNDo}, see \sref{sec:postulates}.
Notice that \eref{eq.math.guidance} is equivalent to \eanderef{eq.joverrho}{eq.guidancelaw}.

By taking the limit $Q \rightarrow 0$ the (classical) Hamilton--Jacobi equation is recovered, from where classical trajectories would be obtained.
Since $Q$ accounts for the quantum (and nonlocal) behavior of the trajectories, it is named the quantum potential, and its magnitude gives an estimation of the deviation of quantum trajectories from their classical counterparts.
Nevertheless, thinking of it as a classical potential can be misleading since it depends on the shape of the wave function (cf. \eref{eq.math.QuantumPotential}). 

The numerical integration of the Hamilton--Jacobi equation is more convoluted than the Schr\"odinger equation.
To begin with, the Hamilton-Jacobi equation is a nonlinear equation (with respect to the modulus of the wave function), and thus numerical instabilities are bound to appear more easily.
Furthermore, the computation of the wave function (modulus and phase) in regions where the modulus is small (for instance, near wave function nodes) should be handled with special care because, depending on the implemented algorithm, the trajectories in those areas can become sparse.
Algorithms such as the derivative propagation and trajectory stability methods~\cite{DPMTSM} were proposed to avoid this kind of problems.

\subsection{Trajectories from complex action}
\label{sec:complex action}

Bohmian mechanics is typically formulated in terms of a set of real
equations.
However, as we may find in classical and semiclassical treatments
\cite{sanz-bk-1}, it can also be recast in a complex form when practical applications are envisioned, in terms of
a complex action, $\bar{S}$, and extended by analytic continuation to
the complex plane.
As seen below, this has been done in the literature 
to develop new efficient quantum computational tools\footnote{To some
extent, this procedure is analogous to considering the complex form
of the classical electric and magnetic fields with the purpose of
determining in a simpler manner solutions to Maxwell's equations.
It is not our intention to discuss the interpretive issues that these complex trajectories may raise,
but only the numerical viability of using them in computations (although this complex formalism is far from the Bohmian postulates presented in \sref{sec:postulates}).}.
This complexification gives rise to alternative dynamical behaviors,
which are specified by a complex-valued time-dependent quantum
Hamilton--Jacobi equation,
\begin{align}
 \frac{\partial \bar{S}}{\partial t} + \frac{(\nabla \bar{S})^2}{2m}
  + V - \frac{i\hbar}{2m}\ \nabla^2 \bar{S} = 0 ,
 \label{cqhje}
\end{align}
where the last term on the left-hand side is the complex quantum potential.
The relationship between $\bar{S}$ and the usual wave function is given by
the transformation relation
\begin{align}
 \bar{S}(\vec{r},t) = \frac{\hbar}{i}\ \ln \Psi(\vec{r},t) .
 \label{cS}
\end{align}
From this equation, one can now define a complex-valued local velocity
vector field,
\begin{align}
 \bar{\vec{v}} = \frac{\nabla\bar{S}}{m} .
 \label{cv}
\end{align}
Taking this expression into account, first we notice that the complex
quantum potential can be expressed in terms of the first spatial
derivative of the complex velocity:
\begin{align}
 \bar{Q} = - \frac{i\hbar}{2m}\ \nabla^2 \bar{S}
  = - \frac{i\hbar}{2}\ \nabla \bar{\vec{v}} .
 \label{cQ}
\end{align}
That is, within this formulation in terms of a complex action, also
known as {\it complex Bohmian mechanics}, there is a direct relationship
between the quantum potential and the local velocity field, thus stressing
the direct role of $\bar{Q}$ on the quantum dynamics.
This is not the case for Bohmian mechanics, where the quantum potential
depends on the spatial derivatives of the probability density, and not on
the phase field.
In this sense, the link between the quantum potential and the dynamics
is not straightforward, as in complex Bohmian mechanics; such a link only
becomes evident if we keep in mind the fact that this potential arises
from the action of the kinetic operator on the wave function.
Second, we also readily find that because $\bar{S}$ is in general a complex
field, the only dynamics compatible with Eq.~(\ref{cv}) has to be complex,
which means that we cannot use the real variable, $\vec{r}$, but a complex
one, $\vec{z}$, obtained by analytical continuation.
This means that the corresponding complex trajectories are obtained after
integration of the (complex) equation of motion
\begin{align}
 \frac{d\vec{z}}{dt} = \bar{\vec{v}} .
 \label{cv2}
\end{align}
A direct correspondence cannot be established between the trajectories
obtained from this equation and the usual Bohmian trajectories in real
space, since a one to one correspondence among them does not exist (nor
the latter correspond to the real part of the former).
Rather each Bohmian trajectory is to be considered as the result of the
crossing of the real axis, at subsequent times, of a continuous set of
complex trajectories \cite{sanz:cpl:2008}.

Although the first evidence of Eq.~(\ref{cqhje}) dates back to 1933, to
Pauli's seminal work {\it ``Die allgemeinen Prinzipien der Wellenmechanik''}
\cite{pauli-hbk-1} (translated into English as {\it ``General Principles
of Quantum Mechanics''} \cite{pauli-hbk-2}), to our knowledge the first
application of the complex quantum Hamilton-Jacobi equation is due to
Leacock and Padgett \cite{leacock:PRL:1983,leacock:PRD:1983}, who in
1983 (fifty years after Pauli!), in the context of the quantum
transformation theory
\cite{jordan:ZPhys:1926,jordan:ZPhys-1:1927,jordan:ZPhys-2:1927,dirac:PRSLA:1927,dirac:PZSov:1933,dirac:RMP:1945},
proposed this equation as a means to obtain bound-state energy levels of
quantum systems with no need to calculate the corresponding eigenfunctions.
According to these authors, the quantum Hamilton-Jacobi equation can be (i)
either stated as a postulate, or (ii) derived
from the Schr\"odinger equation through a simple connection formula.
In the second case, the quantum action is proportional to the natural
(complex) logarithm of the wave function, more specifically the
ansatz (\ref{cS}).
This relationship allows us to easily go from the quantum Hamilton-Jacobi
formulation to the standard one in terms of the time-dependent Schr\"odinger
equation (and vice versa).
Although both the ansatz and the formulation used by Leacock and
Padgett are similar to those considered in the standard (quantum) JWKB
semiclassical approximation
\cite{jeffreys:PLMS:1925,wentzel:ZPhys:1926,kramers:ZPhys:1926,brillouin:ComptRend:1926,sanz-bk-1},
as pointed out by these authors, their approach conceptually differs from
it, since they did not claim to be in a semiclassical regime.
Notice that for nondegenerate eigenstates, $\Psi$ is a real field and
$\bar{S}$ is therefore an imaginary field.
On the other hand, if there is some degeneracy, $\bar{S}$ also acquires
a real part depending on the angular (orbital) part of the eigenstate
(e.g., the three-dimensional harmonic oscillator or the hydrogen atom).

The work developed by Leacock and Padgett was not specifically
oriented towards dynamical issues.
This is a remarkable point, because precisely one of the
``pathologies'' of Bohmian mechanics is the motionless state
assigned to particles associated with nondegenerate eigenstates
\cite{o.Holand1993}, i.e., quantum states characterized by a zero
velocity field in the Bohmian sense.
To circumvent this problem, following independent approaches Floyd since
the early 1980s
\cite{floyd:PRD-1:1982,floyd:PRD-2:1982,floyd:PRD:1984,floyd:PRD:1986,floyd:FPL:1996,floyd:FPL:2000}
and Faraggi and Matone since the late 1990s
\cite{faraggi:PLA:1998,faraggi:PLB:1998,faraggi:PLB-1:1999,faraggi:PLB-2:1999,faraggi:IJMPA:2000}
developed (time-independent) quantum Hamilton-Jacobi-like formulations
starting from (real) bipolar ans\"atze, although they did not claim
full equivalence with standard quantum mechanics regarding their
predictions, and therefore with Bohmian mechanics.
In this latter case, probably the first studies are due to John in the
early 2000s \cite{john:FPL:2002}.
This author proposed a time-dependent complex quantum trajectory
formalism (based on the same connection formula mentioned by Leacock and
Padgett) to study the dynamics associated with some simple analytical
cases, such as the harmonic oscillator or the step barrier.
Later on this modified de Broglie-Bohm approach, as denoted by John, has
also been applied to the analysis of the Born rule and the normalization
conditions of the probability density in the complex plane
\cite{john:AnnPhys:2009,john:AnnPhys:2010}, or the dynamics of coherent
states \cite{john:FoundPhys:2013,fring:PRA:2013}.
Analogous studies carried out to determine different dynamical properties
with this complex Bohmian representation have been carried out extensively
in the literature
\cite{yang:AnnPhys-1:2005,yang:AnnPhys-2:2005,yang:IJQC:2006,yang:CSF-1:2006,yang:CSF-2:2006,yang:AnnPhys:2006,yang:CSF-1:2007,yang:CSF-2:2007,yang:CSF-3:2007,yang:CSF-4:2007,chou:PRA:2007,chou:PRA:2008,chou:JCP-1:2008,chou:JCP-2:2008,chou:JCP-3:2008,chou:PLA:2009,chou:prl:2009,chou:annphys:2010,chou:JCP:2010,chou:PLA:2010,sanz:cpl:2008}.

More recently, the complex version of Bohmian mechanics has been invoked
as an alternative computational tool, the so-called {\it Bohmian mechanics with
complex action}, developed since 2006 by Tannor and coworkers
\cite{tannor:JCP:2006,sanz:JCP:2007,tannor:JCP-1:2007,tannor:JCP-2:2007,tannor:JPCA:2007,tannor:CP:2007,tannor:JCP:2014}
from an earlier, independent derivation of Eq.~(\ref{cqhje})
\cite{tannor-bk}, with extensions to nonadiabatic molecular dynamics
\cite{tannor:JCP-1:2012,tannor:JCP-2:2012}.
Differently from the works mentioned above, Bohmian mechanics with complex
action is aimed at obtaining the wave function directly from the trajectories,
as the {\it quantum trajectory method} \cite{wyatt:prl:1999} or the
{\it derivative propagation methods} \cite{trahan:jcp:2003} do in real space.
In other words, this complexified Bohmian mechanics is also an alternative
synthetic method but in complex space \cite{wyatt-bk}.
With the same spirit and by the same time, an alternative synthetic
approach was also developed by Chou and Wyatt
\cite{chou:JCP:2006,chou:PRE:2006,chou:IJQC:2008}, with applications to
both bound states and scattering systems.

Alternatively, other complex schemes include the complexification of
the usual Bohmian mechanics by formulating it in imaginary time with
the purpose of computing reliable energy eigenstates by means of a
single Bohmian trajectory.
Analytic continuation ($t \to -i\tau$, $p \to -ip$, and $S \to -iS$,
with $\tau>0$) here gives rise to diffusion-like equations, which
benefit from the presence of the quantum potential.
This technique has been implemented and applied to systems ranging from
complex low-dimensional potential functions \cite{makri:MolPhys:2005}
to clusters of 11 atoms (33 degrees of freedom) \cite{garashchuk:TheorChemAcc:2012},
proving to be very efficient in terms of computational cost and
numerical stability.

\subsection{Trajectories from conditional wave functions}
\label{sec:conditional}

Elementary textbooks explain quantum mechanics from a  wave function $\psi(x,y,z,t)$  solution of the  Schr\"odinger equation. The physical space where such a wave function \emph{lives} is the ordinary space $\vec r=\{x,y,z\}$. However, any quantum system of interest implies much more degrees of freedom. For example, a system of $N$ particles, whose positions are $\vec r_1,\vec r_2,\vec r_3, \ldots ,\vec r_N$,  involves a wave function $\Psi(\vec r_1,\vec r_2,\vec r_3, \ldots ,\vec r_N,t)$ that does not live in the physical (ordinary) $3$-dimensional space, but instead in an abstract $3N$-dimensional space plus time. Such an abstract and high-dimensional configuration space, that  plays a central role in the understanding of many quantum phenomena, has been quite \lq\lq{}\emph{indigestible}\rq\rq{}. Einstein remarked  that  \emph{``Schr\"o\-din\-ger is, in the beginning, very captivating.  But the waves in n-dimensional coordinate space are indigestible \ldots''} \cite{o.einstein}. Two particles at $\vec r_1$ and $\vec r_2$, which are very far in physical space, share some common region in such a high-dimensional configuration space. This opens the possibility of instantaneous (i.e. faster than light) interaction between them. Einstein hated this idea. However, Bell\cite{o.Bell1964}, and the posterior experiments of Aspect\cite{o.aspect1982}, showed that quantum phenomena are nonlocal in the sense mentioned before.

From a practical point of view, as we have already mentioned, the Schr\"odinger equation can only be solved with very few degrees of freedom, i.e., we must face the many body problem. Therefore, an \emph{artificial} division has to be done between the degrees of freedom that we will explicitly simulate and those that we will not. The Schr\"odinger equation is only valid for an isolated system where a unitary time-evolution of the wave function takes place. The probability density of finding the particles in the isolated system is conserved because particles have to be somewhere in that region. This is not true for a quantum subsystem.  A standard way to reduce the number of explicitly simulated degrees of freedom, is to \emph{trace out} certain degrees of freedom. This process ends up with what is called the \emph{reduced density matrix} which does no longer describe a pure state, but a mixture of states. In this formalism, the quantum subsystem is no longer described by the Schr\"odinger equation but by the Lindblad equation \cite{o.Linbla,openQsystems}, whose evolution is, in general, irreversible \cite{o.DiVentra2008book}.
There is no way to define a wave function for a quantum subsystem (interacting with the environment) that evolves deterministically in the standard route. In the next subsection we discuss how a Bohmian formulation of quantum mechanics allows us to proceed in a very different way.

\subsubsection{The conditional wave function}

Bohmian mechanics provides an original tool to deal with quantum (sub)systems in a quite general and trivial way. Such tool is the \emph{conditional wave function} 
\cite{o.travis2014,o.oriols2007prl,o.durr1992equilibrium,o.durr2004equilibrium,arXive_GA}. We consider an isolated quantum system. The whole universe, for example. 
The many-particle wave function is $\Psi(\vec r_1,\vec r_2,\vec r_3, \ldots ,\vec r_N\,t)$.  We define $\vec r_a$ as the position of the $a-$particle in the $\mathds{R}^3$ physical space, 
while $\vec r_b=\{\vec r_1, \ldots , \linebreak[4] \vec r_{a-1},\vec r_{a+1}, \ldots ,\vec r_{N}\}$ are the positions of the rest of particles in a $\mathds{R}^{3 (N-1)}$ configuration space. 
The actual particle (Bohmian) trajectories are accordingly denoted by $\{\vec r^\alpha_a(t), \vec r^\alpha_b(t)\}$, where $\{\vec r^\alpha_a(t_0), \vec r^\alpha_b(t_0)\}$ are selected 
according to the quantum equilibrium hypothesis as in \eqref{om.sum_0f_particlesNDo} (the reader is here referred to \sref{sec:postulates}).  
The total wave function cannot be written as a product $\Psi(\vec r) = \psi_a(\vec r_a)\psi_b(\vec r_b)$  if the two subsystems are entangled. However, within Bohmian mechanics, we can define the so called \emph{conditional wave function}\cite{o.travis2014,o.oriols2007prl,o.durr1992equilibrium,o.durr2004equilibrium,arXive_GA}:
\begin{align}
\psi_{a}(\vec r_a,t) \equiv  \Psi(\vec r_a,\vec r_b(t),t),
\label{Conditional}
\end{align}
which constitutes a slice of the whole multi-dimensional wave function $\Psi$. In \eref{Conditional} we omit (for simplicity) the dependence of each conditional wave function $\psi_{a}(\vec r_a,t)$ on $\alpha$. The relevant property extracted from \eref{Conditional} is that the (Bohmian) velocity of the $\vec r_a(t)$ can be equivalently computed from the \emph{big} wave function $\Psi(\vec r_a,\vec r_b,t)$ or from the \emph{little} (conditional) wave function $\psi_{a}(\vec r_a,t)$ as seen:
\begin{eqnarray}
\frac{d \vec r_a(t)}{dt} & = &\frac{\hbar}{m} \, \left. \text{Im}   \frac{\nabla_{\vec r_a} \Psi(\vec r_a, \vec r_b,t)}{\Psi(\vec r_a, \vec r_b,t)} \right|_{\vec r_a = \vec r_a^\alpha(t),\vec r_b = \vec r_b^\alpha(t)}
\nonumber\\  & \equiv & \frac{\hbar}{m} \, \text{Im}
\left. \frac{\nabla_{\vec r_a} \psi_a(\vec r_a,t)}{\psi_a(\vec r_a,t)} \right|_{\vec r_a = \vec r_a^\alpha(t)}\;\;\;\;\;\;\;\;\;\;\;\;\;.
\label{eq-guidance2}
\end{eqnarray}
We would like to stress that, by construction, the little wave function guides the trajectory $\vec r_a(t)$ along exactly the same path as the big wave function. Note that one can deduce one (single-particle) conditional wave for each particle $a=1, \ldots ,N$. The little wave functions $\psi_{a}(\vec r_a,t)$, i.e. the Bohmian conditional wave functions, may thus be regarded as single-particle pilot-waves (propagating in physical space) which guide the motions of the affiliated particles.  In fact, if needed, the technique used in deducing (\ref{eq-guidance2}) can be equivalently developed for any arbitrary particle's subset. For example, we can build the (three particles) conditional wave function $\psi_{1,2,3}(\vec r_1,\vec r_2,\vec r_3,\vec r_4(t), \ldots , \linebreak[4] \vec r_N(t),t)$.

\subsubsection{The nonlinear and nonunitary equation}
\label{nonlinearnonunitary}

Up to know, the reader realizes that these conditional wave functions in physical space are defined in \eref{Conditional} from the universal \emph{big} (many particle) wave function in the huge $\mathds{R}^{3 N}$ configuration space. In order to discuss the practical utility of these conditional wave functions, we need to explore the possibility of defining them independently of the big wave function\cite{o.travis2010fp,o.travis2014,o.oriols2007prl}. We might track the dynamical evolution of a quantum (sub)system \emph{exclusively} in terms of these single-particle (conditional) wave functions, instead of (as we did in their presentation) first solving the many-body Schr\"odinger equation for $\Psi$ and only examining the conditional wave functions $\psi$ after.  It can be demonstrated \cite{o.oriols2007prl} that $\psi_{a}(\vec{r}_a, t)$ obeys the following wave equation\footnote{The relevant point in the development is that the conditional wave function for particle $a$ depends on time in two ways, through the Schr\"
odinger time-evolution of $\Psi$, and also through the time-evolution of the rest of the particles $\vec r_b(t)$:
\begin{eqnarray}
i \hbar \frac{\partial}{\partial t} \psi_a(\vec r_a,t) = i \hbar
\frac{\partial \Psi(\vec r_a,\vec r_b,t)}{\partial t} \Big|_{\vec r_b = \vec r_b(t)} \nonumber\\
+ \sum_{k=1,k \neq a}^{N} i \hbar
\frac{d \vec r_k(t)}{dt} \nabla_{\vec r_k} \Psi(\vec r_a,\vec r_b,t) \Big|_{\vec r_b = \vec r_b(t)\nonumber}
\end{eqnarray}}:
\begin{eqnarray}\label{Conditional_eq}
i\hbar\frac{\partial \Psi_a(\vec r_a,t)}{\partial t}=\Big\{-\frac{\hbar^2}{2m} \nabla^2_a+U_{a}(\vec r_a,\vec r^\alpha_b(t),t) \nonumber \\ +G_{a}(\vec r_a,\vec r^\alpha_b(t),t)+ i J_{a}(\vec r_a,\vec r^\alpha_b(t),t) \Big\} \Psi_{a}(\vec r_a,t).
\end{eqnarray}
The explicit expression of the potentials $G_{a}(\vec r_a,\vec r^\alpha_b(t),t)$ and $J_{a}(\vec r_a,\vec r^\alpha_b(t),t)$ that appear in (\ref{Conditional_eq}) can be found in reference \cite{o.oriols2007prl}. However, their numerical values are in principle unknown and need some educated guesses\cite{o.oriols2011book,o.alarcon2013pcm}. On the other hand, $U_{a}(\vec r_a,\vec r^\alpha_b(t),t)$ is the part of the total potential energy that appears in many-body Schr\"odinger equation with an explicit dependence on $\vec r_a$. One obtains \eref{Conditional_eq} for each particle. From a practical point of view, all quantum trajectories $\vec r^\alpha(t)$ have to be computed simultaneously\cite{time_resolved,o.oriols2011book,o.alarcon2013pcm}.

The computational advantage of the above algorithm using (\ref{Conditional_eq}) instead the many-body Schr\"odinger equation is that, in order to find (approximate) trajectories, $\vec r^\alpha_a(t)$, we do not need to evaluate the wave function and potential energies in the whole configuration space, but only over a smaller number of configuration points, $\{\vec r_a, \vec r^\alpha_b(t)\}$, associated with those trajectories defining the highest probabilities\cite{time_resolved}.

The presence of an imaginary potential $J_{a}(\vec r_a,\vec r^\alpha_b(t),t)$ in (\ref{Conditional_eq}) implies that the Hamiltonian is not Hermitian and the conditional wave function suffers a \emph{nonunitary evolution}. The probability density is not conserved. This is just the obvious consequence of dealing with open quantum (sub)systems. In addition, in a coupled system, the trajectory $\vec r_a(t)$ affects the potential $U_b(\vec r_b,\vec r_a(t),t)$ of the particle $\vec r_b$. In turn, this potential affects $\psi_b(\vec r_b,t)$ that modifies the particle $\vec r_b(t)$. Again, $\vec r_b(t)$ affects the potential $U_b(\vec r_b,\vec r_a(t),t)$ that defines the wave function $\psi_a(\vec r_a,t)$, and the circle starts again. In other words, contrarily to the Schr\"odinger equation, the new equation (\ref{Conditional_eq}) can be \emph{nonlinear}. Let us specify that here the adjective nonlinear refers to the (conditional) wave functions itself, not to the nonlinearity of the modulus discussed in \sref{sec:Hamilton-Jacobi}. In other words, there is no guarantee that the superposition principle satisfied by the \emph{big} wave function (in the \emph{big} configuration space) is also applicable to the \emph{little} (conditional) wave function when dealing with quantum subsystems.

Finally, let us mention that (for spinless electrons) the exchange interaction is \emph{naturally} included in (\ref{Conditional_eq}) through the terms $G_{a}$ and $J_a$. Due to the Pauli exclusion principle, the modulus of the wave function tends to zero, $R(\vec r_a,\vec r^\alpha_b(t),t)\to 0$, in any neighborhood of $\vec r_{a_j}$ such that $|\vec r_{a_j}-\vec r^\alpha_{b_k}(t)|\to 0$ with $j$ and $k$ referring to the individual particles of systems $A$ and $B$ respectively. Thus, both terms, $G_{a}(\vec r_a,\vec r^\alpha_b(t),t)$ and $J_{a}(\vec{r}_{a},\vec r^\alpha_b(t),t)$, have asymptotes at ${\vec{r}_{a_j}}\to {\vec {r}^\alpha_{b_k}}(t)$ that \textit{repel} the $a-$ particle from other electrons. However, in order to exactly compute the terms $G_a$ and $J_a$ we must know the total wave function, which is in principle unknown. There are however a few ways to introduce the symmetry of the wave function without dealing directly with these two coupling terms \cite{o.oriols2007prl,o.alarcon2009pps,o.
alarcon2013pcm}.

The fact that the Bohmian route allows a rigorous definition of the wave function of a quantum (sub)system entangled with the outside is very attractive. The utility of these conditional wave functions and their equations of motion for nonlinear and nonunitary quantum evolutions remains mainly unexplored. See some preliminary example for an approximation of the many-body problem in \sref{sec:manybody}.

\subsection{Expectations values from (pointer) trajectories}
\label{sec:pointer}

Along the four previous subsections we have established the essential ingredients that are necessary to compute both waves and trajectories following different formulations of the same Bohmian theory. 
On the present and next subsections, we focus on how the evaluation of expectation values can be prescribed in terms of Bohmian trajectories. 

\begin{figure}
\resizebox{\columnwidth}{!}{\includegraphics{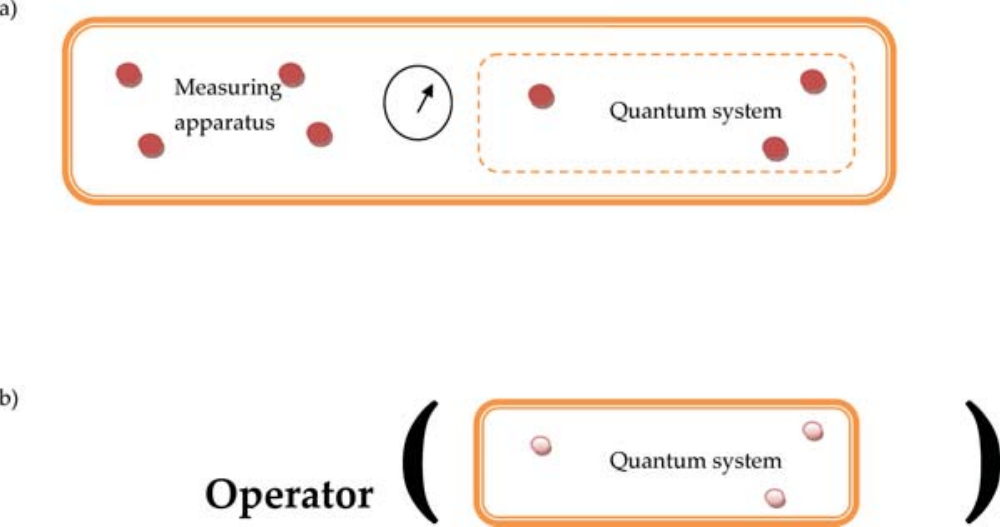}}
\caption{(a) The Bohmian measurement assumes that the quantum system and the measuring apparatus are explicitly simulated. 
(b) The standard measurement assumes that only the quantum system is explicitly simulated, but the measuring apparatus is substituted by a proper operator acting on the wave function of the system.}
\label{om_measure2}
\end{figure}

To describe a measurement process, the active system and the measuring apparatus are usually conceptually treated as separate entities.
This separation is however very different depending on the particular quantum theory that is utilized to compute the expectation values (see \fref{om_measure2}).
The standard prediction of observables, for instance, is described through the use of operators $\hat{G}$ whose \textit{eigenvalues} provide all possible outcomes of the measurement. 
When we measure a particular eigenvalue, the initial wave function is transformed into an \textit{eigenfunction} of the operator, which is known as projective (von Neumman) measurement. Let us remind that in \sref{sec:measurements} we show other types of measurements different from the projective ones. 
Thus, in standard quantum theory, the time evolution of the wave function is governed by two (quite) different laws (see \fref{om_measure2}.b).
The first dynamical evolution is given by the Schr\"odinger equation. This dynamical law is deterministic in the sense that the final wave function of the quantum system is perfectly 
determined when we know the initial wave function and the Hamiltonian of the system.
The second dynamical law is called the \textit{collapse} of the wave function. The collapse is a process that occurs when the wave function interacts with the measuring apparatus. 
The initial wave function before the measurement is substituted by one of the \textit{eigenstates} of the operator $\hat{G}$\!. 
Differently from the dynamical law given by the Schr\"odinger equation, the collapse is not deterministic, since the final wave function is randomly selected among the operator eigenstates.

Contrarily, in the Bohmian theory the measurement process is conceptually treated as any
other interaction event, and hence there is no need to introduce operators
\cite{daumer1996naive,o.durr2009book,o.durr2004equilibrium}.
The entire quantum system (active system plus apparatus) is described by a trajectory plus a wave function, each one obeying its own equation of motion, see \sref{sec:Analytical}, 
independently of whether a measurement process is taking place or not.
Assuming that some kind of pointer indicates the measured quantity\footnote{In
modern electronic measuring devices, the pointer could be
represented by a seven-segment array of light-emitting diode (LED)
displays, each one with two possible states, ON and OFF. When
electrons are present inside the PN interface of one of the LEDs, a
radiative transition of the electrons from the conduction to the
valence band produces light corresponding to the ON state. The
absence of electrons is associated with an OFF state.}, once the Bohmian
trajectories associated with its positions are known,
the expectation value of the observable can be straightforwardly computed by simply averaging over different trajectory realizations of the system. 
Notice that the back-action of the measurement on the wave function is taken into account in a rather natural way. 
It is in this regard that, in Bohmian mechanics, a physical quantum system must be, whenever concerned about observable information, described by a
many-particle Hamiltonian (see \fref{om_measure2}.a). 

Despite their conspicuous conceptual differences, the Bohmian and standard explanations of the measurement process provide the same probabilistic predictions. 
The mathematical implementation of the equations of motion in each case is however quite different. 
To better appreciate these differences, let us consider the measurement of the momentum of 
an electron initially in an energy eigenstate of a square well of size $L$, i.e. $\psi_n(x) = C \; \sin(n \pi x/L)$, $C$ being the normalization constant and
$n$ an integer denoting the vibrational state. Since the wave
function is real, the momentum $p = \partial S(x)/\partial x$ is zero and so the Bohmian particle is at rest.
For high enough
values of $n$, the previous wave function $\psi(x)$ can be roughly
approximated by a sum of two momentum eigenstates with eigenvalues
$\pm n \hbar/L$.
Therefore, when we perform a standard measurement of the momentum, we obtain an outcome $\pm n \hbar/L$ while the
system wave function \textit{collapses} into one of the two
momentum eigenstates.

\begin{figure}
\resizebox{\columnwidth}{!}{\includegraphics{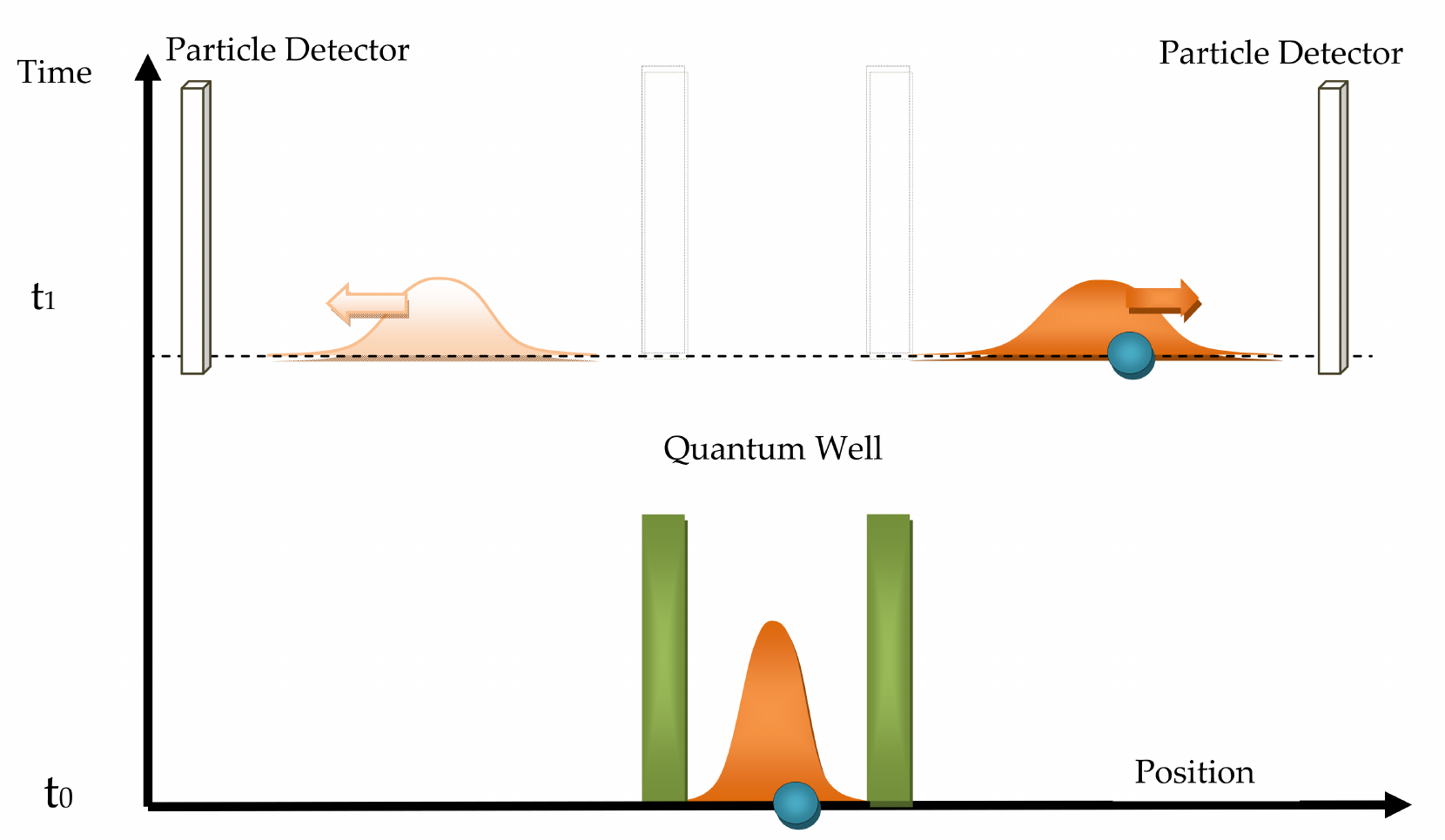}}
\caption
{Schematic explanation of the measurement process as described in Bohmian mechanics. For simplicity, the extra degrees of freedom of the particle detectors $y_R$ and $y_L$ are not included in the scheme. A confined particle within a well at time $t_0$ 
 is released at time $t_1$ in order to measure its momentum. Only the wave packet associated with the Bohmian trajectory (choosen at time $t_0$)
 is later detected (on the right detector) at time $t_1$.}
\label{momentum}
\end{figure}

In Bohmian mechanics the measurement of the momentum is undertaken by
considering how the measurement would take place in a real experiment; for instance, by
removing the walls and detecting the electron
somewhere in a screen far from the initial well (see \fref{momentum}). The first step is considering the right and left position detectors as two new degrees of freedom, $y_R$ and $y_L$. 
The wave function of the particle in the well and two additional particles associated with the detectors is defined in a larger configuration space, $\psi(x,y_R,y_L,t)$. 
The Hamiltonian of this new \emph{big} wave function $\psi(x,y_R,y_L,t)$ needs to include the time-evolution of the barriers and the particle detectors. 
The time interval between removing the walls and detecting the particles allows one
to compute the ``electron velocity''. The time-dependent
process of removing the walls implies that the initial stationary wave
function evolves into two time-dependent wave packets moving on
opposite directions, becoming after a while completely separated in
space. The particle will end up in one wave packet or the other with a
momentum very close to $\pm n \hbar/L$, the sign depending on which
wave packet the initial position of the particle enters (see \fref{momentum})
\cite{om.bohm1952b}. 

In summary, Bohmian mechanics can address quantum phenomena including measurement events just from the Schr\"odinger equation and the equation of motion for the
trajectories.
Thus, the association of operators to ``observables'', which is an indispensable step in the standard formulation of quantum mechanics, is unnecessary in the Bohmian formulation.

\subsection{Expectations values from operators}
\label{sec:operators}

Taking into account the number of degrees of freedom of the measuring apparatus, the many-particle Schr\"odinger equation (describing both the system and the apparatus) is most of the times computationally prohibitive. 
Therefore, approximations to the system-apparatus interaction are usually required to compute expectation values.

A particular route to circumvent this computational problem is based on the use of effective equations of motion for the measuring apparatus, while retaining a full description of the active system. 
A preliminary step in this direction is presented, for example, in \cite{time_resolved}. Approximations to the measurement problem can also be formulated in terms of conditional wave functions, 
which constitute a rigorous way to split a closed system into smaller open pieces (see \sref{sec:conditional}). In fact, the conditional formulation of the Schr\"odinger equation described 
in section \sref{sec:conditional} lends itself as a general starting point to derive approximate equations of motion for the active quantum region under the influence of a measuring apparatus. 

Alternatively, the use of Hermitian operators acting only on the wave function of the active system is probably the most used approach to avoid the computation of the pointer degrees of freedom in practice 
\cite{daumer1996naive,o.durr2009book,o.durr2004equilibrium}. To see how the language of operators can be merged with a trajectory-based formulation of quantum mechanics, one can proceed as follows.
The Hermitian operator $\hat{A}$ and the expectation value $\langle \hat{A} \rangle_{\psi}$ can be always written in the position representation. 
Then, the mean value of this operator over the wave function 
$\psi(\vec{r},t)$ 
is given by:
\begin{align}
\label{om.orthodox_mean_value}
\langle \hat{A} \rangle_{\psi} = \int_{-\infty}^{\infty} \psi^{*}(\vec{r},t) \hat{A} \left( \vec{r},i \hbar \nabla \right) \psi(\vec{r},t) d\vec{r}.
\end{align}
Alternatively, the same mean value can be computed from Bohmian
mechanics by defining a spatial average of a ``local'' magnitude
$A_B(\vec{r},t)$ weighted by $R^2(\vec{r},t)$:
\begin{align}
\label{om.Bohm_mean_value}
\langle \hat{A} \rangle_{\psi} = \int_{-\infty}^{\infty} R^{2}(\vec{r},t) A_B(\vec{r},t) d\vec{r}.
\end{align}
In order to obtain the same result with Eqs. (\ref{om.orthodox_mean_value}) and (\ref{om.Bohm_mean_value}), we can easily identify the local mean value $A_B(\vec{r},t)$ with
\begin{eqnarray}
\label{om.local_Bohm_mean_value}
A_B(\vec{r},t) = 
\mathop{\textrm{Re}} 
\left[\frac {\psi^{*}(\vec{r},t) \hat{A} \left( \vec{r},i\hbar \nabla \right) \psi(\vec{r},t)} {\psi^{*}(\vec{r},t) \psi(\vec{r},t)} \right]_{\psi = R e^{i \frac{S} {\hbar}}}.
\end{eqnarray}
We take only the real part because we know that the mean value of \eref{om.local_Bohm_mean_value} must be real.

For practical purposes, we compute expectation values using \eref{om.Bohm_mean_value} by means of a large $\alpha = 1,\ldots,M$ number of
Bohmian trajectories with initial positions selected according to the quantum equlibrium hypothesis \eqref{om.sum_0f_particlesNDo}. The initial positions $\vec{r}^\alpha(t_0)$ of the trajectories are 
used to rewrite $R^2(\vec{r},t)$ in \eref{om.Bohm_mean_value} as:
\begin{align}
\label{om.meanvalue_discrete}
\langle \hat{A} \rangle_{\psi}=\lim_{M\rightarrow\infty} \frac {1} {M} \sum_{\alpha=1}^{M} A_B(\vec{r}^\alpha(t)).
\end{align}
By construction, in the limit $M\rightarrow\infty$, the value of \eref{om.meanvalue_discrete} is identical to the value of \eref{om.orthodox_mean_value} and \eref{om.Bohm_mean_value}.

A particularly illustrative example of how expectation values can be obtained from operators is the case of the density current.
Let us first notice that the probability density operator can be written as $\ket{\vec{r}}\bra{\vec{r}}$ and its expectation 
value is $\braket{\psi}{\vec{r}} \braket{\vec{r}}{\psi} = \abs{\psi(\vec{r},t)}^2$, or equivalently, in the Bohmian language $\braket{\psi}{\vec{r}} \braket{\vec{r}}{\psi} = R^2(\vec{r},t)$. 
The (Hermitian) current operator can be written as $\hat{J} = 1/(2m) (\ket{\vec{r}}\bra{\vec{r}}\hat{p} + \hat{p}\ket{\vec{r}}\bra{\vec{r}})$, so it can be easily demonstrated that:
\begin{align}
\avg{J}_{\psi}&=J(\vec{r},t)=v(\vec{r},t) R^2(\vec{r},t)\nonumber\\
&=\lim_{M\rightarrow\infty} \frac {1} {M} \sum_{\alpha=1}^{M} v(\vec{r}^\alpha(t)) \delta(\vec{r}-\vec{r}^\alpha(t)).
\end{align}
The average value of the current density depends on the position,
and it is equal to the average Bohmian velocity multiplied by the
square modulus of $\psi(\vec{r},t)$. At a particular position ``$\vec{r}$,'' this
current is just the average of all particle velocities that reside around $\vec{r} = \vec{r}^\alpha(t)$ at time $t$.

It is important to emphasize that the local Bohmian mean values $A_B(\vec{r},t)$ are not the eigenvalues of the operator $\hat{A}$. In
general, the eigenvalues are not position dependent, while $A_B(\vec{r})$
are. The example of the measurement of the momentum described in section \ref{sec:pointer} sheds light into the differences between
eigenvalues and local Bohmian operators. Consider again the particle
between two walls, separated by a distance $L$, whose wave function
is $\psi(x) = C \; \sin(n\pi x/L)$ within the walls (and zero
elsewhere). Both the local Bohmian momentum $p_B(x) =
{\partial S(x,t)}/{\partial x}$ and the mean value in
\eref{om.Bohm_mean_value} are zero. Alternatively, the wave
function can be written as:
\begin{align}
\psi(x) \approx C' \left(e^{-i\pi x/L}-e^{i\pi x/L} \right).
\end{align}
The eigenvalues $a_1 = n \hbar
\pi/L$ and $a_2 = -n \hbar \pi/L$ of the momentum operator have identical probabilities and thus the
mean value of the momentum computed from \eref{om.orthodox_mean_value} is again zero.
Therefore, in general, $A_B(\vec{r})$ cannot be identified with $a_i$.
However, by construction, the mean values computed from $a_i$ and
$A_B(\vec{r})$ are identical. A similar discussion about the evaluation of
the mean value of the  quantum power can be found in \cite{om.quantumpower}. In any case, notice that according to what we mention in \sref{sec:measurements} on the measurement of local 
velocities, these local (Bohmian) mean values $A_B(\vec{r},t)$ are nowadays experimentally accessible with weak measurements.

\subsection{Summary of the formalism and interpretations}
\label{sec:postulates}

Despite its several formulations, some of them addressed in the previous subsections, here we present the theory of Bohmian mechanics through a small set of very short and simple working postulates\footnote{These Bohmian postulates are only valid for a nonrelativistic quantum world, where the number of particles does not change with time. The generalization of these postulates in a system described by quantum field theory is far from the scope of this review, which only focuses on the practical utility of the Bohmian theory for nonrelativistic quantum scenarios. }.  
In what follows, we assume a many-particle wave function without spin, so any global symmetry of the wave function comes from its orbital part.

\textbf{First Postulate} (dynamics of a quantum system): \textit{The dynamics of a nonrelativistic quantum system of $N$ particles comprises a many-particle wave function $\Psi(\vec r, t)$, 
defined in the configuration space $\vec r = (\vec r_1,\vec r_2,\ldots,\vec r_N)$ and time $t$, and a many-particle trajectory $\vec r(t) = (\vec r_1(t)$, $\vec r_2(t),\ldots,\vec r_N(t))$ 
that evolves continuously under the guidance of the wave function.}

\textit{The wave function is a solution of the Schr\"odinger equation}:
\begin{eqnarray}\label{TDSE}
i \hbar \frac{\partial \Psi(\vec r, t)}{\partial t} &=& \left( \sum_{k = 1}^N -\frac{\hbar^2}{2m_k}
\nabla^2_k + V(\vec r,t) \right) \Psi(\vec r, t)
\end{eqnarray}
\textit{where $-\hbar^2\nabla^2_k/2m_k$ is the kinetic energy operator of particle $k$ (with mass $m_k$) and the potential $V(\vec r,t)$ includes all interactions in the system (internal and with an arbitrary external scalar potential)}\footnote{Strictly speaking, and in order for Eq. \eqref{TDSE} to be exact, the quantum system should be completely isolated such that the potential $V(\vec r,t)$ only accounts for internal interactions, i.e. $V(\vec r,t) = V(\vec r)$. 
A rigorous description of the dynamics of an open quantum system would require the use of the reduced density matrix equations including Lindblad operators \cite{o.Linbla,openQsystems} or the conditional wave function formalism (see Sects.~\ref{sec:manybody} and \ref{sec:conditional}).
Nevertheless, from a practical point of view it is often a very good approximation to assume the quantum system to be closed even if it is exposed to an external
potential, in this way making \eref{TDSE} suitable to describe more general systems. Notice finally that in order to account for both scalar and vector potentials, Eqs. (\ref{TDSE}) and (\ref{TDCD}) 
should be replaced respectively by \eref{eq.schrodingervectorpotential} and each current density component in \eref{fluxvector}.}.

\textit{Each component $\vec r_k(t)$ of the trajectory is obtained by time-integrating the particle velocity $\vec v_k(t) = \vec v_k(\vec r(t),t)$ defined through the velocity field}
\begin{align}
\vec v_k(\vec r,t) = \frac{\vec J_k(\vec r,t)} {|\Psi(\vec r,t)|^2},
\end{align}
\textit{where $\vec J_k(\vec r,t)$ is the $k$-th particle current density}
\begin{align}\label{TDCD}
\vec J_k(\vec r,t) = \frac {\hbar} {m_k} \mathop{\textrm{Im}}\left[\Psi(\vec r,t)^* \nabla_{k} \Psi(\vec r,t)\right]. 
\end{align}

\textbf{Second Postulate} (quantum equilibrium hypothesis): \textit{The initial position $\vec{r}(t_0)$ of the trajectory $\vec{r}(t)$ cannot be known with certainty, and it is randomly 
distributed according to the quantum probability density $|\Psi(\vec r,t_0)|^2$.
Its initial velocity is then determined by} 
\begin{align}
\vec v_k(t_0) = \vec v_k(\vec r(t_0),t_0).
\end{align}

\textbf{Third Postulate} (symmetrization postulate of quantum mechanics): \textit{If the variables $\vec r_i$ and $\vec r_j$ refer to two identical particles of the system, then the many-particle wave function is either symmetric:
\begin{align}
\label{om.spinbos}
\Psi(.,\vec r_i,.,\vec r_j,.,t) = \Psi(.,\vec r_j,.,\vec r_i,.,t) 
\end{align}
if the particles are bosons, or antisymmetric:
\begin{align}
\label{om.spinfer} 
\Psi(.,\vec r_i,.,\vec r_j,.,t) = -\Psi(.,\vec r_j,.,\vec r_i,.,t)
\end{align} 
if the particles are fermions. In \eref{om.spinbos} and \eref{om.spinfer} it is understood that all other degrees of freedom of 
the other particles remain unchanged.
For general wave functions, this postulate implies more complicated restrictions on the possible orbital and spin components of the wave functions.}

Some authors claim that it is not necessary to postulate the quantum equilibrium hypothesis and that it is only a requirement to yield the same experimental results as standard quantum mechanics\footnote{Quantum equilibrium makes the trajectory positions of a reduced system to follow Born's statistical law in the same manner that, regardless of the initial conditions, the distribution of velocities of a gas in thermal equilibrium typically follow the Maxwell--Boltzmann law.}.
Furthermore, as Bohm already discussed in his original papers~\cite{o.Bohm1952b}, it is also claimed that quantum equilibrium can be derived. See Ref. ~\cite{o.durr2009book} and also \sref{sec:relax}.
Nevertheless, in practical scenarios the initial position and velocity of a particular many-particle trajectory $\vec{r}(t)$ cannot be known with certainty.
When an experiment is repeated several times, the initial positions of an ensemble of trajectories associated with the same wave function, $\{\vec{r}^\alpha(t)\}$, have to be distributed according to 
the quantum equilibrium hypothesis, i.e. following the initial probability density $|\Psi(\vec r,t_0)|^2$.
This condition can be written mathematically as:
\begin{align}
\label{om.sum_0f_particlesNDo}
|\Psi(\vec{r},t_0)|^2 = \lim_{M\rightarrow\infty} \frac {1} {M} \sum_{\alpha = 1}^{M} \prod_{k = 1}^{N}\delta(\vec r_{k}-\vec r_{k}^\alpha(t_0))
\end{align}
Notice the presence of two indices, $\alpha = 1,\ldots,M$ denotes an infinite ensemble of trajectories accounting for the initial uncertainity and $k = 1,\ldots,N$ accounts for the total number $N$ of 
particles.
The initial velocity of the trajectory $\vec{r}^\alpha(t)$ is then determined by
\begin{align}
\vec v_k^\alpha(t_0) = 
\vec v_k(\vec r^\alpha(t_0),t_0).
\end{align}
Similarly, the symmetrization postulate in \eanderef{om.spinbos}{om.spinfer} is also assumed by some authors as a direct consequence of dealing with trajectories\footnote{According to references \cite{o.brown1999,o.durr2009book},  Bohmian mechanics for identical particles can be described in a  ``reduced'' space $\mathds{R}^{3N}/S_N$, with $S_N$ the permutation space of $N-$particles.}. 
We also want to note also that Bohmian mechanics does not require an additional (\emph{ad hoc}) postulate for the measurement since it is treated as a particular case of the interaction between 
(pointer and system) particles, as discussed in \sref{sec:pointer}.
In any case, the previous three postulates must be interpreted here as a summary of the basic ingredients necessary to obtain (the Bohmian) predictions for the many-particle (nonrelativistic) systems covered in this review.

Finally, let us make a brief comment on the interpretation of the Bohmian theory. We have emphasized along this review that the mathematical formalism behind the Bohmian theory 
is enough to ensure that predictions reproduce the experimental results (for single-particle or many-body problems, with or without measuring apparatus). 
Certainly, behind a formalism, one can infer an interpretation on how nature is built. There are several quantum theories (Copenhagen, Bohmian, spontaneous collapse, many worlds, etc) which are 
empirically equivalent, while providing a radically different understanding of nature\footnote{Some readers can be surprised that, instead of its unquestionable success, there is still a vivid debate 
on how to interpret the quantum mechanical wave function. See, for example, the recent paper of Pusey et al.\cite{o.nature2012} where they discuss whether the wave function is only information or 
corresponds directly to reality.}. Among all these
possible interpretations, the Bohmian theory provides a quite trivial (and empirically correct!) understanding of any type of quantum experiment in terms of (point) particles guided by 
waves\footnote{The Bohmian theory is not free from slightly different interpretations. For example, David Albert proposes that there is a unique trajectory in the high-dimensional configuration space, 
``the marvellous point'' moving under the influence of a wave in the high-dimensional configuration space\cite{o.Albert1996}. Additionally, for example, D\"urr, Goldstein and Zangh\`i  write 
``\emph{We propose} [...] \emph{that the wave function is a component of physical law rather than of the reality described by the law.''}\cite{o.law1996}. 
Others, for example T. Norsen, emphasize the ability of the conditional wave function to interpret quantum mechanics in real space\cite{o.travis2010fp,o.travis2014}}. 
In this regard, since the role of physics is providing understandable explanations on what seemed incomprehensible at the beginning, the simpler an explanation is, the more understandable it becomes. 
Therefore, the Bohmian route seems a very \emph{healthy and beautiful} path to take while traveling through the quantum territory, and we certainly recommend it.

\section{Final remarks}
\label{sec:conclusions}

In 1924, Louis de Broglie\cite{o.debroglie1923} and later, in 1952, David Bohm\cite{o.Bohm1952a,o.Bohm1952b} proposed an explanation of quantum phenomena in terms of particles guided by waves. In spite of being fully compatible with all empirical results\cite{o.durr2004equilibrium,o.oriols2011book,o.durr2009book,o.bell2004book,o.Holand1993,bohm-hiley-bk,o.durr2012book}, since its origin almost a century ago till quite recently, the Bohmian theory provoked only metaphysical discussions\footnote{Bohmian mechanics is a counterexample that disproves von Neumann's conclusions\cite{o.impossibility_proofs}, in the sense that it is possible to obtain the predictions of standard quantum mechanics with a \emph{hidden-variables} theory \cite{o.oriols2011book,o.durr2009book,o.bell2004book,o.Holand1993,bohm-hiley-bk,o.durr2012book}. This was perhaps the first utility of Bohm's work.}. During the last decades the scientific community has started to consider seriously the Bohmian formalism as an additional route to compute quantum phenomena.

In this review we have shown the utility of Bohmian mechanics with different \emph{practical examples} rather than with \emph{abstract discussions}.
We have presented a brief explanation on the Bohmian formalism and a large list of nonrelativistic spinless quantum problems solved with it.
In this manner, we wanted to let the readers to evaluate for themselves whether what has already been done with Bohmian mechanics is attractive enough for them to consider using it.
In any case, we want to apologize to researchers whose work may have not been included in this review.

The formalism and the interpretation of any theory belong to different planes. In this review, we have focused on the Bohmian formalism, avoiding metaphysical (or interpretive) discussions about the Bohmian theory.
One can use the Bohmian route even if one \emph{dislikes} its interpretation of quantum phenomena. Equivalently, some people can be interested on Feynman's path integrals or Heisenberg's matrices as a mathematical tool, while others can try to relate how these elements explain the intrinsic structure of nature (whatever this means).

It was in the late 1990s, with the works of Wyatt and co-workers\cite{wyatt:prl:1999,wyatt-bk}, that the chemistry community started to study seriously the practical utility of the Bohmian formalism in their everyday research. The first step was analyzing the ability of these Bohmian trajectories to reproduce the (unitary) evolution of a wave function. At this stage, the Bohmian route and the quantum hydrodynamic route due to Madelung\cite{o.Madelung} are perfectly equivalent.

Although it is not possible to obtain the Bohmian trajectories in a single measurement, we want to remark the relevance of the (average) trajectories measured in Ref. \cite{kocsis:Science:2011} (and plotted in \fref{Kocsis}). 
The measurement of trajectories opens relevant and unexplored possibilities for the understanding of quantum phenomena through the quantitative comparison between simulated and measured (Bohmian or hydrodynamic) trajectories, instead of using the wave function and its related parameters.
In any case, along this review, we have emphasized many times that Bohmian mechanics is much more than reproducing the (unitary) time-evolution of a (single-particle) wave function with trajectories. This is certainly an ability of the Bohmian formalism, but it is \emph{only the tip of the iceberg}.

Some people erroneously consider that the Bohmian formalism is some kind of semiclassical approach to the quantum theory. Bohmian mechanics can indeed be very useful in studying the quantum-to-classical transition, but it is a theory that, by construction, provides an explanation for all experimental results of quantum phenomena, involving nonlocality, entanglement, superposition, etc \cite{o.durr2004equilibrium,o.oriols2011book,o.durr2009book,o.bell2004book,o.Holand1993,bohm-hiley-bk,o.durr2012book}.
Nowadays, as have been shown in this review, the Bohmian formalism is also starting to be used to tackle many-body problems where the computation of the \emph{big} wave function is unaccessible. We have emphasized that the Bohmian theory is a very enlightening route to study nonunitary and nonlinear quantum evolutions. Such type of evolutions appears in open quantum systems coupled to the environment,  to a measuring apparatus, etc. 
The field of quantum computing is a paradigmatic example. Most theoretical quantum computing algorithms are based on unitary and linear manipulations of states. However, as we discussed in \sref{sec:conditional}, these properties are strictly valid only for closed quantum systems. Any experimental implementation of quantum bits will deal with open quantum systems where the states are initialized, manipulated and measured from outside. Then, unitary evolutions and linear superpositions of states are not fully guaranteed. The Bohmian route offers a natural way of finding a single-particle wave function defined in physical (open) space, while still capturing many-particle features of a larger (closed) space. The so-called conditional wave function is a \emph{natural} bridge between the high-dimensional \emph{indigestible} configuration space and the physical space.

Let us explain the dichotomy between open and closed quantum systems  with different words. The standard formalism assumes an \emph{intrinsic} separation between the so-called quantum (sub)system and the rest of the world (meaning the environment, the apparatus, etc). The interaction of the rest of the world with the (sub)system is introduced in terms of operators. Such an \emph{intrinsic} separation and the operators are not needed in the Bohmian formalism\footnote{J. S. Bell illustrated this point with the following sentence: \emph{``Bohm's 1952 papers  on quantum  mechanics were for me a revelation. The elimination of indeterminism was very striking. But more important, it seemed to me, was the elimination  of any need for a vague division of the world into 'system' on the one hand, and 'apparatus' or 'observer' on the other. I have always felt since that people who have not grasped the ideas of those papers [\ldots] and unfortunately  they remain the majority [\ldots] are handicapped in any discussion of the meaning of quantum mechanics''} \cite{o.bell2004book}.} that, in principle, 
includes all degrees of freedom. Bohmian mechanics allows us to \emph{look inside the operators} with a microscopic vision of the interaction between the quantum (sub)system and the rest of the world. To be fair, none of the quantum routes (the Bohmian one included) seems able to escape the fact that the \emph{big} wave function, living in the high-dimensional configuration space, is inaccessible. Therefore, any practical Bohmian simulation, needs also a separation between those degrees of freedom that will be effectively simulated and those that will not. In this case, such (computational) separation is very often completely arbitrary and it depends, for example, on the capabilities of our computers to deal with algorithms involving a large number of variables. Therefore, the Bohmian route does not provide any magical computational solution, but it shows a different route to tackle the (nonlinear and nonunitary) evolution of quantum subsystems. It allows a fresh view on many-body problems, on novel types of (weak) measurements, on the  quantum-to-classical transitions, etc.

As we have already indicated, the serious analysis of the practical usefulness of Bohmian mechanics started about fifteen years ago.  Since then, step by step, the interest of the Bohmian formalism among the scientific community as an additional route to travel through all corners of the quantum territory is growing. However, a lot of work is still needed!  We hope this review encourages more researchers to add the Bohmian route to their collection of tools (not as the only one, but as a useful alternative in some problems) to help in their forefront research.

\begin{acknowledgement}
We gratefully acknowledge Damiano Marian, Travis Norsen, Nino Zangh\`i, Ward Struyve, and Albert Sol\'e for very useful discussions.
A.B. acknowledges financial support from the Okinawa Institute of Science and Technology Graduate University.
G.A. acknowledges support from the Beatriu de Pin\'os program through the project 2010BP-A00069.
A.S. acknowledges support from the Ministerio de Econom\'ia y Competitividad (Spain) under Project No.~FIS2011-29596-C02-01 and a ``Ram\'on y Cajal'' Grant, and from the COST Action MP1006 ``Fundamental Problems in Quantum Physics''.
J.M. acknowledges support from the Spanish MICINN contract FIS2011-23719 and the Catalan Government contract SGR2009-00347.
X.O. acknowledges support from the ``Ministerio de Ciencia e Innovaci\'{o}n'' through the Spanish Project TEC2012-31330, the Catalan Government contract SGR-2009-783 and by the Grant agreement no: 604391 of the Flagship ``Graphene-Based Revolutions in ICT and Beyond''. 
\end{acknowledgement}

\bibliographystyle{epj}
\bibliography{ABM_bib}

\begin{thebibliography}{426}

\bibitem{o.goldstein2014book}
H.~Goldstein, C.P. {Poole Jr.}, J.L. Safko, \emph{Classical Mechanics},
  3rd~edn. (Pearson New International Edition, United States of America, 2014)

\bibitem{o.Born1926}
M.~Born, Z. Phys. \textbf{37}, 863 (1926)

\bibitem{o.Heisenber1925c}
M.~Born, W.~Heisenberg, P.~Jordan, Z. Phys. \textbf{35}, 557 (1925), english
  translation in Ref. \cite{waerden}.

\bibitem{o.Heisenber1925b}
M.~Born, P.~Jordan, Z. Phys. \textbf{34}, 858 (1925), english translation in
  Ref. \cite{waerden}.

\bibitem{o.valentini2009Solvay}
G.~Bacciagaluppi, A.~Valentini, \emph{Quantum Theory at the Cross-roads:
  Reconsidering the 1927 Solvay Conference} (Cambridge University Press,
  Cambridge, 2009)

\bibitem{o.cohen1978book}
C.~Cohen-Tannoudji, B.~Diu, F.~Lalo\"e, \emph{Quantum Mechanics}, Vol. I and II
  (John Wiley \& Sons, Paris, France, 1978)

\bibitem{o.feynmann1965book}
R.P. Feynman, A.R. Hibbs, \emph{Quantum Mechanics and Path Integrals}
  (McGraw-Hill, New York, 1965)

\bibitem{o.Bohmian1996book}
J.T. Cushing, A.~Fine, S.~Goldstein, eds., \emph{Bohmian Mechanics and Quantum
  Theory: An Appraisal} (Kluwer Academic Publishers, Dordrecht, 1996)

\bibitem{o.oriols2011book}
X.~Oriols, J.~Mompart, eds., \emph{Applied Bohmian Mechanics: From Nanoscale
  Systems to Cosmology} (Pan Stanford Publishing, Singapore, 2011)

\bibitem{o.durr2009book}
D.~D{\"{u}}rr, S.~Teufel, \emph{Bohmian Mechanics: The Physics and Mathematics
  of Quantum Theory} (Spinger, Germany, 2009)

\bibitem{o.durr2012book}
D.~D{\"{u}}rr, S.~Goldstein, N.~Zangh\`i, \emph{Quantum Physics Without Quantum
  Philosophy} (Spinger, Germany, 2012)

\bibitem{o.dB_AnnPhys}
L.~{de Broglie}, Annales de Physique \textbf{3}, 22 (1925)

\bibitem{o.debroglie1927b}
L.~{de Broglie}, Journal de Physique et du Radium \textbf{8}, 225 (1927)

\bibitem{o.bell2004book}
J.S. Bell, \emph{Speakable and Unspeakable in Quantum Mechanics} (Cambridge
  University Press, United Kingdom, 2004)

\bibitem{o.Bohm1952a}
D.~Bohm, Phys. Rev. \textbf{85}, 166 (1952)

\bibitem{o.Bohm1952b}
D.~Bohm, Phys. Rev. \textbf{85}, 180 (1952)

\bibitem{o.Bohm1953b}
D.~Bohm, Phys. Rev. \textbf{89}, 458 (1953)

\bibitem{o.Holand1993}
P.R. Holland, \emph{The Quantum Theory of Motion: An account of the de
  Broglie-Bohm Causal Interpretation of Quantum mechanics} (Cambridge
  University Press, Cambridge, 1993)

\bibitem{bohm-hiley-bk}
D.~Bohm, B.J. Hiley, \emph{The Undivided Universe} (Routledge, New York, 1993)

\bibitem{o.Weinberg}
Private exchange of letters between S. Goldstein and S. Weinberg; See
  {http://www.mathematik.uni-muenchen.de/~bohmmech/BohmHome/weingold.htm}

\bibitem{o.Madelung}
E.~Madelung, Z. Phys. \textbf{40}, 322 (1926)

\bibitem{o.durr2004equilibrium}
D.~D{\"{u}}rr, S.~Goldstein, N.~Zangh\`i, J. Stat. Phys. \textbf{116}, 959
  (2004)

\bibitem{parker_uv_2009}
J.S. Parker, G.S.J. Armstrong, M.~Boca, K.T. Taylor, J. Phys. B \textbf{42},
  134011 (2009)

\bibitem{schlosser_scalable_2011}
M.~Schlosser, S.~Tichelmann, J.~Kruse, G.~Birkl, Quantum Inf. Process.
  \textbf{10}, 907 (2011)

\bibitem{greiner_quantum_2002}
M.~Greiner, O.~Mandel, T.~Esslinger, T.W. H{\"a}nsch, I.~Bloch, Nature
  \textbf{415}, 39 (2002)

\bibitem{lewenstein_ultracold_2012}
M.~Lewenstein, A.~Sanpera, V.~Ahufinger, \emph{Ultracold atoms in optical
  lattices: Simulating quantum many-body systems} (Oxford University Press,
  Oxford, 2012)

\bibitem{benseny_need_2012}
A.~Benseny, J.~Bagud{\`a}, X.~Oriols, J.~Mompart, Phys. Rev. A \textbf{85},
  053619 (2012)

\bibitem{eckert_three-level_2004}
K.~Eckert, M.~Lewenstein, R.~Corbal{\'a}n, G.~Birkl, W.~Ertmer, J.~Mompart,
  Phys. Rev. A \textbf{70}, 023606 (2004)

\bibitem{bergmann_coherent_1998}
K.~Bergmann, H.~Theuer, B.W. Shore, Rev. Mod. Phys. \textbf{70}, 1003 (1998)

\bibitem{rab_spatial_2008}
M.~Rab, J.H. Cole, N.G. Parker, A.D. Greentree, L.C.L. Hollenberg, A.M. Martin,
  Phys. Rev. A \textbf{77}, 061602 (2008)

\bibitem{leavens_are_1998}
C.~Leavens, R.~Sala~Mayato, Ann. Phys. (Berlin) \textbf{7}, 662 (1998)

\bibitem{rel.Tausk2010}
D.V. Tausk, R.~Tumulka, J. Math. Phys. \textbf{51}, 122306 (2010)

\bibitem{Huneke2013}
J.~Huneke, G.~Platero, S.~Kohler, Phys. Rev. Lett. \textbf{110}, 036802 (2013)

\bibitem{benseny_atomtronics_2010}
A.~Benseny, S.~Fern{\'a}ndez-Vidal, J.~Bagud{\`a}, R.~Corbal{\'a}n,
  A.~Pic{\'o}n, L.~Roso, G.~Birkl, J.~Mompart, Phys. Rev. A \textbf{82}, 013604
  (2010)

\bibitem{o.oriols2007prl}
X.~Oriols, Phys. Rev. Lett. \textbf{98}, 066803 (2007)

\bibitem{AIMD}
M.E. Tuckerman, J. Phys.-Condens. Matter \textbf{14}, R1297 (2002)

\bibitem{Nature}
S.A. Harich, D.~Dai, C.C. Wang, X.~Yang, S.~Der~Chao, R.T. Skodje, Nature
  \textbf{419}, 281 (2002)

\bibitem{Perspective}
J.C. Tully, J. Chem. Phys. \textbf{137}, 22 (2012)

\bibitem{Schiffer}
S.~Hammes-Schiffer, A.V. Soudackov, J. Phys. Chem. B \textbf{112}, 14108
  (2008), pMID: 18842015

\bibitem{MolecularNonadiabatic}
A.~Aviram, M.A. Ratner, Chem. Phys. Lett. \textbf{29}, 277 (1974)

\bibitem{Ehrenfest1}
A.~McLachlan, Mol. Phys. \textbf{8}, 39 (1964)

\bibitem{Ehrenfest2}
D.A. Micha, J. Chem. Phys. \textbf{78}, 7138 (1983)

\bibitem{Ehrenfest3}
Z.~Kirson, R.B. Gerber, A.~Nitzan, M.A. Ratner, Surf. Sci. \textbf{137}, 527
  (1984)

\bibitem{Ehrenfest4}
S.I. Sawada, A.~Nitzan, H.~Metiu, Phys. Rev. B \textbf{32}, 851 (1985)

\bibitem{TSH1}
J.C. Tully, R.K. Preston, J. Chem. Phys. \textbf{55}, 562 (1971)

\bibitem{TSH2}
J.C. Tully, J. Chem. Phys. \textbf{93}, 1061 (1990)

\bibitem{tunneling}
Y.~Arasaki, K.~Takatsuka, K.~Wang, V.~McKoy, Phys. Rev. Lett. \textbf{90},
  248303 (2003)

\bibitem{decoherence}
B.R. Landry, J.E. Subotnik, J. Chem. Phys. \textbf{137}, 22A513 (2012)

\bibitem{interferences}
I.~Horenko, C.~Salzmann, B.~Schmidt, C.~Sch\"{u}tte, J. Chem. Phys.
  \textbf{117}, 11075 (2002)

\bibitem{MCTDH}
M.H. Beck, A.~J{\"a}ckle, G.~Worth, H.D. Meyer, Phys. Rep. \textbf{324}(1), 1
  (2000)

\bibitem{BurghardtCederbaum}
I.~Burghardt, H.D. Meyer, L.~Cederbaum, J. Chem. Phys. \textbf{111}(7), 2927
  (1999)

\bibitem{Martinez}
T.J. Martinez, M.~Ben-Nun, R.D. Levine, J. Phys. Chem. \textbf{100}(19), 7884
  (1996)

\bibitem{wyatt:prl:1999}
C.L. Lopreore, R.E. Wyatt, Phys. Rev. Lett. \textbf{82}, 5190 (1999)

\bibitem{Wyatt2}
R.E. Wyatt, C.L. Lopreore, G.~Parlant, J. Chem. Phys. \textbf{114}, 5113 (2001)

\bibitem{Wyatt3}
C.L. Lopreore, R.E. Wyatt, J. Chem. Phys. \textbf{116}, 1228 (2002)

\bibitem{Tavernelli1}
B.F. Curchod, I.~Tavernelli, U.~Rothlisberger, Phys. Chem. Chem. Phys.
  \textbf{13}, 3231 (2011)

\bibitem{Tavernelli2}
I.~Tavernelli, B.F.E. Curchod, A.~Laktionov, U.~Rothlisberger, J. Chem. Phys.
  \textbf{133}, 194104 (2010)

\bibitem{Meier1}
E.~Gindensperger, C.~Meier, J.A. Beswick, J. Chem. Phys. \textbf{113}, 9369
  (2000)

\bibitem{Meier2}
C.~Meier, Phys. Rev. Lett. \textbf{93}, 173003 (2004)

\bibitem{Prezhdo}
O.V. Prezhdo, C.~Brooksby, Phys. Rev. Lett. \textbf{86}, 3215 (2001)

\bibitem{PrezhdoBook}
S.~Garashchuk, V.~Rassolov, O.~Prezhdo, in \emph{Reviews in Computational
  Chemistry, volume 27}, edited by K.B. Lipkowitz (John Wiley \& Sons, Inc.,
  New York, 2010), chap. Semiclassical Bohmian Dynamics, pp. 287--368

\bibitem{Christov}
I.P. Christov, J. Chem. Phys. \textbf{129}, 214107 (2008)

\bibitem{tannor:JCP-1:2012}
N.~Zamstein, D.J. Tannor, J. Chem. Phys. \textbf{137}, 22A517 (2012)

\bibitem{tannor:JCP-2:2012}
N.~Zamstein, D.J. Tannor, J. Chem. Phys. \textbf{137}, 22A518 (2012)

\bibitem{BPoirier2012}
J.~Schiff, B.~Poirier, J. Chem. Phys. \textbf{136}(3), 031102 (2012)

\bibitem{arXive_GA}
G.~Albareda, H.~Appel, I.~Franco, A.~Abedi, A.~Rubio, Phys. Rev. Lett.
  \textbf{113}, 083003 (2014)

\bibitem{plaja_attosecond_2013}
L.~Plaja, R.~Torres, A.~Za\"ir, eds., \emph{Attosecond Physics}, Springer
  Series in Optical Sciences (Springer Berlin Heidelberg, 2013)

\bibitem{milosevic_above-threshold_2006}
D.B. Milo\v{s}evi\'{c}, G.G. Paulus, D.~Bauer, W.~Becker, J. Phys. B
  \textbf{39}, R203 (2006)

\bibitem{winterfeldt_colloquium:_2008}
C.~Winterfeldt, C.~Spielmann, G.~Gerber, Rev. Mod. Phys. \textbf{80}, 117
  (2008)

\bibitem{altucci_single_2011}
C.~Altucci, J.~Tisch, R.~Velotta, J. Mod. Opt. \textbf{58}, 1585 (2011)

\bibitem{popmintchev_bright_2012}
T.~Popmintchev, M.C. Chen, D.~Popmintchev, P.~Arpin, S.~Brown,
  S.~Ali\v{s}auskas, G.~Andriukaitis, T.~Bal\v{c}iunas, O.D. M\"{u}cke,
  A.~Pugzlys et~al., Science \textbf{336}, 1287 (2012)

\bibitem{vozzi_generalized_2011}
C.~Vozzi, M.~Negro, F.~Calegari, G.~Sansone, M.~Nisoli, S.~De~Silvestri,
  S.~Stagira, Nat. Phys. \textbf{7}, 822 (2011)

\bibitem{ruiz_lithium_2005}
C.~Ruiz, L.~Plaja, L.~Roso, Phys. Rev. Lett. \textbf{94}, 063002 (2005)

\bibitem{parker_single-ionization_2007}
J.S. Parker, K.J. Meharg, G.A. {McKenna}, K.T. Taylor, J. Phys. B \textbf{40},
  1729 (2007)

\bibitem{lai_above-threshold_2009}
X.Y. Lai, Q.Y. Cai, M.S. Zhan, Eur. Phys. J. D \textbf{53}, 393 (2009)

\bibitem{lai_bohmian_2010}
X.Y. Lai, Q.Y. Cai, M.S. Zhan, Chin. Phys. B \textbf{19}, 020302 (2010)

\bibitem{benseny_hydrogen_2012}
A.~Benseny, A.~Pic\'on, J.~Mompart, L.~Plaja, L.~Roso, in \emph{Applied Bohmian
  mechanics: From nanoscale systems to cosmology}, edited by X.~Oriols,
  J.~Mompart (Pan Stanford Publishing, Singapore, 2012), chap. Hydrogen
  photoionization with strong lasers

\bibitem{lai_quantum_2009}
X.Y. Lai, Q.Y. Cai, M.S. Zhan, New J. Phys. \textbf{11}, 113035 (2009)

\bibitem{faisal_Broglie--Bohm_1998}
F.H.M. Faisal, U.~Schwengelbeck, Pramana \textbf{51}, 585 (1998)

\bibitem{botheron_self-consistent_2010}
P.~Botheron, B.~Pons, Phys. Rev. A \textbf{82}, 021404 (2010)

\bibitem{christov_time_2010}
I.P. Christov, \emph{Time dependent quantum Monte Carlo: Principles and
  perspectives}, in \emph{{AIP} Conference Proceedings} ({AIP} Publishing,
  2010), Vol. 1228, pp. 379--392

\bibitem{christov_polynomial-time-scaling_2009}
I.P. Christov, J. Phys. Chem. A \textbf{113}, 6016 (2009)

\bibitem{christov_time-dependent_2007}
I.P. Christov, New J. Phys. \textbf{9}, 70 (2007)

\bibitem{christov_correlated_2006}
I.P. Christov, Opt. Express \textbf{14}, 6906 (2006)

\bibitem{christov_time-dependent_2007-1}
I.P. Christov, J. Chem. Phys. \textbf{127}, 134110 (2007)

\bibitem{christov_correlated_2011}
I.P. Christov, J. Chem. Phys. \textbf{135}, 044120 (2011)

\bibitem{christov_dynamic_2008}
I.P. Christov, J. Chem. Phys. \textbf{128}, 244106 (2008)

\bibitem{christov_exploring_2012}
I.P. Christov, J. Chem. Phys. \textbf{136}, 034116 (2012)

\bibitem{song_investigation_2012}
Y.~Song, F.M. Guo, S.Y. Li, J.G. Chen, S.L. Zeng, Y.J. Yang, Phys. Rev. A
  \textbf{86}, 033424 (2012)

\bibitem{wu_local_2013}
J.~Wu, B.B. Augstein, C.~Figueira~de Morisson~Faria, Phys. Rev. A \textbf{88},
  023415 (2013)

\bibitem{wu_bohmian-trajectory_2013}
J.~Wu, B.B. Augstein, C.~Figueira~de Morisson~Faria, Phys. Rev. A \textbf{88},
  063416 (2013)

\bibitem{picon_photoionization_2010}
A.~Pic\'on, J.~Mompart, J.R. V\'azquez~de Aldana, L.~Plaja, G.F. Calvo,
  L.~Roso, Opt. Express \textbf{18}, 3660 (2010)

\bibitem{picon_transferring_2010}
A.~Pic\'on, A.~Benseny, J.~Mompart, J.R. V\'azquez~de Aldana, L.~Plaja, G.F.
  Calvo, L.~Roso, New J. Phys. \textbf{12}, 083053 (2010)

\bibitem{takemoto_multiple_2010}
N.~Takemoto, A.~Becker, Phys. Rev. Lett. \textbf{105}, 203004 (2010)

\bibitem{takemoto_visualization_2011}
N.~Takemoto, A.~Becker, J. Chem. Phys. \textbf{134}, 074309 (2011)

\bibitem{sawada_analysis_2013}
R.~Sawada, T.~Sato, K.L. Ishikawa, Phys. Rev. A \textbf{90}, 023404 (2014)

\bibitem{landauer}
R.~Landauer, IBM J. Res. Dev. \textbf{1}, 223 (1957)

\bibitem{o.DiVentra2008book}
M.~Di~Ventra, \emph{Electrical transport in nanoscale systems} (Cambridge
  University Press, 2008)

\bibitem{vortex1}
J.~Barker, R.~Akis, D.~Ferry, Superlattice. Microst. \textbf{27}, 319  (2000)

\bibitem{vortex2}
H.~Wu, D.~Sprung, Phys. Lett. A \textbf{196}, 229  (1994)

\bibitem{quantumFerry}
L.~Shifren, R.~Akis, D.~Ferry, Phys. Lett. A \textbf{274}, 75  (2000)

\bibitem{dissipation}
T.~Lundberg, E.~Sj\"oqvist, K.F. Berggren, J. Phys.-Condens. Matter
  \textbf{10}, 5583 (1998)

\bibitem{MCBOHM}
X.~Oriols, J.J. Garc\'{i}a-Garc\'{i}a, F.~Mart\'{i}n, J.~Su\~{n}\'{e},
  T.~Gonz\'{a}lez, J.~Mateos, D.~Pardo, Appl. Phys. Lett. \textbf{72}, 806
  (1998)

\bibitem{o.oriols1999sst}
X.~Oriols, J.J. Garcia-Garcia, F.~Mart\'{i}n, J.Su{\~n}{\'e}, J.~Mateos,
  T.~Gonz\'{a}lez, D.P.O. Vanbesien, Semicond. Sci. Tech. \textbf{14}, 532
  (1999)

\bibitem{o.oriols2001apl}
X.~Oriols, F.~Mart\'{i}n, J.~Su{\~n}{\'e}, Appl. Phys. Lett. \textbf{79}, 1703
  (2001)

\bibitem{o.oriols2002apl}
X.~Oriols, F.~Mart\'{i}n, J.~Su{\~n}{\'e}, Appl. Phys. Lett. \textbf{80}, 4048
  (2002)

\bibitem{o.oriols2003ted}
X.~Oriols, IEEE Trans. Electron Devices \textbf{50}, 1830 (2003)

\bibitem{o.oriols2004apl}
X.~Oriols, A.~Trois, G.~Blouin, Appl. Phys. Lett. \textbf{85}, 3596 (2004)

\bibitem{AC_quantum}
M.~B{\"u}ttiker, H.~Thomas, A.~Pr{\^e}tre, Phys. Lett. A \textbf{180}, 364
  (1993)

\bibitem{o.oriols2005prb}
X.~Oriols, A.~Alarc\'{o}n, E.~Fern\`{a}ndez-D\'{i}az, Phys. Rev. B \textbf{71},
  245322 (2005)

\bibitem{Backaction1}
M.~Schlosshauer, Rev. Mod. Phys. \textbf{76}, 1267 (2005)

\bibitem{Backaction2}
A.G. Kofman, S.~Ashhab, F.~Nori, Phys. Rep. \textbf{520}, 43  (2012)

\bibitem{higher_moments1}
B.~Reulet, J.~Senzier, D.E. Prober, Phys. Rev. Lett. \textbf{91}, 196601 (2003)

\bibitem{higher_moments2}
Y.~Bomze, G.~Gershon, D.~Shovkun, L.S. Levitov, M.~Reznikov, Phys. Rev. Lett.
  \textbf{95}, 176601 (2005)

\bibitem{quant_class1}
D.K. Ferry, A.M. Burke, R.~Akis, R.~Brunner, T.E. Day, R.~Meisels, F.~Kuchar,
  J.P. Bird, B.R. Bennett, Semicond. Sci. Tech. \textbf{26}, 043001 (2011)

\bibitem{quant_class2}
N.~Lambert, C.~Emary, Y.N. Chen, F.~Nori, Phys. Rev. Lett. \textbf{105}, 176801
  (2010)

\bibitem{Legget1}
A.J. Leggett, A.~Garg, Phys. Rev. Lett. \textbf{54}, 857 (1985)

\bibitem{Legget2}
S.~Gr{\"o}blacher, T.~Paterek, R.~Kaltenbaek, {\v{C}}.~Brukner,
  M.~\.{Z}ukowski, M.~Aspelmeyer, A.~Zeilinger, Nature \textbf{446}, 871 (2007)

\bibitem{Legget3}
J.~Romero, J.~Leach, B.~Jack, S.~Barnett, M.~Padgett, S.~Franke-Arnold, New J.
  Phys. \textbf{12}, 123007 (2010)

\bibitem{beyond_DC}
X.~Oriols, D.~Ferry, J. Comput. Electron. \textbf{12}, 317 (2013)

\bibitem{2_quant1}
Y.M. Blanter, M.~B{\"u}ttiker, Phys. Rep. \textbf{336}, 1 (2000)

\bibitem{2_quant2}
M.~B{\"u}ttiker, Phys. Rev. Lett. \textbf{65}, 2901 (1990)

\bibitem{2_quant3}
M.~B{\"u}ttiker, Phys. Rev. B \textbf{46}, 12485 (1992)

\bibitem{time_resolved}
G.~Albareda, D.~Marian, A.~Benali, S.~Yaro, N.~Zangh\`{i}, X.~Oriols, J.
  Comput. Electron. \textbf{12}, 405 (2013)

\bibitem{albaredaPRB2}
G.~Albareda, H.~L{\'o}pez, X.~Cartoix{\`a}, J.~Su{\~n}{\'e}, X.~Oriols, Phys.
  Rev. B \textbf{82}, 085301 (2010)

\bibitem{albaredaPRB1}
G.~Albareda, J.~Su{\~n}{\'e}, X.~Oriols, Phys. Rev. B \textbf{79}, 075315
  (2009)

\bibitem{JCE_boundaries1}
H.~L{\'o}pez, G.~Albareda, X.~Cartoix{\`a}, J.~Su{\~n}{\'e}, X.~Oriols, J.
  Comput. Electron. \textbf{7}, 213 (2008)

\bibitem{JCE_boundaries2}
G.~Albareda, A.~Benali, X.~Oriols, J. Comput. Electron. \textbf{12}, 730 (2013)

\bibitem{RSP1}
W.~Shockley, J. Appl. Phys. \textbf{9}, 635 (1938)

\bibitem{RSP2}
S.~Ramo, Proc. IRE \textbf{27}, 584 (1939)

\bibitem{RSP3}
B.~Pellegrini, Phys. Rev. B \textbf{34}, 5921 (1986)

\bibitem{RSP4}
B.~Pellegrini, Nuovo Cimento D \textbf{15}, 881 (1993)

\bibitem{RSP5}
B.~Pellegrini, Nuovo Cimento D \textbf{15}, 855 (1993)

\bibitem{albaredaFNL}
G.~Albareda, F.~Traversa, A.~Benali, X.~Oriols, Fluct. Noise Lett. \textbf{11}
  (2012)

\bibitem{alarcon2009computation}
A.~Alarc{\'o}n, X.~Oriols, J. Stat. Mech.-Theory E \textbf{2009}, P01051 (2009)

\bibitem{o.oriols2007see}
X.~Oriols, E.~Fern\`{a}ndez-D\'{i}az, A.~Alvarez, A.~Alarc\'{o}n, Solid State
  Electron. \textbf{51}, 306 (2007)

\bibitem{AbdelilahAPL}
A.~Benali, F.~Traversa, G.~Albareda, M.~Aghoutane, X.~Oriols, Appl. Phys. Lett.
  \textbf{102}, 173506 (2013)

\bibitem{albaredaJAP}
G.~Albareda, X.~Saura, X.~Oriols, J.~Su{\~n}{\'e}, J. Appl. Phys. \textbf{108},
  043706 (2010)

\bibitem{albaredaJSTAT}
G.~Albareda, D.~Jim{\'e}nez, X.~Oriols, J. Stat. Mech.-Theory E \textbf{2009},
  P01044 (2009)

\bibitem{AbdelilahFNL}
A.~Benali, F.~Traversa, G.~Albareda, A.~Alarcon, M.~Aghoutane, X.~Oriols,
  Fluct. Noise Lett. \textbf{11} (2012)

\bibitem{AlbaredaMC}
G.~Albareda, F.L. Traversa, A.~Benali, X.~Oriols, in \emph{Applications of
  Monte Carlo Method in Science and Engineering}, edited by S.~Mordechai
  (Intech Pub., 2011), chap. Many-particle Monte Carlo Approach to Electron
  Transport

\bibitem{o.alarcon2009pps}
A.~Alarc\'{o}n, X.~Cartoix\`{a}, X.~Oriols, Phys. Status Solidi \textbf{7},
  2636 (2010)

\bibitem{RTD}
M.~Buttiker, IBM J. Res. Dev. \textbf{32}, 63 (1988)

\bibitem{analogue}
H.~De~Los~Santos, K.~Chui, D.~Chow, H.~Dunlap, IEEE Microw. Wirel. Co.
  \textbf{11}, 193 (2001)

\bibitem{digital}
P.~Mazumder, S.~Kulkarni, M.~Bhattacharya, J.P. Sun, G.~Haddad, Proc. IEEE
  \textbf{86}, 664 (1998)

\bibitem{o.oriols1996ssc}
X.~Oriols, F.~Mart\'{i}n, J.~Su{\~n}{\'e}, Solid State Commun. \textbf{99}, 123
  (1996)

\bibitem{RTDMB1}
R.C. Bowen, G.~Klimeck, R.~Lake, W.~Frensley, T.~Moise, J. Appl. Phys.
  \textbf{81}, 3207 (1997)

\bibitem{RTDMB2}
R.~Lake, S.~Datta, Phys. Rev. B \textbf{45}, 6670 (1992)

\bibitem{RTDnoise1}
Y.M. Blanter, M.~B\"uttiker, Phys. Rev. B \textbf{59}, 10217 (1999)

\bibitem{RTDnoise2}
G.~Iannaccone, G.~Lombardi, M.~Macucci, B.~Pellegrini, Phys. Rev. Lett.
  \textbf{80}, 1054 (1998)

\bibitem{fabioIEEE}
F.L. Traversa, E.~Buccafurri, A.~Alarcon, G.~Albareda, R.~Clerc, F.~Calmon,
  A.~Poncet, X.~Oriols, IEEE T. Electron. Dev. \textbf{58}, 2104 (2011)

\bibitem{spin.Bell1971}
J.~Bell, Proceedings of the International School of Physics `Enrico Fermi',
  course IL: Foundations of Quantum Mechanics, New York, Academic pp. 171--81
  (1971)

\bibitem{spin.Dewdney1986}
C.~Dewdney, P.~Holland, A.~Kyprianidis, Phys. Lett. \textbf{119}, 259 (1986)

\bibitem{rel.Durr2013}
D.~D\"urr, S.~Goldstein, T.~Norsen, W.~Struyve, N.~Zangh\`i, Proc. R. Soc. A
  \textbf{470}, 20130699 (2013)

\bibitem{rel.Horton2004}
G.~Horton, C.~Dewdney, J. Phys. A \textbf{37}, 11935 (2004)

\bibitem{rel.Berndl1996}
K.~Berndl, D.~D\"urr, S.~Goldstein, N.~Zangh\`i, Phys. Rev. A \textbf{53}, 2062
  (1996)

\bibitem{rel.Horton2001}
G.~Horton, C.~Dewdney, J. Phys. A \textbf{34}, 9871 (2001)

\bibitem{rel.Dewdney2002}
C.~Dewdney, G.~Horton, J. Phys. A \textbf{35}, 10117 (2002)

\bibitem{o.nikolic2005}
H.~Nikoli\'{c}, Found. Phys. Lett. \textbf{18}, 549 (2005)

\bibitem{o.nikolic2009}
H.~Nikoli\'{c}, Int. J. Quantum Inf. \textbf{7}, 595 (2009)

\bibitem{o.nikolic2011}
H.~Nikoli\'{c}, Int. J. Quantum Inf. \textbf{9}, 367 (2011)

\bibitem{rel.Tumulka2007}
R.~Tumulka, J. Phys. A \textbf{40}, 3245 (2007)

\bibitem{qft.dur2004}
D.~D\"urr, S.~Goldstein, R.~Tumulka, N.~Zangh\`i, Phys. Rev. Lett.
  \textbf{93}(9), 090402 (2004)

\bibitem{qft.col2007}
S.~Colin, W.~Struyve, J. Phys. A \textbf{40}, 7309 (2007)

\bibitem{qft.str2011}
W.~Struyve, J. Phys. Conf. Ser. \textbf{306}(1), 012047 (2011)

\bibitem{qft.str2010}
W.~Struyve, Rev. Prog. Phys. \textbf{73}, 106001 (2010)

\bibitem{takabayasi:ProgTheorPhys:1952}
T.~Takabayasi, Prog. Theor. Phys. \textbf{8}, 143 (1952)

\bibitem{qft.Holland1988}
P.~Holland, Phys. Lett. A \textbf{128}, 9 (1988)

\bibitem{qft.Valentini1996}
A.~Valentini, in \emph{Bohmian Mechanics and quantum Theory: An Appraisal},
  edited by A.F. J.T.~Cushing, S.~Goldstein (Kluwer, Dordrecht, 1996), chap.
  Pilot-wave theory of fields, gravitation and cosmology, pp. 45--66

\bibitem{qft.Struyve2007}
W.~Struyve, H.~Westman, Proc. R. Soc. A \textbf{463}, 3115 (2007)

\bibitem{qc.Callender1994}
C.~Callender, R.~Weingard, PSA: Proceedings of the Biennial Meeting of the
  Philosophy of Science Association 1994 pp. 218--227 (1994)

\bibitem{qc.Pinto2005}
N.~Pinto-Neto, Found. Phys. \textbf{35}, 577 (2005)

\bibitem{qc.Shojai2007}
F.~Shojai, S.~Molladovoudi, Gen. Rel. Grav. \textbf{39}, 795 (2007)

\bibitem{qc.deBarros1998}
J.A. de~Barros, N.~Pinto-Neto, M.A. Sagioro-Leal, Phys. Lett. A \textbf{241},
  229 (1998)

\bibitem{qc.Shojai1998}
F.~Shojai, M.~Golshani, Int. J. Mod. Phys. A \textbf{13}, 677 (1998)

\bibitem{qc.Pinto2005b}
N.~Pinto-Neto, E.S. Santini, F.T. Falciano, Phys. Lett. A \textbf{344}, 131
  (2005)

\bibitem{kocsis:Science:2011}
S.~Kocsis, B.~Braverman, S.~Ravets, M.J. Stevens, R.P. Mirin, L.K. Shalm, A.M.
  Steinberg, Science \textbf{332}, 1170 (2011)

\bibitem{o.orefice2009}
A.~Orefice, R.~Giovanelli, D.~Ditto, Found. Phys. \textbf{39}, 256 (2009)

\bibitem{mazur:JCP:1959}
J.~Mazur, R.J. Rubin, J. Chem. Phys. \textbf{31}, 1395 (1959)

\bibitem{goldberg:AJP:1967}
A.~Goldberg, H.M. Schey, J.L. Schwartz, Am. J. Phys. \textbf{35}, 177 (1967)

\bibitem{mccullough:JCP:1969}
E.A. McCullough, R.E. Wyatt, J. Chem. Phys. \textbf{51}, 1253 (1969)

\bibitem{mccullough:JCP:1971-1}
E.A. McCullough, R.E. Wyatt, J. Chem. Phys. \textbf{54}, 3578 (1971)

\bibitem{marcus:JCP:1966}
R.A. Marcus, J. Chem. Phys. \textbf{45}, 4493 (1966)

\bibitem{hirsch:JCP:1974-2}
J.O. Hirschfelder, C.J. Goebel, L.W. Bruch, J. Chem. Phys. \textbf{61}, 5456
  (1974)

\bibitem{hirsch:JCP:1977}
J.O. Hirschfelder, J. Chem. Phys. \textbf{67}, 5477 (1977)

\bibitem{hirsch:JCP:1976-1}
J.O. Hirschfelder, K.T. Tang, J. Chem. Phys. \textbf{64}, 760 (1976)

\bibitem{wyatt:JCP:1999}
R.E. Wyatt, J. Chem. Phys. \textbf{111}, 4406 (1999)

\bibitem{sanz:cpl:2009}
A.S. Sanz, X.~Gim\'enez, J.M. Bofill, S.~Miret-Art\'es, Chem. Phys. Lett.
  \textbf{478}, 89 (2009)

\bibitem{sanz:cpl:2010E}
A.S. Sanz, X.~Gim\'enez, J.M. Bofill, S.~Miret-Art\'es, Chem. Phys. Lett.
  \textbf{488}, 235 (2010)

\bibitem{sanz:CP:2011}
A.S. Sanz, D.~L\'opez-Dur\'an, T.~Gonz\'alez-Lezana, Chem. Phys. \textbf{399},
  151 (2012)

\bibitem{hirsch:JCP:1974-1}
J.O. Hirschfelder, A.C. Christoph, W.E. Palke, J. Chem. Phys. \textbf{61}, 5435
  (1974)

\bibitem{hiley:foundphys:1982}
C.~Dewdney, B.J. Hiley, Found. Phys. \textbf{12}, 27 (1982)

\bibitem{wyatt-bk}
R.E. Wyatt, \emph{Quantum Dynamics with Trajectories} (Springer, New York,
  2005)

\bibitem{hirsch:JCP:1976-2}
J.O. Hirschfelder, K.T. Tang, J. Chem. Phys. \textbf{65}, 470 (1976)

\bibitem{abraham:AJP:1984}
I.~Galbraith, Y.S. Ching, E.~Abraham, Am. J. Phys. \textbf{52}, 60 (1984)

\bibitem{sanz:jcp:2005}
A.S. Sanz, S.~Miret-Art\'es, J. Chem. Phys. \textbf{122}, 014702 (2005)

\bibitem{efthymio:IJBC:2012}
N.~Delis, C.~Efthymiopoulos, G.~Contopoulos, Int. J. Bifur. Chaos \textbf{22},
  1250214 (2012)

\bibitem{efthymio:AnnPhys:2012}
C.~Efthymiopoulos, N.~Delis, G.~Contopoulos, Ann. Phys. \textbf{327}, 438
  (2012)

\bibitem{muga:PhysRep:2000}
J.G. Muga, C.R. Leavens, Phys. Rep. \textbf{338}, 353 (2000)

\bibitem{muga-bk-1:LNP:2002}
J.G. Muga, R.S. Mayato, I.L. Egusquiza, eds., \emph{Time in Quantum Mechanics},
  Vol.~72 of \emph{Lecture Notes in Physics} (Springer, Berlin, 2002)

\bibitem{muga-bk-2:LNP:2009}
J.G. Muga, A.~Ruschhaupt, A.~del Campo, eds., \emph{Time in Quantum Mechanics -
  Vol.~2}, Vol. 789 of \emph{Lecture Notes in Physics} (Springer, Berlin, 2009)

\bibitem{sanz:prb:2000}
A.S. Sanz, F.~Borondo, S.~Miret-Art\'es, Phys. Rev. B \textbf{61}, 7743 (2000)

\bibitem{sanz:EPL:2001}
A.S. Sanz, F.~Borondo, S.~Miret-Art\'es, Europhys. Lett. \textbf{55}, 303
  (2001)

\bibitem{sanz:JPCM:2002}
A.S. Sanz, F.~Borondo, S.~Miret-Art\'es, J. Phys.-Condens. Matter \textbf{14},
  6109 (2002)

\bibitem{sanz:jcp:2004}
A.S. Sanz, F.~Borondo, S.~Miret-Art\'es, J. Chem. Phys. \textbf{120}, 8794
  (2004)

\bibitem{sanz:prb:2004}
A.S. Sanz, F.~Borondo, S.~Miret-Art\'es, Phys. Rev. B \textbf{69}, 115413
  (2004)

\bibitem{takabayasi:ProgTheorPhys:1953}
T.~Takabayasi, Prog. Theor. Phys. \textbf{9}, 187 (1953)

\bibitem{dewdney:NuovoCimB:1979}
C.~Philippidis, C.~Dewdney, B.J. Hiley, Nuovo Cimento \textbf{52B}, 15 (1979)

\bibitem{sanz:JPA:2008}
A.S. Sanz, S.~Miret-Art\'es, J. Phys. A: Math. Theor. \textbf{41}, 435303
  (2008)

\bibitem{prosser:ijtp:1976-1}
R.D. Prosser, Int. J. Theor. Phys. \textbf{15}, 169 (1976)

\bibitem{herrmann:AJP:2002}
T.~W\"unscher, H.~Hauptmann, F.~Herrmann, Am. J. Phys. \textbf{70}, 599 (2002)

\bibitem{sanz:AnnPhysPhoton:2010}
A.S. Sanz, M.~Davidovi\'c, M.~Bo\v{z}i\'c, S.~Miret-Art\'es, Ann. Phys.
  \textbf{325}, 763 (2010)

\bibitem{sanz:JCP-Talbot:2007}
A.S. Sanz, S.~Miret-Art\'es, J. Chem. Phys. \textbf{126}, 234106 (2007)

\bibitem{o.dirac1929}
P.A.M. Dirac, Proc. R. Soc. Lond. \textbf{A123}, 714 (1929)

\bibitem{o.nightingale1999}
M.P. Nightingale, C.J. Umrigar, \emph{Quantum Monte Carlo Methods in Physics
  and Chemistry} (Springer, 1999)

\bibitem{o.Hartree2}
D.R. Hartree, Math. Proc. Cambridge \textbf{24 (01)}, 111 (1928)

\bibitem{o.fock1930}
V.Fock, Z. Phys. \textbf{61}, 126 (1930)

\bibitem{o.Kohn1}
W.~Kohn, R. Mod. Phys. \textbf{71}, 1253  (1998)

\bibitem{o.kohn1964}
P.~Hohenberg, W.~Kohn, Phys. Rev. \textbf{136}, B864 (1964)

\bibitem{o.kohn1965}
W.~Kohn, L.J. Sham, Phys. Rev. \textbf{140}, A1133 (1965)

\bibitem{o.Siesta}
J.M. Soler, E.~Artacho, J.D. Gale, A.~Garcia, J.~Junquera, P.~Ordejon,
  D.~Sanchez-Portal, J. Phys.-Condens. Matter \textbf{14}, 2745  (2002)

\bibitem{o.Runge1984}
E.~Runge, E.K.U. Gross, Phys. Rev. Lett. \textbf{52}, 997  (1984)

\bibitem{o.dmitry2000}
J.H.F. Dmitry~Nerukh, Chem. Phys. Lett. \textbf{332}, 145 (2000)

\bibitem{o.dellesite2004}
L.D. Site, Physica B \textbf{349}, 218 (2004)

\bibitem{o.donoso2001}
A.~Donoso, C.C. Martens, Phys. Rev. Lett. \textbf{87}, 223202 (2001)

\bibitem{o.drassolov2005}
V.A. Rassolov, S.~Garashchuk, Phys. Rev. A \textbf{71}, 032511 (2005)

\bibitem{o.alarcon2013pcm}
A.~Alarc\'{o}n, S.~Yaro, X.~Cartoix\`{a}, X.~Oriols, J. Phys.-Condens. Matter
  \textbf{25}, 325601 (2013)

\bibitem{o.travis2010fp}
T.~Norsen, Found. Phys. \textbf{40}, 1858 (2010)

\bibitem{o.travis2014}
T.~Norsen, D.~Marian, X.~Oriols, submitted  (2014)

\bibitem{o.bell1990pw}
J.S. Bell, Physics World \textbf{3}, 33 (1990)

\bibitem{o.AVV1988}
Y.~Aharonov, D.Z. Albert, L.~Vaidman, Phys. Rev. Lett. \textbf{60}, 1351 (1988)

\bibitem{o.wiseman2007}
H.M. Wiseman, New J. Phys. \textbf{9(6)}, 165 (2007)

\bibitem{o.Braverman_prl2013}
B.~Braverman, C.~Simon, Phys. Rev. Lett. \textbf{110}, 060406 (2013)

\bibitem{o.traversa_con2013}
F.L. Traversa, G.~Albareda, M.~Ventra, X.~Oriols, Phys. Rev. A \textbf{87},
  052124 (2013)

\bibitem{o.DGZ2009}
D.~D\"{u}rr, S.~Goldstein, N.~Zangh\`{i}, J. Stat. Phys. \textbf{134}, 1023
  (2009)

\bibitem{o.lundeen_prl2012}
J.~Lundeen, C.~Bamber, Phys. Rev. Lett. \textbf{108}, 070402 (2012)

\bibitem{o.lundeen_nature2011}
J.S. Lundeen, B.~Sutherland, A.~Patel, C.~Stewart, C.~Bamber, Nature
  \textbf{474}, 188 (2011)

\bibitem{o.Travis_con2013}
T.~Norsen, W.~Struyve, Ann. Phys. \textbf{350}, 166 (2014)

\bibitem{pauli-hbk-1}
W.~Pauli, \emph{Handbuch der Physik}, Vol. 24/1, 2nd~edn. (Springer, Berlin,
  1933)

\bibitem{pauli-hbk-2}
W.~Pauli, \emph{General Principles of Quantum Mechanics} (Springer, Berlin,
  1980)

\bibitem{olkhovsky:NuovoCim:1974}
V.S. Olkhovsky, E.~Recami, A.J. Gerasimchuk, Nuovo Cimento \textbf{22}, 263
  (1974)

\bibitem{olkhovsky:AMP:2009}
V.S. Olkhovsky, Adv. Math. Phys. \textbf{2009}, 859710 (2009)

\bibitem{wang:AnnPhys:2007}
Z.Y. Wang, C.D. Xiong, Ann. Phys. \textbf{322}, 2304 (2007)

\bibitem{leavens:SSC:1990-1}
C.R. Leavens, Solid State Comm. \textbf{74}, 923 (1990)

\bibitem{leavens:SSC:1990-2}
C.R. Leavens, Solid State Comm. \textbf{76}, 253 (1990)

\bibitem{leavens:PhysLettA:1993}
C.R. Leavens, Phys. Lett. A \textbf{178}, 27 (1993)

\bibitem{leavens:FoundPhys:1995}
C.R. Leavens, Solid State Comm. \textbf{25}, 229 (1995)

\bibitem{o.vona2013}
N.~Vona, G.~Hinrichs, D.~D\"urr, Phys. Rev. Lett. \textbf{111}, 220404 (2013)

\bibitem{o.vona2014}
N.~Vona, D.~D\"urr, in \emph{The Message of Quantum Science - Attempts Towards
  a Synthesis}, edited by P.~Blanchard, J.~Fr\"ohlich (Springer, Berlin, 2014),
  chap. The role of the probability current for time measurements

\bibitem{berry:PhysScr:1989}
M.V. Berry, Phys. Scr. \textbf{40}, 335 (1989)

\bibitem{zurek:PRL:1994}
W.H. Zurek, J.P. Paz, Phys. Rev. Lett. \textbf{72}, 2508 (1994)

\bibitem{sanz:PRE:2012}
A.S. Sanz, Y.~Elran, P.~Brumer, Phys. Rev. E \textbf{85}, 036218 (2012)

\bibitem{gutzwiller:1990}
M.C. Gutzwiller, \emph{Chaos in Classical and Quantum Mechanics}
  (Springer-Verlag, New York, 1990)

\bibitem{reichl:1992}
L.E. Reichl, \emph{The Transition to Chaos in Conservative Classical Systems:
  Quantum Manifestations} (Springer, Berlin, 1992)

\bibitem{bohigas:PRL:1984}
O.~Bohigas, M.J. Giannoni, C.~Schmit, Phys. Rev. Lett. \textbf{52}, 1 (1984)

\bibitem{mehta:1991}
M.L. Mehta, \emph{Random Matrix Theory} (Academic Press, New York, 1991)

\bibitem{zaslavskii:ZhETF:1973}
G.M. Zaslavskii, N.N. Filonenko, Zh. Eksp. Teor. Fiz. \textbf{65}, 643 (1973)

\bibitem{zaslavskii:SPJETP:1974}
G.M. Zaslavskii, N.N. Filonenko, Sov. Phys. JETP \textbf{38}, 317 (1974)

\bibitem{mcdonald:PRL:1979}
S.W. McDonald, A.N. Kaufman, Phys. Rev. Lett. \textbf{42}, 1189 (1979)

\bibitem{heller:PRL:1984}
E.J. Heller, Phys. Rev. Lett. \textbf{53}, 1515 (1984)

\bibitem{duerr:JStatPhys:1992}
D.~D\"urr, S.~Goldstein, N.~Zangh\`i, J. Stat. Phys. \textbf{68}, 259 (1992)

\bibitem{frisk:PLA:1997}
H.~Frisk, Phys. Lett. A \textbf{227}, 139 (1997)

\bibitem{parmenter:PLA:1995}
R.~Parmenter, R.W. Valentine, Phys. Lett. A \textbf{201}, 1 (1995)

\bibitem{parmenter:PLA:1996}
R.~Parmenter, R.W. Valentine, Phys. Lett. A \textbf{213}, 319 (1996)

\bibitem{makowski:PLA:2000}
A.J. Makowski, P.~Pep{\l}oswski, S.T. Dembi\'nski, Phys. Lett. A \textbf{266},
  241 (2000)

\bibitem{makowski:ActaPhysPolB:2001}
A.J. Makowski, M.F. ackowiak, Acta Physica Polonica B \textbf{32}, 2831 (2001)

\bibitem{bonfin:PRE:1998}
O.F. de~Alcantara~Bonfim, J.~Florencio, F.C.S. Barreto, Phys. Rev. E
  \textbf{58}, R2693 (1998)

\bibitem{bonfin:PLA:2000}
O.F. de~Alcantara~Bonfim, J.~Florencio, F.C.S. Barreto, Phys. Rev. E
  \textbf{277}, 129 (2000)

\bibitem{baker-bk:1996}
G.L. Baker, J.P. Gollub, \emph{Chaotic Dynamics: An Introduction}, 2nd~edn.
  (Cambridge University Press, Cambridge, 1996)

\bibitem{faisal:PLA-1:1995}
U.~Schwengelbeck, F.H.M. Faisal, Phys. Lett. A \textbf{199}, 281 (1995)

\bibitem{faisal:PLA-2:1995}
F.H.M. Faisal, U.~Schwengelbeck, Phys. Lett. A \textbf{207}, 31 (1995)

\bibitem{iacomelli:PLA:1996}
G.~Iacomelli, M.~Pettini, Phys. Lett. A \textbf{212}, 29 (1996)

\bibitem{chattaraj:PLA:1996}
S.~Sengupta, P.K. Chattaraj, Phys. Lett. A \textbf{215}, 119 (1996)

\bibitem{chattaraj:CurrentSci:1996}
P.K. Chattaraj, S.~Sengupta, Curr. Sci. \textbf{71}, 134 (1996)

\bibitem{chattaraj:CurrentSci:1998}
P.K. Chattaraj, S.~Sengupta, A.~Poddar, Curr. Sci. \textbf{74}, 758 (1998)

\bibitem{polavieja:PRA:1996}
G.G. de~Polavieja, Phys. Rev. A \textbf{53}, 2059 (1996)

\bibitem{polavieja:PRE:1997}
G.G. a~de Polavieja, M.S. Child, Phys. Rev. E \textbf{55}, 1451 (1997)

\bibitem{polavieja:PLA:1996}
G.G. de~Polavieja, Phys. Lett. A \textbf{220}, 303 (1996)

\bibitem{makowski:PLA:1998}
S.~Konkel, A.J. Makowski, Phys. Lett. A \textbf{238}, 95 (1998)

\bibitem{makowski:ActaPhysPolB:2002}
A.J. Makowski, Acta Physica Polonica B \textbf{33}, 583 (2002)

\bibitem{sprung:PLA:1999}
H.~Wu, D.W.L. Sprung, Phys. Lett. A \textbf{261}, 150 (1999)

\bibitem{wisniacki:EPL:2005}
D.A. Wisniacki, E.R. Pujals, Europhys. Lett. \textbf{71}, 159 (2005)

\bibitem{wisniacki:JPA:2007}
D.A. Wisniacki, E.R. Pujals, F.~Borondo, J. Phys. A \textbf{40}, 14353 (2007)

\bibitem{efthymio:JPA:2007}
C.~Efthymiopoulos, C.~Kalapotharakos, G.~Contopoulos, J. Phys. A \textbf{40},
  12945 (2007)

\bibitem{efthymio:CMDynAstr:2008}
G.~Contopoulos, C.~Efthymiopoulos, Celest. Mech. Dyn. Astron. \textbf{102}, 219
  (2008)

\bibitem{efthymio:PRE:2009}
C.~Efthymiopoulos, C.~Kalapotharakos, G.~Contopoulos, Phys. Rev. E \textbf{79},
  036203 (2009)

\bibitem{efthymio:JPA:2012}
G.~Contopoulos, N.~Delis, C.~Efthymiopoulos, J. Phys. A \textbf{45}, 165301
  (2011)

\bibitem{valentini:PRSA:2005}
A.~Valentini, H.~Westman, Proc. R. Soc. A \textbf{461}, 253 (2005)

\bibitem{valentini:PRSA:2012}
M.D. Towler, N.J. Russell, A.~Valentini, Proc. R. Soc. A \textbf{468}, 990
  (2012)

\bibitem{schlegel:PLA:2008}
K.G. Schlegel, S.~F\"orster, Phys. Lett. A \textbf{372}, 3620 (2008)

\bibitem{bennett:JPA:2010}
A.~Bennett, J. Phys. A \textbf{43}, 5304 (2010)

\bibitem{struyve:NJP:2010}
S.~Colin, W.~Struyve, New J. Phys. \textbf{12}, 043008 (2010)

\bibitem{brumer-elran:JCP:2013}
Y.~Elran, P.~Brumer, J. Chem. Phys. \textbf{138}, 234308 (2013)

\bibitem{sanz:AnnPhys:2013}
A.S. Sanz, S.~Miret-Art\'es, Ann. Phys. \textbf{339}, 11 (2013)

\bibitem{maddox:JCP:2001}
J.B. Maddox, E.R. Bittner, J. Chem. Phys. \textbf{115}, 6309 (2001)

\bibitem{maddox:PRE:2002}
J.B. Maddox, E.R. Bittner, Phys. Rev. E \textbf{65}, 026143 (2002)

\bibitem{holloway:JCP:2001}
Z.S. Wang, G.R. Darling, S.~Holloway, J. Chem. Phys. \textbf{115}, 10373 (2001)

\bibitem{burghardt:JCP-1:2001}
I.~Burghardt, L.S. Cederbaum, J. Chem. Phys. \textbf{115}, 10303 (2001)

\bibitem{burghardt:JCP-2:2001}
I.~Burghardt, L.S. Cederbaum, J. Chem. Phys. \textbf{115}, 10312 (2001)

\bibitem{burghardt:IJQC:2004}
I.~Burghardt, K.B. M{\o}ller, G.~Parlant, L.S. Cederbaum, E.R. Bittner, Int. J.
  Quantum Chem. \textbf{100}, 1153 (2004)

\bibitem{burghardt:JCP:2004}
I.~Burghardt, G.~Parlant, J. Chem. Phys. \textbf{120}, 3055 (2004)

\bibitem{burghardt:JPCA:2007}
K.H. Hughes, S.M. Parry, G.~Parlant, I.~Burghardt, J. Phys. Chem. A
  \textbf{111}, 10269 (2007)

\bibitem{garashchuk:JCP:2013}
S.~Garashchuk, V.~Dixit, B.~Gu, J.~Mazzuca, J. Chem. Phys. \textbf{138}, 054107
  (2013)

\bibitem{bittner:JCP:2003}
E.R. Bittner, J. Chem. Phys. \textbf{119}, 1358 (2003)

\bibitem{makri:JCP:2003}
Y.~Zhao, N.~Makri, J. Chem. Phys. \textbf{119}, 60 (2003)

\bibitem{makri:JPCA-1:2004}
J.~Liu, N.~Makri, J. Phys. Chem. A \textbf{108}, 806 (2004)

\bibitem{makri:JPCA-2:2004}
J.~Liu, N.~Makri, J. Phys. Chem. A \textbf{108}, 5408 (2004)

\bibitem{rosen:AJP-1:1964}
N.~Rosen, Am. J. Phys. \textbf{32}, 377 (1964)

\bibitem{rosen:AJP-2:1964}
N.~Rosen, Am. J. Phys. \textbf{32}, 597 (1964)

\bibitem{rosen:AJP:1965}
N.~Rosen, Am. J. Phys. \textbf{33}, 146 (1965)

\bibitem{rosen:FoundPhys:1985}
N.~Rosen, Found. Phys. \textbf{16}, 687 (1985)

\bibitem{berry:LesHouches:1991}
M.V. Berry, in \emph{Chaos and Quantum Physics}, edited by M.J. Giannoni,
  A.~Voros, J.~Zinn-Justin (North-Holland, Amsterdam, 1996), Les Houches
  Lecture Series LII (1989), chap. Some quantum-to-classical asymptotics, pp.
  251--304

\bibitem{sanz:SSR:2004}
R.~Guantes, A.S. Sanz, J.~Margalef-Roig, S.~Miret-Art\'es, Surf. Sci. Rep.
  \textbf{53}, 199 (2004)

\bibitem{drezet:ch:2014}
A.~Drezet, in \emph{Protective Measurements and Quantum Reality: Toward a New
  Understanding of Quantum Mechanics}, edited by S.~Gao (Cambridge University
  Press, Cambridge, 2014), chap. Implications of protective measurements on de
  Broglie-Bohm trajectories, \texttt{arXiv:1402.7256}

\bibitem{sanz-bk-2}
A.S. Sanz, S.~Miret-Art\'es, \emph{A Trajectory Description of Quantum
  Processes. II. Applications}, Vol. 831 of \emph{Lecture Notes in Physics}
  (Springer, Berlin, 2013)

\bibitem{child-bk:1974}
M.S. Child, \emph{Molecular Collision Theory} (Academic Press, London, 1974)

\bibitem{sanz-bk-1}
A.S. Sanz, S.~Miret-Art\'es, \emph{A Trajectory Description of Quantum
  Processes. I. Fundamentals}, Vol. 850 of \emph{Lecture Notes in Physics}
  (Springer, Berlin, 2012)

\bibitem{o.oriols1996pra}
X.~Oriols, F.~Mart\'{i}n, J.~Su{\~n}{\'e}, Phys. Rev. A \textbf{54}, 2594
  (1996)

\bibitem{makowski:PRA-1:2002}
A.J. Makowski, Phys. Rev. A \textbf{65}, 032103 (2002)

\bibitem{makowski:PRA-2:2002}
A.J. Makowski, K.J. G\'orska, Phys. Rev. A \textbf{66}, 062103 (2002)

\bibitem{holland-cushing-bk:1996}
P.R. Holland, in \emph{Bohmian Mechanics and Quantum Theory: An Appraisal},
  edited by J.T. Cushing, A.~Fine, S.~Goldstein (Kluwer, Dordrecht, 1996),
  chap. Is quantum mechanics universal?, pp. 99--220

\bibitem{bolivar:CanJPhys:2003}
A.O. Bolivar, Can. J. Phys. \textbf{81}, 971 (2003)

\bibitem{bolivar-bk:2004}
A.O. Bolivar, \emph{Quantum-Classical Correspondence. Dynamical Quantization
  and the Classical Limit} (Springer, Berlin, 2004)

\bibitem{matzkin:PLA:2007}
A.~Matzkin, Phys. Lett. A \textbf{361}, 294 (2007)

\bibitem{matzkin:SHPMP:2008}
A.~Matzkin, V.~Nurock, Studies in History and Philosophy of Modern Physics B
  \textbf{39}, 17 (2008)

\bibitem{matzkin:JPCS:2009}
A.~Matzkin, J. Phys.: Conf. Ser. \textbf{174}, 012039 (2009)

\bibitem{matzkin:FoundPhys:2009}
A.~Matzkin, V.~Nurock, Found. Phys. \textbf{39}, 903 (2009)

\bibitem{giulini-bk}
D.~Giulini, E.~Joos, C.~Kiefer, J.~Kupsch, I.O. Stamatescu, H.D. Zeh,
  \emph{Decoherence and the Appearance of a Classical World in Quantum Theory},
  2nd~edn. (Springer, 1996, Berlin)

\bibitem{schlosshauer-bk:2007}
M.~Schlosshauer, \emph{Decoherence and the Quantum-to-Classical Transition}
  (Springer, Berlin, 2007)

\bibitem{dewdney:FoundPhys:1988}
C.~Dewdney, Found. Phys. \textbf{18}, 867 (1988)

\bibitem{dewdney:PLA:1990}
M.M. Lam, C.~Dewdney, Phys. Lett. A \textbf{150}, 127 (1990)

\bibitem{hiley:FoundPhys:1999}
O.~Maroney, B.J. Hiley, Found. Phys. \textbf{29}, 1403 (1999)

\bibitem{golshani:JPA:2001}
M.~Golshani, O.~Akhavan, J. Phys. A \textbf{34}, 5259 (2001)

\bibitem{ghose:PRAMANA:2002}
P.~Ghose, PRAMANA \textbf{59}, 417 (2002)

\bibitem{marchildon:JModOpt:2003}
L.~Marchildon, J. Mod. Opt. \textbf{50}, 873 (2003)

\bibitem{marchildon:JPA:2003}
E.~Guay, L.~Marchildon, J. Phys. A \textbf{36}, 5617 (2003)

\bibitem{struyve:JPA:2003}
W.~Struyve, W.D. Baere, J.D. Neve, S.D. Weirdt, J. Phys. A \textbf{36}, 1525
  (2003)

\bibitem{durt:PRA:2002}
T.~Durt, Y.~Pierseaux, Phys. Rev. A \textbf{66}, 052109(1 (2002)

\bibitem{na:PLA:2002}
K.~Na, R.E. Wyatt, Phys. Lett. A \textbf{306}, 97 (2002)

\bibitem{na:PhysScr:2003}
K.~Na, R.E. Wyatt, Phys. Scr. \textbf{67}, 169 (2003)

\bibitem{breuer-bk:2002}
H.P. Breuer, F.~Petruccione, \emph{The Theory of Open Quantum Systems} (Oxford
  University Press, Oxford, 2002)

\bibitem{allori:JOptB:2002}
V.~Allori, D.~D\"urr, N.~Zangh{\'\i}, S.~Goldstein, J. Opt. B \textbf{4}, 482
  (2002)

\bibitem{gindensperger:JCP:2002-1}
E.~Gindensperger, C.~Meier, J.A. Beswick, J. Chem. Phys. \textbf{116}, 8 (2002)

\bibitem{gindensperger:JCP:2002-2}
E.~Gindensperger, C.~Meier, J.A. Beswick, J. Chem. Phys. \textbf{116}, 10051
  (2002)

\bibitem{gindensperger:AdvQuantChem:2004}
E.~Gindensperger, C.~Meier, J.A. Beswick, Adv. Quant. Chem. \textbf{47}, 331
  (2004)

\bibitem{meier:JCP:2004}
C.~Meier, J.A. Beswick, J. Chem. Phys. \textbf{121}, 4550 (2004)

\bibitem{daumer1996naive}
M.~Daumer, D.~D{\"u}rr, S.~Goldstein, N.~Zangh{\`i}, Erkenntnis \textbf{45},
  379 (1996)

\bibitem{DPMTSM}
R.E. Wyatt, in \emph{Quantum Dynamics with Trajectories} (Springer, New York,
  2005), chap. Derivative Propagation Along Quantum Trajectories, pp. 235--253

\bibitem{sanz:cpl:2008}
A.S. Sanz, S.~Miret-Art\'es, Chem. Phys. Lett. \textbf{458}, 239 (2008)

\bibitem{leacock:PRL:1983}
R.A. Leacock, M.J. Padgett, Phys. Rev. Lett. \textbf{50}, 3 (1983)

\bibitem{leacock:PRD:1983}
R.A. Leacock, M.J. Padgett, Phys. Rev. D \textbf{28}, 2491 (1983)

\bibitem{jordan:ZPhys:1926}
P.~Jordan, Z. Phys. \textbf{38}, 513 (1926)

\bibitem{jordan:ZPhys-1:1927}
P.~Jordan, Z. Phys. \textbf{40}, 809 (1927)

\bibitem{jordan:ZPhys-2:1927}
P.~Jordan, Z. Phys. \textbf{44}, 1 (1927)

\bibitem{dirac:PRSLA:1927}
P.A.M. Dirac, Proc. R. Soc. Lond. \textbf{113A}, 621 (1927)

\bibitem{dirac:PZSov:1933}
P.A.M. Dirac, Phys. Z. Sowjetunion \textbf{3}, 64 (1933)

\bibitem{dirac:RMP:1945}
P.A.M. Dirac, Rev. Mod. Phys. \textbf{17}, 195 (1945)

\bibitem{jeffreys:PLMS:1925}
K.~Gottfried, Proc. Lond. Math. Soc. \textbf{23}, 428 (1925)

\bibitem{wentzel:ZPhys:1926}
G.~Wentzel, Z. Phys. \textbf{38}, 518 (1926)

\bibitem{kramers:ZPhys:1926}
H.A. Kramers, Z. Phys. \textbf{39}, 828 (1926)

\bibitem{brillouin:ComptRend:1926}
L.~Brillouin, Compt. Rend. \textbf{183}, 24 (1926)

\bibitem{floyd:PRD-1:1982}
E.R. Floyd, Phys. Rev. D \textbf{25}, 1547 (1982)

\bibitem{floyd:PRD-2:1982}
E.R. Floyd, Phys. Rev. D \textbf{26}, 1339 (1982)

\bibitem{floyd:PRD:1984}
E.R. Floyd, Phys. Rev. D \textbf{29}, 1842 (1984)

\bibitem{floyd:PRD:1986}
E.R. Floyd, Phys. Rev. D \textbf{34}, 3246 (1986)

\bibitem{floyd:FPL:1996}
E.R. Floyd, Found. Phys. Lett. \textbf{9}, 489 (1996)

\bibitem{floyd:FPL:2000}
E.R. Floyd, Found. Phys. Lett. \textbf{13}, 235 (2000)

\bibitem{faraggi:PLA:1998}
A.E. Faraggi, M.~Matone, Phys. Lett. A \textbf{249}, 180 (1998)

\bibitem{faraggi:PLB:1998}
A.E. Faraggi, M.~Matone, Phys. Lett. B \textbf{437}, 369 (1998)

\bibitem{faraggi:PLB-1:1999}
A.E. Faraggi, M.~Matone, Phys. Lett. B \textbf{445}, 77 (1999)

\bibitem{faraggi:PLB-2:1999}
A.E. Faraggi, M.~Matone, Phys. Lett. B \textbf{450}, 34 (1999)

\bibitem{faraggi:IJMPA:2000}
A.E. Faraggi, M.~Matone, Int. J. Mod. Phys. A \textbf{15}, 1869 (2000)

\bibitem{john:FPL:2002}
M.V. John, Found. Phys. Lett. \textbf{15}, 329 (2002)

\bibitem{john:AnnPhys:2009}
M.V. John, Ann. Phys. \textbf{324}, 220 (2009)

\bibitem{john:AnnPhys:2010}
M.V. John, Ann. Phys. \textbf{325}, 2132 (2010)

\bibitem{john:FoundPhys:2013}
M.V. John, K.~Mathew, Found. Phys. \textbf{43}, 859 (2013)

\bibitem{fring:PRA:2013}
S.~Dey, A.~Fring, Phys. Rev. A \textbf{88}, 022116 (2013)

\bibitem{yang:AnnPhys-1:2005}
C.D. Yang, Ann. Phys. \textbf{319}, 399 (2005)

\bibitem{yang:AnnPhys-2:2005}
C.D. Yang, Ann. Phys. \textbf{319}, 445 (2005)

\bibitem{yang:IJQC:2006}
C.D. Yang, Int. J. Quantum Chem. \textbf{106}, 1620 (2006)

\bibitem{yang:CSF-1:2006}
C.D. Yang, Chaos Soliton. Fract. \textbf{30}, 41 (2006)

\bibitem{yang:CSF-2:2006}
C.D. Yang, Chaos Soliton. Fract. \textbf{30}, 342 (2006)

\bibitem{yang:AnnPhys:2006}
C.D. Yang, Ann. Phys. \textbf{321}, 2876 (2006)

\bibitem{yang:CSF-1:2007}
C.D. Yang, Chaos Soliton. Fract. \textbf{32}, 274 (2007)

\bibitem{yang:CSF-2:2007}
C.D. Yang, Chaos Soliton. Fract. \textbf{32}, 312 (2007)

\bibitem{yang:CSF-3:2007}
C.D. Yang, Chaos Soliton. Fract. \textbf{33}, 1073 (2007)

\bibitem{yang:CSF-4:2007}
C.D. Yang, C.H. Wei, Chaos Soliton. Fract. \textbf{33}, 118 (2007)

\bibitem{chou:PRA:2007}
C.C. Chou, R.E. Wyatt, Phys. Rev. A \textbf{76}, 012115 (2007)

\bibitem{chou:PRA:2008}
C.C. Chou, R.E. Wyatt, Phys. Rev. A \textbf{78}, 044101 (2007)

\bibitem{chou:JCP-1:2008}
C.C. Chou, R.E. Wyatt, J. Chem. Phys. \textbf{128}, 154106 (2008)

\bibitem{chou:JCP-2:2008}
C.C. Chou, R.E. Wyatt, J. Chem. Phys. \textbf{128}, 234106 (2008)

\bibitem{chou:JCP-3:2008}
C.C. Chou, R.E. Wyatt, J. Chem. Phys. \textbf{129}, 124113 (2008)

\bibitem{chou:PLA:2009}
C.C. Chou, R.E. Wyatt, J. Chem. Phys. \textbf{373}, 1811 (2009)

\bibitem{chou:prl:2009}
C.C. Chou, A.S. Sanz, S.~Miret-Art\'es, R.E. Wyatt, Phys. Rev. Lett.
  \textbf{102}, 250401 (2009)

\bibitem{chou:annphys:2010}
C.C. Chou, A.S. Sanz, S.~Miret-Art\'es, R.E. Wyatt, Ann. Phys. \textbf{325},
  2193 (2010)

\bibitem{chou:JCP:2010}
C.C. Chou, R.E. Wyatt, J. Chem. Phys. \textbf{132}, 134102 (2010)

\bibitem{chou:PLA:2010}
C.C. Chou, R.E. Wyatt, J. Chem. Phys. \textbf{374}, 2608 (2010)

\bibitem{tannor:JCP:2006}
Y.~Goldfarb, I.~Degani, D.J. Tannor, J. Chem. Phys. \textbf{125}, 231103 (2006)

\bibitem{sanz:JCP:2007}
A.S. Sanz, S.~Miret-Art\'es, J. Chem. Phys. \textbf{127}, 197101 (2007)

\bibitem{tannor:JCP-1:2007}
Y.~Goldfarb, I.~Degani, D.J. Tannor, J. Chem. Phys. \textbf{127}, 197102 (2007)

\bibitem{tannor:JCP-2:2007}
Y.~Goldfarb, D.J. Tannor, J. Chem. Phys. \textbf{127}, 161101 (2007)

\bibitem{tannor:JPCA:2007}
Y.~Goldfarb, J.~Schiff, D.J. Tannor, J. Chem. Phys. \textbf{111}, 10416 (2007)

\bibitem{tannor:CP:2007}
Y.~Goldfarb, I.~Degani, D.J. Tannor, Chem. Phys. \textbf{338}, 106 (2007)

\bibitem{tannor:JCP:2014}
N.~Zamstein, D.J. Tannor, J. Chem. Phys. \textbf{140}, 041105 (2014)

\bibitem{tannor-bk}
D.J. Tannor, \emph{Introduction to Quantum Mechanics: A Time-Dependent
  Perspective} (University Science Books, Sausalito, CA, 2006)

\bibitem{trahan:jcp:2003}
C.J. Trahan, K.~Hughes, R.E. Wyatt, J. Chem. Phys. \textbf{118}, 9911 (2003)

\bibitem{chou:JCP:2006}
C.C. Chou, R.E. Wyatt, J. Chem. Phys. \textbf{125}, 174103 (2006)

\bibitem{chou:PRE:2006}
C.C. Chou, R.E. Wyatt, Phys. Rev. E \textbf{74}, 066702 (2006)

\bibitem{chou:IJQC:2008}
C.C. Chou, R.E. Wyatt, Int. J. Quantum Chem. \textbf{108}, 238 (2008)

\bibitem{makri:MolPhys:2005}
J.~Liu, N.~Makri, Mol. Phys. \textbf{103}, 1083 (2005)

\bibitem{garashchuk:TheorChemAcc:2012}
S.~Garashchuk, Theor. Chem. Acc. \textbf{131}, 1083 (2012)

\bibitem{o.einstein}
A.I.M. (Ed.), \emph{\emph{Sixty-Two Years of Uncertainty}} (New York: Plenum
  Press., New York, 1990), chapter `Nicht Sein Kann Was Nicht Sein Darf,' or
  the Prehistory of EPR, 1909-1935: Einstein's Early Worries about the Quantum
  Mechanics of Composite Systems. from {Howard}, D. (pp. 61-111).

\bibitem{o.Bell1964}
J.S. Bell, Physics \textbf{1}, 195 (1964)

\bibitem{o.aspect1982}
A.~Aspect, P.~Grangier, G.~Roger, Phys. Rev. Lett. \textbf{49 (2)}, 91 (1982)

\bibitem{o.Linbla}
G.~Lindblad, Commun. Math. Phys. \textbf{48}, 119  (1976)

\bibitem{openQsystems}
H.P. Breuer, F.~Petruccione, \emph{The theory of open quantum systems} (Oxford
  university press, 2002)

\bibitem{o.durr1992equilibrium}
D.~D{\"{u}}rr, S.~Goldstein, N.~Zangh\`i, J. Stat. Phys. \textbf{67}, 843
  (1992)

\bibitem{om.bohm1952b}
D.~Bohm, Phys. Rev. \textbf{85}, 180 (1952)

\bibitem{om.quantumpower}
D.H. Kobe, J. Phys. A \textbf{40}, 5155 (2007)

\bibitem{o.brown1999}
H.R. Brown, E.~Sj\"{o}qvist, G.~Bacciagaluppi, Phys. Lett. A \textbf{251}, 229
  (1999)

\bibitem{o.nature2012}
M.~F.Pusey, J.~Barrettt, T.~Rudolph, Nature \textbf{8}, 475 (2012)

\bibitem{o.Albert1996}
D.~Albert, in \emph{Bohmian mechanics and quantum theory: an appraisal}, edited
  by J.~Cushing, A.~Fine, S.~Goldstein (Kluwer, 1996), chap. Elementary quantum
  metaphysics, pp. 277--284

\bibitem{o.law1996}
D.~D\"urr, S.~Goldstein, N.~Zangh\`i, in \emph{Experimental Metaphysics --
  Quantum Mechanical Studies in Honor of Abner Shimony}, edited by R.S. Cohen,
  M.~Horne, J.~Stachel (Kluwer Academic Publishers, Dordrecht, 1997), Vol.~I,
  chap. Bohmian mechanics and the meaning of the wave function, pp. 25--38

\bibitem{o.debroglie1923}
L.~{de Broglie}, Ph.D. thesis, University of Paris (1924)

\bibitem{o.impossibility_proofs}
J.~{von Neumann}, \emph{Mathematische Grundlagen der Quantenmechanik} (Springer
  Verlag, Berlin, 1932), english translation by: R.T. Beyer, Mathematical
  Foundations of Quantum Mechanics (Princeton University Press, Princeton,
  1955)

\bibitem{waerden}
B.L. van~der Waerden, \emph{Sources of Quantum Mechanics} (Dover Publications,
  New York, 1968)

\end{thebibliography}

\end{document}